\documentclass[twocolumn]{aastex61}

\usepackage{xspace,graphics,graphicx,amsmath}
\usepackage{natbib,amsbsy}
\usepackage{appendix}
\def \lya {Ly$\alpha$ }
\def \mkms {{\rm \; km\;s^{-1}}}

\def \zprimus {$z^{\rm PR}$}
\def \mrperp {R_{\perp}}

\newcommand{\msun}{~M_{\odot}}
\newcommand{\overbar}[1]{\mkern 1.5mu\overline{\mkern-1.5mu#1\mkern-1.5mu}\mkern 1.5mu}
\newcommand{\refPaperII}{Paper II}

\shorttitle{Galaxies Probing Galaxies in PRIMUS -- I.}
\shortauthors{Rubin et al.}
\accepted{to ApJ September 22, 2017}

\begin{document}

\title{Galaxies Probing Galaxies in PRIMUS -- I. Sample, Spectroscopy,
  and Characteristics of the $\lowercase{z}\sim0.5$ \ion{Mg}{2}-Absorbing Circumgalactic Medium}
\author{Kate H. R. Rubin}
\affiliation{San Diego State University, Department of Astronomy,
  San Diego, CA 92182, USA} 

\author{Aleksandar M. Diamond-Stanic}
\affiliation{Bates College, Department of Physics \& Astronomy, 44
Campus Ave, Carnegie Science Hall, Lewiston, ME 04240, USA}, 

\author{Alison L. Coil}
\affiliation{Center for Astrophysics and Space Sciences,
  Department of Physics, University of California, San Diego, 9500 Gilman Drive, La Jolla, CA 92093, USA}

\author{Neil H. M. Crighton}
\affiliation{Australian Competition and Consumer Commission,
  Melbourne, VIC 3001; Center for Astrophysics and Supercomputing,
  Swinburne University of Technology, Hawthorn, VIC 3122, Australia}

\author{John Moustakas}
\affiliation{Department of Physics and Astronomy, Siena College, 515
Loudon Road, Loudonville, NY 12211, USA}

\correspondingauthor{Kate H. R. Rubin}
\email{krubin@sdsu.edu}

\begin{abstract}
Spectroscopy of background QSO sightlines passing close to foreground
galaxies is a potent technique for studying the circumgalactic medium (CGM).
QSOs are effectively point sources, however, limiting their
potential to constrain the size of circumgalactic gaseous structures.
Here we present the first large Keck/LRIS and VLT/FORS2 spectroscopic survey of 
bright ($B_{\rm AB}<22.3$) background galaxies whose lines of sight
probe \ion{Mg}{2}
$\lambda\lambda2796,2803$ absorption from the CGM around close projected
foreground galaxies at transverse distances $10~\mathrm{kpc}<
R_{\perp} <150~\mathrm{kpc}$.
Our sample of 72 projected pairs, 
drawn from the PRIsm MUlti-object
Survey (PRIMUS), 
includes 48 background galaxies which do not host bright AGN, and 
both star-forming and quiescent foreground
galaxies with stellar masses $9.0<\log M_{*}/M_{\odot}<11.2$ at redshifts 
$0.35<z_{\rm f/g}<0.8$.  
We detect \ion{Mg}{2}
absorption
associated with these foreground
galaxies with equivalent widths
$0.25~\mathrm{\AA}<W_{2796}<2.6~\mathrm{\AA}$ at $>2\sigma$ significance
in 20 individual background sightlines passing within $R_{\perp}<50$ kpc, and
place $2\sigma$ upper limits on $W_{2796}$ of $\lesssim0.5~\mathrm{\AA}$ in an
additional 11 close sightlines.  
Within $R_{\perp}<50$ kpc, $W_{2796}$ is anticorrelated
with $R_{\perp}$, consistent with analyses of 
\ion{Mg}{2} absorption detected along background QSO
sightlines.  Subsamples of these foreground hosts 
divided at $\log M_*/M_{\odot}=9.9$ exhibit statistically
inconsistent $W_{2796}$ distributions at $30~\mathrm{kpc} <
  R_{\perp} < 50~\mathrm{kpc}$, with the higher-$M_*$ galaxies
yielding a larger median $W_{2796}$ by $0.9~\mathrm{\AA}$.
Finally, we demonstrate that foreground galaxies with similar stellar masses exhibit
the same median $W_{2796}$ at a given $R_{\perp}$ to within $<0.2~\mathrm{\AA}$ toward both
background galaxies and toward QSO sightlines drawn from the literature.
Analysis of  these datasets 
constraining the spatial coherence scale of  circumgalactic
\ion{Mg}{2} absorption is presented in a companion paper.
\end{abstract}

\section{Introduction}\label{sec.intro}


QSO absorption line spectroscopy has for nearly half a century been
our principal and most powerful tool for the study of diffuse
baryons. From measurement of the incidence and metallicity of material
in the most rarefied intergalactic regions 
\citep{Simcoe2004,Lehner2007,DanforthShull2008}
to detailed constraints on the kinematics, ionization state, metal
content, and mass of highly metal-enriched superwind ejecta close to
massive galaxies 
\citep[e.g.,][]{Tripp2011},
spectroscopy of bright QSOs has revealed the nature of ``dark''
gaseous 
material in virtually all galactic and intergalactic environments. 
Experiments which search the sky for QSO sightlines close in
projection to foreground galaxies have been designed to assess the
properties of gas in the 
circumgalactic medium (CGM) surrounding a wide variety of galaxy
hosts, from sub-luminous dwarfs 
\citep{Prochaska2011,Burchett2015,Bordoloi2014,Rubin2015}
to luminous red galaxies and massive QSO hosts 
\citep{QPQ5,Farina2014,Zhu2014,Johnson2015}.
The assembly of large samples of such background QSO - foreground
galaxy pairs 
provides a statistical picture of the absorption 
exhibited by the targeted foreground galaxy environment 
\citep[e.g.,][]{Churchill2000,Chen2001,Adelberger2005,Chen2010,Werk2013}.

At $z\lesssim2$, the vast majority of QSO-galaxy pair studies have
focused primarily 
(and in many cases, exclusively) on measurement of the absorption
strength of the 
\ion{Mg}{2} $\lambda \lambda 2796, 2803$ doublet. This transition, 
accessible from the ground at $z\gtrsim0.2$, arises from cool,
photoionized gas 
at a temperature $T \sim 10^4$ K 
\citep{BergeronStasinska1986},
and is sensitive enough to yield large equivalent widths 
($W_{2796} > 0.1$ \AA) in sightlines having neutral hydrogen column
densities as low as $N$(\ion{H}{1}) $\gtrsim 10^{16.5} \rm cm^{−2}$ 
\citep[e.g.,][]{Lehner2013}.
Such studies have demonstrated that 
$\sim L^*$ galaxies at $z\lesssim0.5$ are enveloped by
\ion{Mg}{2}-absorbing 
gas extending to radii $R_{\rm MgII} \sim 100$ kpc \citep{Chen2010}.
Within this region, \ion{Mg}{2} absorbers having 
$W_{2796} > 0.1$ \AA\ arise with an incidence
$C_f \sim 80\%$. 
Recent work has suggested a higher incidence of stronger absorption
around 
galaxies with higher stellar masses 
\citep{Chen2010,Churchill2013}.
Finally, experiments leveraging quantitative morphological
measurements 
of the targeted foreground systems suggest the strongest \ion{Mg}{2} 
absorbers occur toward QSOs located close to (i.e., within $\Phi
\lesssim 45^{\circ}$ of)
the minor axis of the host galaxy 
\citep{Bouche2012,Kacprzak2012,Nielsen2015}.

While such studies have proven very fruitful, they have nevertheless
relied 
exclusively on a single tool: the $10^{-3} - 10^{-2}$ pc beam of
UV-bright 
continuum emitted by the accretion disks powering luminous QSOs
\citep{SS1973,Frank2007}.
Due to the extremely small scale of this beam, QSO spectroscopy cannot 
directly distinguish between compact clouds with volumes less than a cubic
parsec and elongated filaments stretching over many kiloparsecs -- and
hence cannot be used to bring such geometrical constraints to bear on
the physical origin of this material.

There is ample evidence demonstrating that cool gas traced by
\ion{Mg}{2} absorption is launched away from star-forming
regions in galactic winds
\citep{Weiner2009,RubinTKRS2009,Martin2012,Rubin2013}; however,
the relation between this material and the \ion{Mg}{2}-absorbing
structures detected at 
projected distances
$\mrperp>10$ kpc has not been established.
Moreover, there is a strong theoretical expectation of the presence of
an additional, hotter gas phase ($T\sim10^6$ K) filling the same extended halos,
fed by the virial shock front formed by accreting material \citep{ReesOstriker1977,Keres2005}.
If this hot phase is indeed ubiquitous, a cool cloud
passing through it will be destroyed on a timescale which is
approximately  linearly
dependent on its size \citep{Schaye2007,Crighton2014,McCourt2015}.
Constraints on the scale of this material therefore in principle also constrain
the structure lifetime as a function of its relative velocity.  Such
estimates may be used to test the viability of several presumptive
origins for the cool CGM, including the cool winds described above
\citep[e.g.,][]{Bond2001,Steidel2010,Bouche2012},
infalling cold streams \citep[e.g.][]{Keres2005,Kacprzak2012,Bouche2013,Crighton2013}, or condensation of the hot
halo material via thermal instability \citep{MB2004,Binney2009}.

Unlike surveys relying purely on QSO sightlines,
absorption spectroscopy toward background probes having a wide range
in the projected spatial extent of their UV continuum emission can
reveal the small-scale structure of the cool CGM. 
Galaxies, with typical sizes $> \rm kpc^2$, are now being used as
bright background sources in a growing number of studies 
\citep{Adelberger2005,Barger2008,Rubin2010,Steidel2010,Lee2014,Bordoloi2011,ADS2015,CookeOMeara2015,Lee2016}.
Spatially-resolved spectroscopy of such extended
background beams 
probe variations in absorption strength and kinematics along multiple
independent 
sightlines. As we will demonstrate in the second paper of
this series (Rubin et al.\ 2017, in prep; hereafter cited as \refPaperII), even if
it is not possible to resolve the beam, an analysis comparing the
properties of absorbers observed along integrated spectra of
background 
galaxies and QSOs provides a direct constraint on the coherence scale
of the 
cool absorption.

To facilitate these experiments, large spectroscopic galaxy redshift
surveys 
may be searched for projected pairs of systems in analogy to targeted 
QSO-galaxy pair searches. While the spectral signal-to-noise achieved
in the vast majority of redshift surveys is insufficient to assess
foreground absorber properties along individual background galaxy 
sightlines, many of the studies listed above have coadded background 
galaxy spectra in the rest-frame of the foreground system in each pair
for constraints on the mean foreground absorption equivalent width 
\citep[e.g.,][]{Steidel2010,Bordoloi2011}.
Only a handful of studies have achieved the S/N required to securely 
detect foreground absorption in individual sightlines, and each of
these works report on just one or two projected pairs 
\citep{Adelberger2005,Barger2008,Rubin2010,ADS2015,CookeOMeara2015}.
Indeed, the sizes of these samples have been severely limited by the
scarcity of galaxies which are both sufficiently bright to obtain S/N
$\gtrsim5~\rm \AA^{-1}$ spectroscopy in the near-UV, and which are
located within $\lesssim100$ projected kpc of a foreground galaxy whose
redshift is known \emph{a priori}. 
In principle, however, a redshift survey covering a large sky volume
at high density can yield significant numbers of such pairs.
The background sightlines may
then be reobserved with UV-sensitive instrumentation to achieve
high-S/N constraints on foreground absorbers, 
the vast majority of which arise due to \ion{H}{1} and metal-line transitions 
at rest wavelengths blueward of 3000 \AA.

Such a high-volume, high-density redshift survey is now available in
 PRIMUS, the PRIsm MUlti-object Survey 
\citep{Coil2011primus}.
Here we present high-S/N Keck/LRIS and VLT/FORS2 
rest-frame near-UV spectroscopy of 72
projected pairs 
of galaxies having impact parameters 
$R_{\perp} < 150$ kpc identified in the PRIMUS redshift catalog. 
Our galaxy pair sample, spanning the redshift range $0.4 \lesssim z
\lesssim 1.0$, includes 49 pairs with projected separations 
$R_{\perp} < 50$ kpc, thoroughly sampling the ``inner'' CGM which 
typically gives rise to strong \ion{Mg}{2} absorbers having 
$W_{2796} > 0.3$ \AA\ 
\citep[e.g.,][]{Chen2010}.
The foreground galaxies in our sample span the star-forming 
sequence to a stellar mass limit $\gtrsim 10^9 \msun$, and at high
stellar masses 
($M_* \gtrsim 10^{10.5} \msun$) include both star-forming and quiescent
systems. These data permit the first investigation of the absorption
strength of the 
\ion{Mg}{2} $\lambda 2796$ transition to a limiting $W_{2796} \gtrsim
0.5$ \AA\ associated with foreground galaxy halos 
in a statistical sample of individual background galaxy sightlines. 
We explore the dependence of this $W_{2796}$ on 
intrinsic galaxy properties (i.e., star formation rate, $M_*$) as a
function of $R_{\perp}$, and 
compare these measurements to those 
drawn from 
QSO-galaxy pair studies 
in the literature. 
In a companion paper (\refPaperII), we take advantage of all of these data to develop 
direct constraints on the spatial extent of the cool material 
giving rise to the observed \ion{Mg}{2} absorption, and use this
analysis to address the lifetime and fate of these structures.

We describe our sample selection in Section~\ref{sec.sample}, and
describe our observations and data reduction
procedures in Section~\ref{sec.obs}. 
Section~\ref{sec.analysis} details our methods of redshift 
estimation and absorption line analysis. We present salient 
properties of our foreground and background galaxy samples in 
Sections~\ref{sec.fg} and \ref{sec.bg}. Section~\ref{sec.results_cgm} describes our results on the relationship 
between the \ion{Mg}{2} absorption strength in the CGM and 
the intrinsic host galaxy properties, and compares these findings to
the results of previous QSO-galaxy and galaxy-galaxy pair studies. 
We present a brief summary in Section~\ref{sec.summary}.
Throughout, we adopt a $\Lambda$CDM cosmology with $H_0 = 70~\rm
km~s^{-1}~Mpc^{-1}$, $\Omega_{\rm M} = 0.3$, and  $\Omega_{\Lambda} =
0.7$.  
Magnitudes are quoted in the AB system.

\section{Sample Selection}\label{sec.sample}

\subsection{PRIMUS Galaxy Pairs}

Our galaxy pair sample is drawn from PRIMUS, a spectroscopic survey of galaxies with redshifts in the
range $0 < z < 1.2$ \citep{Coil2011primus,Cool2013}.  Using the IMACS
instrument on the Magellan Baade Telescope
\citep{BigelowDressler2003}, PRIMUS obtained redshifts for
$\sim120,000$ galaxies over 9.1 deg$^2$ to a magnitude limit $i \sim
23$.  
The PRIMUS sample is distributed over seven ``science'' fields
selected to have existing ancillary multi-wavelength imaging: 
the Chandra Deep Field South-SWIRE field (CDFS-SWIRE;
\citealt{Giacconi2001}), the DEEP2 fields at 23hr and 02hr \citep{Davis2003},
the COSMOS field \citep{Ilbert2009}, the {\it XMM}-Large Scale Structure Survey
field (XMM-LSS; \citealt{Pierre2004}),  
the European Large Area ISO Survey-South 1
field \citep{Oliver2000}, and the Deep Lens Survey F5 field \citep{Wittman2002}.

We used four main criteria to select our primary sample of projected 
galaxy pairs from this parent catalog for follow-up spectroscopy in the near-UV:

\begin{enumerate}
\item  First, we considered all galaxies having a PRIMUS
redshift \zprimus\ $\ge 0.35$ with high confidence (i.e., a redshift
confidence flag $Q=3$ or $4$) to ensure spectral
coverage of \ion{Mg}{2} $\lambda \lambda 2796, 2803$ absorption 
within the wavelength range at which LRIS and FORS2 have optimum
sensitivity ($\lambda \gtrsim 3700$ \AA).  The low-dispersion
prism used to carry out the PRIMUS survey yields a redshift accuracy
of $\sigma_z / (1+z) = 0.0051$ for objects assigned these confidence flags, with an outlier rate of
objects having $\Delta_z/(1+z) > 0.03$ of $8\%$ \citep{Coil2011primus}.

\item We then selected pairs of these objects with projected separations 
$\mrperp \le 50$ kpc at the \zprimus\ value of the foreground (f/g)
galaxy, and further required that the redshift offset between the
f/g and background (b/g) galaxies  satisfies $z_{\rm b/g}^{\rm PR} - z_{\rm
  f/g}^{\rm PR} \ge 0.02$ (corresponding to a velocity difference $\gtrsim 3500\mkms)$.

\item We required that the b/g galaxy have an apparent $B$-band
  magnitude sufficient to yield a $3\sigma$ $W_{2796}$ detection limit
  of 0.5 \AA\ at $z_{\rm f/g}^{\rm PR}$ in an exposure time $<2.5$
  hours with LRIS or FORS2.  All b/g galaxies satisfying this
  criterion have $B_{\rm AB} < 22.3$.

\item We finally demanded that each f/g galaxy have an apparent
  $B$-band magnitude sufficient to yield a $3\sigma$ $W_{2796}$
  detection limit of 1.5 \AA\ toward its own stellar continuum within
  2.5 hours of exposure time.  This
  corresponds to an approximate magnitude limit of $B_{\rm AB}\lesssim23.3$,
  and permits both higher-resolution spectroscopic confirmation of $z_{\rm f/g}^{\rm PR}$
  as well as detailed analysis of ``down-the-barrel'' absorption for
  comparison with halo gas kinematics observed toward the b/g galaxy.
\end{enumerate}

In the five PRIMUS science fields that we targeted in this study (the
two DEEP2 fields, the XMM-LSS field, the COSMOS field, and the
CDFS-SWIRE field), 
there are 78 pairs of galaxies that satisfy these criteria.
We selected 59 pairs from among this
sample to observe in the  rest-frame near-UV, prioritizing brighter pairs, 
those having $\delta z^{\rm PR} = z_{\rm b/g}^{\rm PR} - z_{\rm f/g}^{\rm PR} \ge
0.1$, and pairs which are close on the sky such that they 
could be observed simultaneously in multislit
mode.  These objects are listed in
Table~\ref{tab.pairinfo}, along with their redshifts, apparent
magnitudes, and angular separations, and are indicated with three-digit
identification numbers.  
Due to the occasional underestimation of the 
uncertainty in the redshifts determined from the low-dispersion PRIMUS discovery spectra, five of
these pairs having $\delta z^{\rm PR} \approx 0.03-0.13$ were
identified as physical (i.e., not projected) in our followup
observations,  with $\lvert z_{\rm b/g} - z_{\rm f/g} \rvert <
  0.003$ (where $z_{\rm b/g}$ and $z_{\rm f/g}$ are galaxy
  redshifts estimated from
  our LRIS and FORS2 spectroscopy as described in \S\ref{sec.redshifts}).
Two additional pairs were 
found to include stellar (Galactic) sources.

We also obtained spectra of 32 
serendipitous pairs, most
of which have
larger (50 kpc $< \mrperp < 150$ kpc) projected separations.  These
pairs were targeted for their
exceptionally bright b/g objects, or where they could be included on
the same slitmask with a primary (close) pair target.  These objects
are indicated in Table~\ref{tab.pairinfo} with identification numbers greater than 1000.

\subsection{QSO-Galaxy Comparison Pairs}\label{sec.sample_qsocomp}

 In the analysis to follow, 
we also draw on published
samples of galaxies for which the circumgalactic
\ion{Mg}{2} absorption has been well-characterized using background
QSO sightlines.  We select these QSO-galaxy pair measurements based on
the experimental design of the work in which they are reported.  
That is, we require these samples to be designed using a methodology
as similar as possible to that of our PRIMUS pairs experiment.
Because our PRIMUS b/g sightlines are selected without prior knowledge
of f/g \ion{Mg}{2} absorption, the selected QSO-galaxy samples must
also be designed without such prior knowledge.   
Including absorption-selected systems would
tend to yield higher overall $W_{2796}$ profiles, and so would introduce
a bias into our comparison of these datasets.
We choose to include measurements from the two largest available
QSO-galaxy pair studies  with the appropriate experimental design: the
69 pairs probing ``isolated'' galaxies studied in \citet{Chen2010},
and the 39 pairs composing the COS-Halos sample 
\citep{Werk2013}.  Both of these works focus on the gaseous environments of
$\sim L^*$ galaxies at low redshift, and hence offer a comparison
sample with quite similar f/g galaxy properties to those selected from
PRIMUS (see \S\ref{sec.fg} for further detail).


\section{Spectroscopic Observations and Supplementary Data}\label{sec.obs}

Our follow-up spectroscopy was carried out with two instruments: the
Low Resolution Imaging Spectrometer (LRIS) on Keck 1
\citep{Cohen1994}, and the FOcal Reducer/low dispersion Spectrograph 2
(FORS2) on VLT-UT1 \citep{Appenzeller1998}.  Primary pair targets requiring
exposure times longer than 1 hr were observed in multislit mode in
most cases.  We chose longslit mode for the remaining targets.  

\subsection{Keck/LRIS Spectroscopy}

Keck/LRIS observations were carried out 
during three observing runs on
2011 Oct 1 UT, 2012 Jan 20-21 UT, and 2012 Dec 13-15 UT.
Seeing conditions on these dates varied over the range FWHM $\sim 0.5-1.5\arcsec$.
A slit width of $1\arcsec$ was used for both  multislit 
and longslit
observations.  We used the $600~\rm l~mm^{-1}$ grism blazed at 4000
\AA\ on the blue side and the $600~\rm l~mm^{-1}$ grating blazed at
7500 \AA\ on the red side with the D560 dichroic, obtaining full
wavelength coverage between $\sim3200$ \AA\ and $\sim8450$ \AA.
The velocity resolution of the spectra is FWHM $\sim160\mkms$ at
8000 \AA, degrading to 
FWHM $\sim 345\mkms$ at 3500 \AA.

All fields observed in multislit mode are listed in
Table~\ref{tab.maskobs}, along with the ID of the pairs on each slitmask
and the date of observation.  Total exposure times ranged between 0.4
and 3.2 hrs.  Individual exposures were typically $\sim1200-1800$ sec 
in length on the blue side and $\sim400-900$ sec in length on the
red.  Pairs observed in longslit mode are listed in
Table~\ref{tab.longobs}, with integration times ranging from 0.4 to
1.8 hrs.  


\subsection{VLT/FORS2 Spectroscopy}

Our VLT/FORS2 program (with ESO program IDs 088.A-0529A and 090.A-0485A)
was carried out in visitor mode 
over three nights on 2011 Nov 25 UT and
2012 Nov 14-15 UT.  Seeing conditions were excellent for two of these
nights (FWHM $\sim0.5-0.7\arcsec$) and varied between $1$ and
$2\arcsec$ on 2012 Nov 14.  A slit width of $1\arcsec$ was chosen in
both longslit and multislit (MXU) mode.  We observed with the blue-sensitive
E2V CCDs, using the GRIS$\_$1200B grism to obtain coverage between 
3670 and 5120 \AA.  We additionally observed each mask and longslit
pointing with one of two red-sensitive grisms: GRIS$\_$600V or GRIS$\_$600RI.
These latter setups cover from 4530 to 7510 \AA\ and from 5150
to 8470 \AA, respectively, providing spectroscopy of nebular emission
lines and Balmer absorption at rest-frame wavelengths $3700-5010$ \AA.
The GRIS$\_$1200B grism yields a velocity resolution FWHM $\sim155\mkms$
near 5000 \AA\ and $\sim185\mkms$ at 3670 \AA, while both of the red
grisms provide a median resolution FWHM $\sim250\mkms$.

The fields observed in MXU mode are listed toward the bottom of
Table~\ref{tab.maskobs}.  Exposure times for spectra taken with the
GRIS$\_$1200B grism are listed in the 5th column, and range between
0.7 and 3 hrs.  The length of exposures taken with the red grisms in place are
listed in the 6th column, and are all between 15 and 30 min.  The
three pairs observed with the FORS2 longslit are included in
Table~\ref{tab.longobs}, and were observed for 0.4-0.9 hrs and 15-30
min with the GRIS$\_$1200B and GRIS$\_$600V setups, respectively.

Both LRIS and FORS2 data were reduced using the XIDL
LowRedux\footnote{http://www.ucolick.org/${\sim}$xavier/LowRedux/}
data reduction pipeline.  The pipeline includes bias subtraction,
flat-fielding, slit finding, wavelength calibration, object
identification, sky subtraction, and relative flux calibration.
Wavelength calibrations were adjusted for flexure by applying an
offset estimated from the cross-correlation of the sky spectrum with a
sky spectral template.  Wavelengths for the final, coadded
one-dimensional spectra are in vacuum and have been corrected to the
heliocentric frame.

\begin{figure*}
\begin{center}
\includegraphics[angle=0,width=7in,trim=50 100 70 100,clip=]{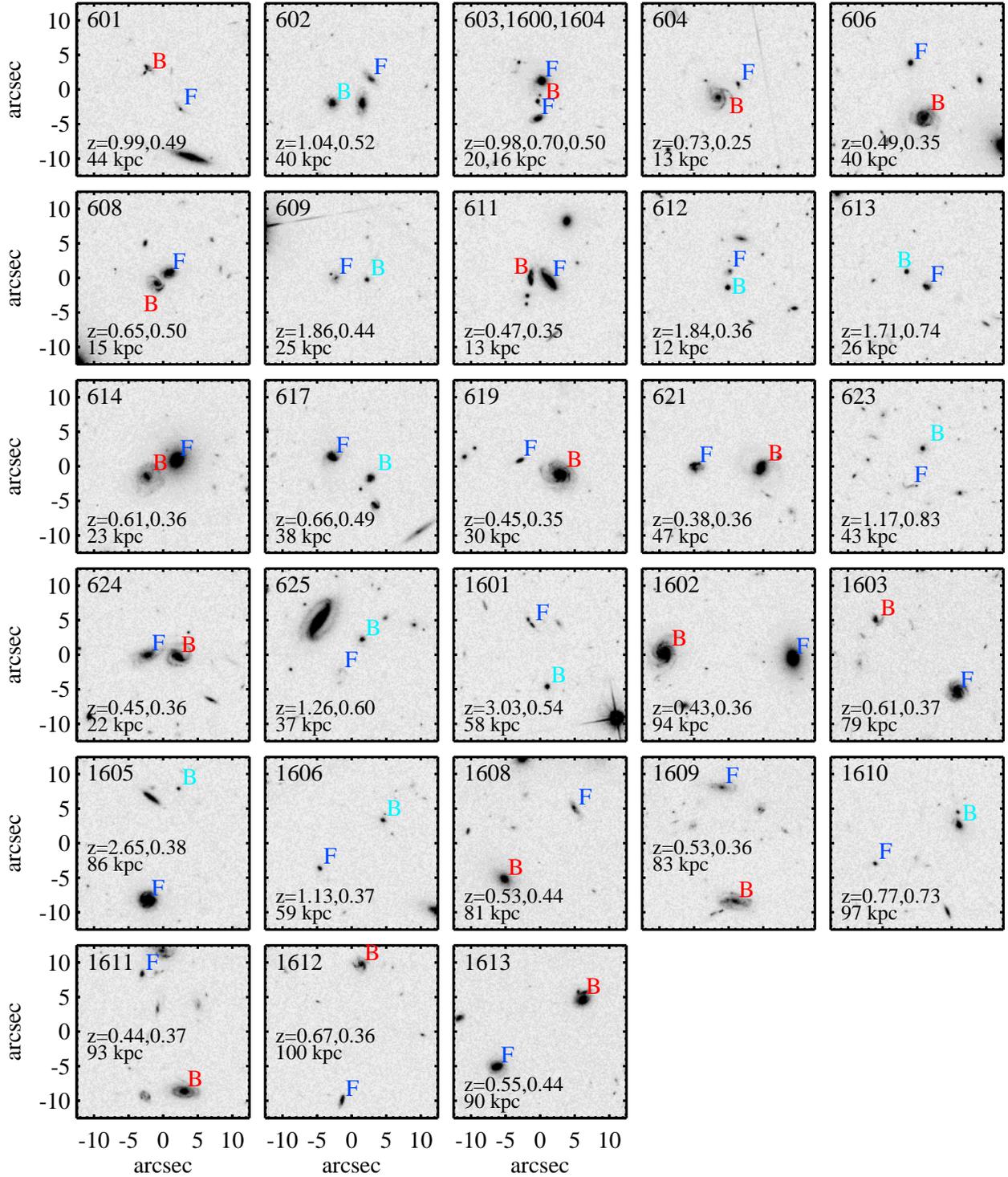}
\caption{{\it HST}/ACS F814W-band imaging of projected pairs of
  galaxies for which we have obtained deep near-UV spectroscopy in the
  COSMOS field \citep{Scoville2007}.  Each panel is $25\arcsec \times25\arcsec$.  Background objects are
  indicated with a cyan ``B'' if they exhibit broad-line AGN emission
  and are marked with a red ``B'' in the remaining cases.  All
  foreground objects are indicated with a blue ``F''.  The images are
  labeled with the corresponding pair IDs at upper left, with the
  galaxy redshifts and projected separation of each pair listed at 
  lower left.  
\label{fig.cosmosims}}
\end{center}
\end{figure*}

\subsection{Supplementary Data}\label{sec.supplementary}

As detailed in \citet{Coil2011primus},  the PRIMUS survey fields have been
deeply imaged in multiple broad passbands.  
The XMM-LSS, CDFS-SWIRE, DEEP2 02h, DEEP2 23h, and COSMOS fields have all been
observed in the near- and far-UV by {\it GALEX}, and with the
exception of the DEEP2 23h field have also been imaged in 
four {\it Spitzer}/IRAC passbands.  Each of these fields has been
imaged over the full optical range from ground-based facilities.  

Together with PRIMUS survey redshifts, \citet{Moustakas2013} have
assembled these photometric measurements to produce broadband 
spectral energy distributions (SEDs) for the full PRIMUS galaxy
sample.  These authors have also developed 
a custom suite of routines (\texttt{iSEDfit}\footnote{ http://www.sos.siena.edu/${\sim}$jmoustakas/isedfit/}) to model these SEDs
via stellar population synthesis,
calculating rest-frame magnitudes and colors and constraining star
formation histories.  In brief, \texttt{iSEDfit} adopts the
Flexible Stellar Population Synthesis models of \citet{ConroyGunn2010}
with a \citet{Chabrier2003} IMF.  Star formation histories are assumed
to be exponentially declining with added stochastic bursts.  The
resulting library of model SEDs covers a broad range of ages, 
metallicities, burst timing and strength, and dust attenuation.  This
parameter space is then sampled to calculate the marginalized probability distribution
functions for stellar mass ($M_*$) and star formation rate (SFR).  In the present work, we use the
same photometric catalogs and procedures described above (and in full
in \citealt{Moustakas2013})
to constrain these quantities, 
adopting the galaxy redshifts estimated from 
our LRIS and FORS2 spectroscopy as described in 
\S\ref{sec.redshifts}.   
 The typical uncertainties on the values of $\log M_{*}/M_{\odot}$
  and $\log~\rm SFR$ estimated for the parent PRIMUS sample using this
  method are $0.08$
dex and $0.2$ dex, respectively \citep{Mendez2016}.

The COSMOS field, in which 30 of the observed pairs in our
sample are located, has also been imaged by the {\it HST}
Advanced Camera for Surveys (ACS) in the F814W filter to a $5\sigma$ depth of 
$I_{\rm AB} = 28$ mag for point sources
\citep{Scoville2007}.  We use the publicly available mosaic imaging
provided by the COSMOS team with a pixel scale $0.03\arcsec \rm
pix^{-1}$.  Small ($25\arcsec \times 25\arcsec$) sections of this
imaging showing each of the pairs for which our spectroscopy of the
b/g object covers \ion{Mg}{2} at the foreground redshift $z_{\rm f/g}$ (and for which $c
(z_{\rm b/g} - z_{\rm f/g}) / (1 + z_{\rm pair}) > 1000\mkms$, with
$z_{\rm pair} = (z_{\rm b/g} + z_{\rm f/g})/2$), are shown in Figure~\ref{fig.cosmosims}.



\section{Spectroscopic Analysis}\label{sec.analysis}

\subsection{Redshifts}\label{sec.redshifts}

The following analysis  associates 
\ion{Mg}{2}-absorbing material observed along a background galaxy or
QSO sightline with a nearby ``host'' galaxy.  We draw this
association based on the relative kinematics of the galaxies and
absorbers, and thus require precise measurements of the foreground
galaxy redshifts (with uncertainties $c \sigma_z / (1+z) < 100 \mkms$).  
As the accuracy of PRIMUS redshifts is $c \sigma_z / (1+z) \approx
1500 \mkms$,  and because there is a
non-negligible outlier rate for the sample of interest (in particular
because galaxies which are close on the sky are more likely to be
physically associated than a pair of galaxies selected at random), 
we analyze our follow-up
spectroscopy to improve the precision of both our 
f/g and b/g galaxy redshift estimates.

We use IDL routines adapted from the DEEP2 data reduction pipeline\footnote{http://deep.ps.uci.edu/spec2d}
 to
perform these redshift measurements.  
For this analysis, we enable the pipeline to accept a
redshift estimated by eye as an initial guess.
The code then determines the best-fit offset
between the observed spectrum and a linear combination of a 
nebular emission-line template, an early-type
galaxy spectrum,  a post-starburst galaxy spectrum, and the spectrum
of a broad-line AGN (as in \citealt{Coil2011winds}).
This best-fit offset
is determined from the blue and red spectrum of each object
independently.  The portion of each spectrum blueward of
$\lambda_{\rm rest} < 3000$ \AA\ was masked prior to
fitting for all foreground galaxies to prevent intrinsic kinematic
offsets (due to, e.g., winds or inflows) from biasing the
measurements.  Redshifts estimated from the red spectra are adopted in
most cases, with blue spectra providing redshifts for a few objects
with low S/N red coverage.  The dispersion in 
redshifts measured from the red vs.\ blue  spectra for our
foreground galaxy sample (i.e., the dispersion in the quantity $c
(z_{\mathrm blue} - z_{\mathrm red})/(1+z_{\mathrm red})$) is
$86\mkms$.  We consider this a conservative upper bound on our
redshift measurement uncertainty, as these offsets are systematically affected by the
large difference in spectral coverage as well as occasional significant differences in
S/N in the spectra taken from the two cameras/grisms of LRIS and FORS2.


\subsection{Spectroscopic Data Quality}

Among the 59 primary sample pairs observed, there are 3 pairs for which the S/N
of the foreground galaxy spectrum is insufficient to yield a
high-quality redshift measurement (pairs 216, 419, and 605).
There are seven more primary sample pairs for which at least one of the PRIMUS redshift
estimates was in error, and yielded physically-associated systems or
stellar sources.
This leaves a
sample of 49 {\it bona fide} projected galaxy pairs having $\mrperp <
50$ kpc in our primary sample with high-quality spectroscopic redshifts.  There are an
additional 25 serendipitous pairs, 23 at larger impact parameters (50 kpc $< \mrperp <
150$ kpc), and 2 of which have $\mrperp < 50$ kpc (pairs 1600 and
1604),  which our spectra confirm to be extragalactic objects in projection.

\begin{figure*}
\begin{center}
\includegraphics[angle=0,width=7in,trim=0 0 20 0,clip=]{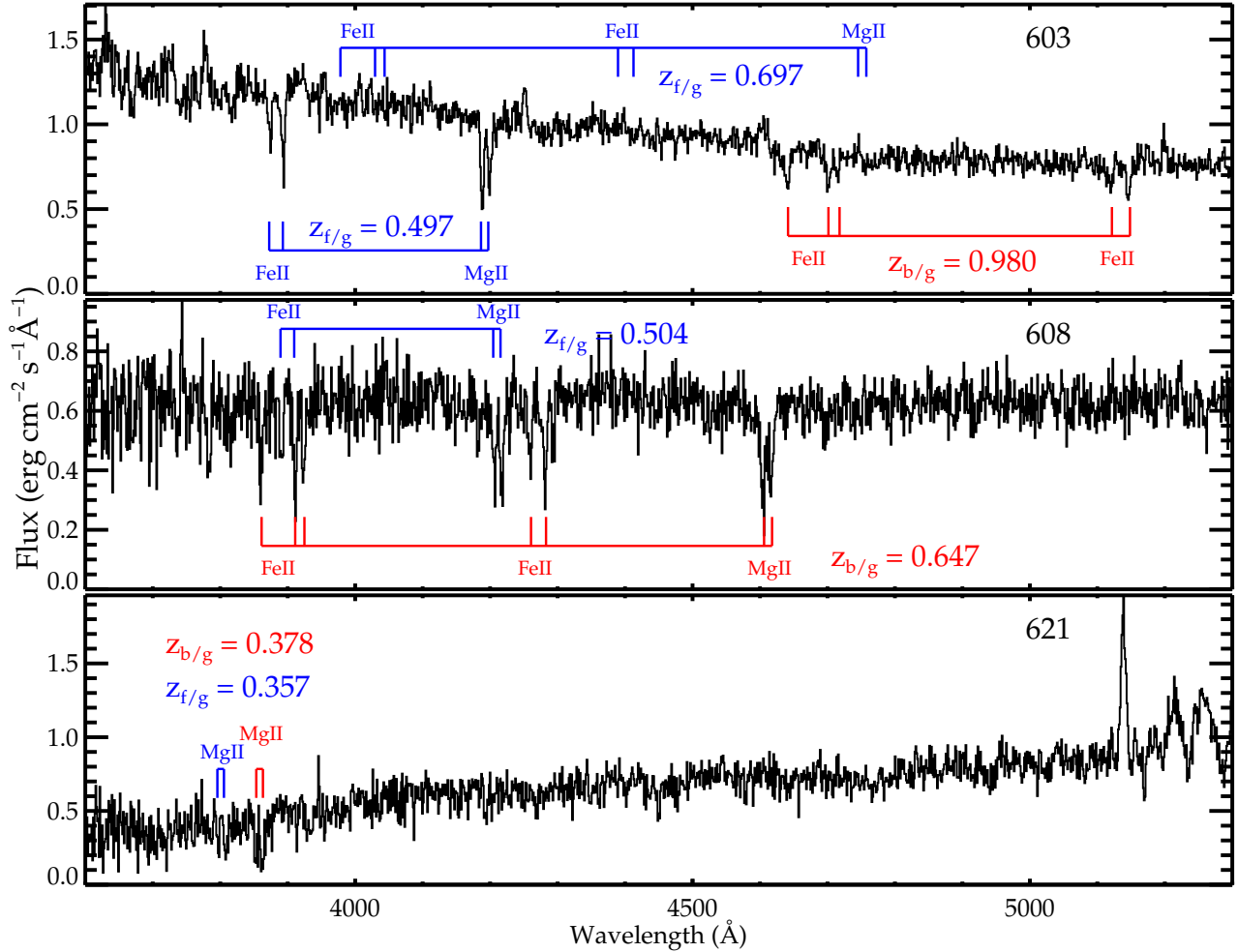}
\caption{Spectroscopy of the b/g galaxy in three of the close
  ($\mrperp < 50$ kpc) pairs in our sample, with pair ID numbers
  indicated at upper right.  Subsets of the
  transitions \ion{Mg}{2} $\lambda \lambda 2796, 2803$, \ion{Fe}{2}
  $\lambda \lambda 2586, 2600$, and \ion{Fe}{2} $\lambda \lambda 2344,
  2374, 2382$ at the systemic velocity of the b/g galaxy are marked in
  red.  Wherever spectroscopic coverage is available, the same
  transitions are marked in the rest-frame of the corresponding f/g
  galaxy in blue.   The spectrum shown in the top panel probes f/g
  systems within 50 kpc at two redshifts ($z_{\rm f/g} = 0.697$ and
  $z_{\rm f/g} = 0.497$ in pairs 603 and 1600, respectively).  
  The spectra span the range in S/N at $\lambda_{2796}^{\rm f/g}$ of
 the b/g galaxy spectroscopy in our sample, with S/N(\ion{Mg}{2}) $=
 19.1~\rm \AA^{-1}$ (top), $15.8~\rm \AA^{-1}$ (middle), and $6.4~\rm \AA^{-1}$ (bottom).
\label{fig.primus_specdemo}}
\end{center}
\end{figure*}

The spectroscopy of the b/g galaxies in two of the $\mrperp > 50$
kpc pairs does not extend blueward of $2800$ \AA\ at the systemic
velocity of the corresponding f/g object measured as described above,
and so must be expunged from the sample.   Hence, our dataset includes
a total of 72 projected pairs (51 of them having $\mrperp < 50$ kpc)
with spectroscopic coverage of the \ion{Mg}{2} doublet at $z_{\rm f/g}$.
Figure~\ref{fig.primus_specdemo} shows representative spectroscopy of
three of the sample b/g galaxies in close pairs, with strong \ion{Mg}{2} and
\ion{Fe}{2} transitions at the systemic velocities of the b/g and f/g
galaxies marked in red and blue, respectively. 

\begin{figure}
\begin{center}
\includegraphics[angle=0,width=\columnwidth,trim=20 80 20 30,clip=]{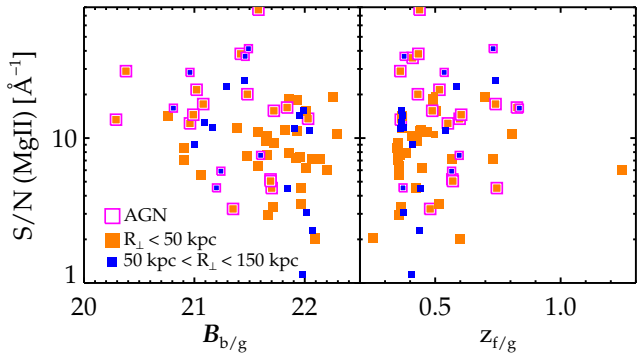}
\caption{{\it Left:} S/N measured in each b/g galaxy spectrum within $\delta v <
  500\mkms$ of the \ion{Mg}{2} 2796 transition in the rest frame of
  the close projected f/g galaxy ($\lambda_{2796}^{\rm f/g}$) vs.\ the
  apparent $B$-band magnitude of the b/g object.  Large orange squares indicate pairs with
  impact parameters $< 50$ kpc, and small blue squares mark pairs with
  larger separations.  Symbols outlined in magenta correspond to b/g
  sightlines dominated by emission from a broad-line AGN. 
{\it Right:} S/N measured at $\lambda_{2796}^{\rm f/g}$ vs.\ $z_{\rm
  f/g}$.  Symbol shapes and colors are consistent with those in the
left-hand panel.  The S/N of our spectroscopy ranges from
$\sim2-40~\rm \AA^{-1}$, with the median S/N for those b/g objects
without dominant broad-line AGN $\approx 9.0 ~\rm \AA^{-1}$.
\label{fig.primus_s2n}}
\end{center}
\end{figure}

The median S/N measured in a velocity window $\delta v\pm 500\mkms$
around the observed wavelength of the \ion{Mg}{2} $\lambda 2796$
transition at $z_{\rm f/g}$
($\lambda_{2796}^{\rm  f/g}$) in our b/g galaxy spectra is shown in
Figure~\ref{fig.primus_s2n} versus the apparent $B$-band magnitude of
the b/g object (left).  Close pairs are
indicated with large orange squares, and pairs having $\mrperp > 50$ kpc 
are marked with small blue squares.  Those b/g galaxies for which
the best-fitting redshift template spectrum was that of the broad-line
AGN (and those exhibiting any broad-line \ion{Mg}{2} emission obvious
in a visual inspection) are outlined in magenta.  The S/N of this
spectroscopy ranges from $\sim2-40~\rm \AA^{-1}$, and tends to
increase with the brightness of the b/g object.  The spectra of the
objects hosting broad-line AGN have overall higher S/N, with a median
S/N $= 16.1~\rm \AA^{-1}$ (vs.\ a median S/N $=9.0~\rm \AA^{-1}$ for
the remaining b/g galaxies).  

We compare this S/N with $z_{\rm f/g}$ in the right-hand panel of
Figure~\ref{fig.primus_s2n}.  The redshift distribution of the f/g
galaxy sample peaks toward the lower limit of our selection criterion
for $z^{\rm PR}$, with the median $z_{\rm f/g} = 0.44$ for both
the close pair sample and the full sample of pairs.  Moreover, the S/N in the
background sightlines is uncorrelated with $z_{\rm f/g}$, indicating
that the drop in efficiency of the spectrographs blueward of 4000 \AA\
is not significantly affecting our sensitivity to foreground
absorption for the lower-$z_{\rm f/g}$ portion of the sample.

\subsection{Absorption Equivalent Widths and Velocity Centroids}

Prior to making measurements of the absorption strength in rest-frame
UV transitions arising in the CGM of the f/g galaxy sample, we normalize each background galaxy
spectrum to the level of the continuum.  
The majority of our b/g objects are dominated by stellar continuum
emission blueward of $\lambda_{\rm rest}^{\rm b/g}\sim 3700$ \AA, such
that their spectra are relatively flat and featureless 
in the wavelength range of interest.  In these cases, the continuum
level is 
determined via a linear fit to the data in the spectral regions on
either side of each feature.  In particular, a fit to the data in the windows
$(2765 - 2785) \times (1 + z_{\rm f/g})$\AA\ and $(2810-2830) \times (1 + z_{\rm f/g})$ \AA\ 
is assumed to describe the continuum level at $\lambda_{2796}^{\rm
  f/g}$.  
We visually inspected these regions in each b/g galaxy spectrum to
ensure they do not include strong emission or absorption features
associated with the b/g object, and made small adjustments to their
boundaries to avoid such features in several cases.
For each spectrum, we also generate 1000 Monte Carlo realizations of this
continuum fit by first adding Gaussian random noise with a dispersion given by the
median error in the data to the original fit, and then
fitting a line to each of these continuum realizations.  This allows
us to assess the degree of uncertainty in the continuum level.  We
find that the dispersion in the $W_{2796}$ values (calculated as described
below) measured after adopting this set of continuum fits is
typically only $\sim40\%$ of the formal uncertainty in $W_{2796}$ given by the square
root of the sum of variances in each absorption-line pixel. 

Twelve of the b/g objects in our sample 
are host to bright QSOs, and
their continua are dominated by the complex broad emission line
features typical of such systems.  
In these cases, the continuum level
was determined using a custom routine available in the XIDL software
package\footnote{www.ucolick.org/${\sim}$xavier/IDL}.  The routine (x$\_$continuum)
facilitates a by-eye spline fit to the full QSO spectrum.  Previous
studies invoking this technique have found the average uncertainty in
the resulting continua is $\lesssim5\%$ outside of the \lya forest \citep{QPQ6}.

After continuum normalization, we search the spectral region
within $\delta v \pm 300\mkms$ of $\lambda_{2796}^{\rm f/g}$ and
$\lambda_{2803}^{\rm f/g}$ by eye to identify absorption associated
with each transition.  We select the velocity range spanning the
absorption profiles by hand, and then perform a boxcar sum over this
range to calculate $W_{2796}$ and $W_{2803}$.   We also flag any
profiles which are affected by blending with absorption transitions
associated with the b/g galaxy itself, or which (in the case of two of our
serendipitous pairs)
are blended with the \lya forest.
We additionally calculate the flux-weighted wavelength centroid of
each \ion{Mg}{2} 2796 profile, $\left< \lambda_{2796} \right>= \sum_i (1-f_i) \lambda_i
/ \sum_i (1-f_i)$, with $f_i$ and $\lambda_i$ the normalized flux and
wavelength of each pixel within the line.   Our measurements of
$W_{2796}$, $W_{2803}$, and the velocity offset between $\left<
  \lambda_{2796} \right>$ and $z_{\rm f/g}$ ($\left<
 \delta v_{2796} \right>$) are listed in Table~\ref{tab.galewinfo}.
Our spectroscopic coverage of \ion{Mg}{2} at $z_{\rm f/g}$ along each
background galaxy sightline is shown in Appendix~\ref{sec.appendix} for reference.
We detect unblended \ion{Mg}{2}
$\lambda 2796$ absorption securely (at $>2\sigma$ significance)
in 20 individual background sightlines having $R_{\perp}<50$
kpc, with $W_{2796}$ values in the range
$0.25~\mathrm{\AA}<W_{2796}<2.6~\mathrm{\AA}$.
We place $2\sigma$ upper limits on $W_{2796}$ of $\lesssim0.5~\mathrm{\AA}$ in an
additional 11 close sightlines.


\section{Foreground Galaxy Physical Parameters}\label{sec.fg}

\begin{figure*}
\begin{center}
\includegraphics[angle=0,width=\columnwidth,trim=30 180 20 120,clip=]{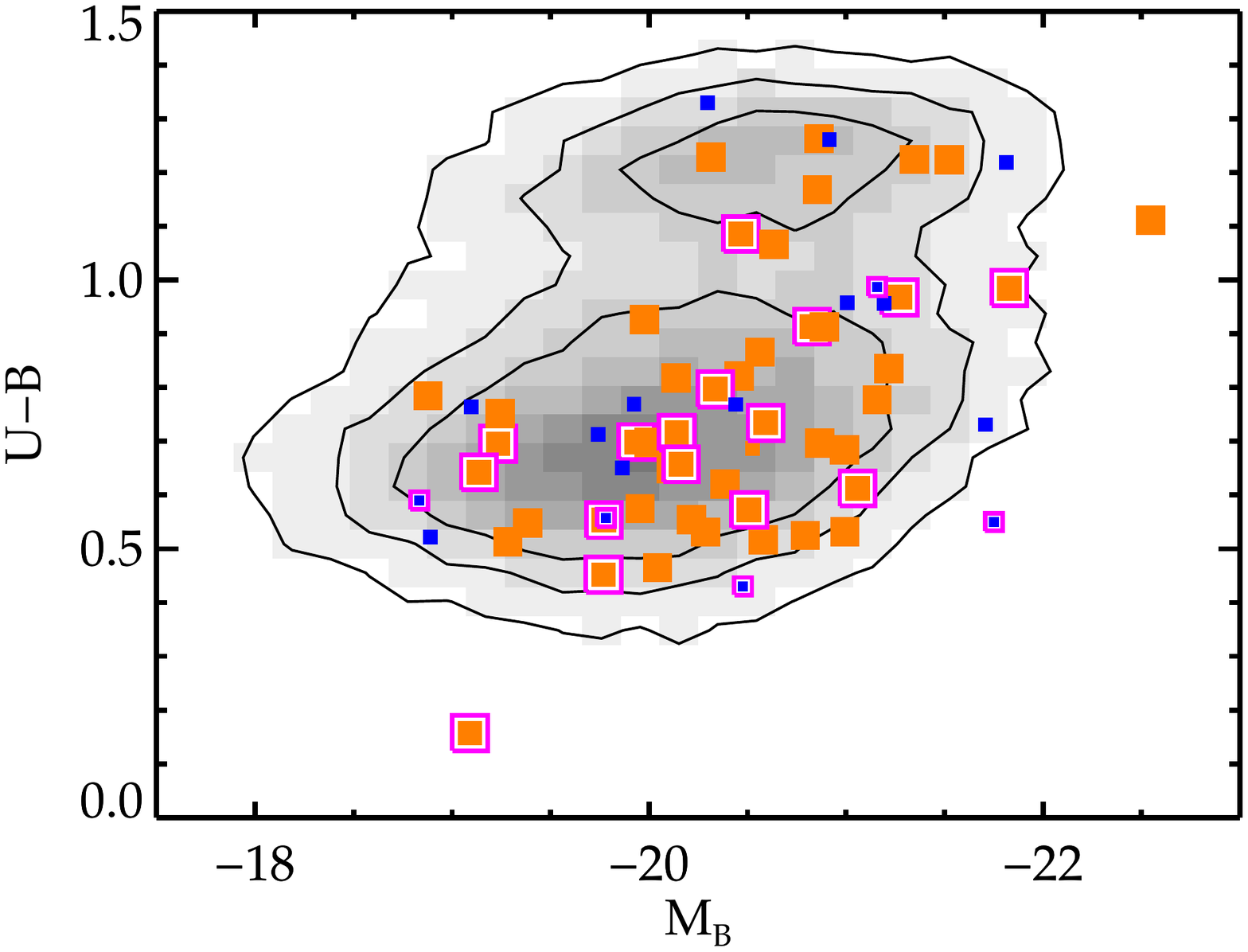}
\includegraphics[angle=0,width=\columnwidth,trim=30 180 20 120,clip=]{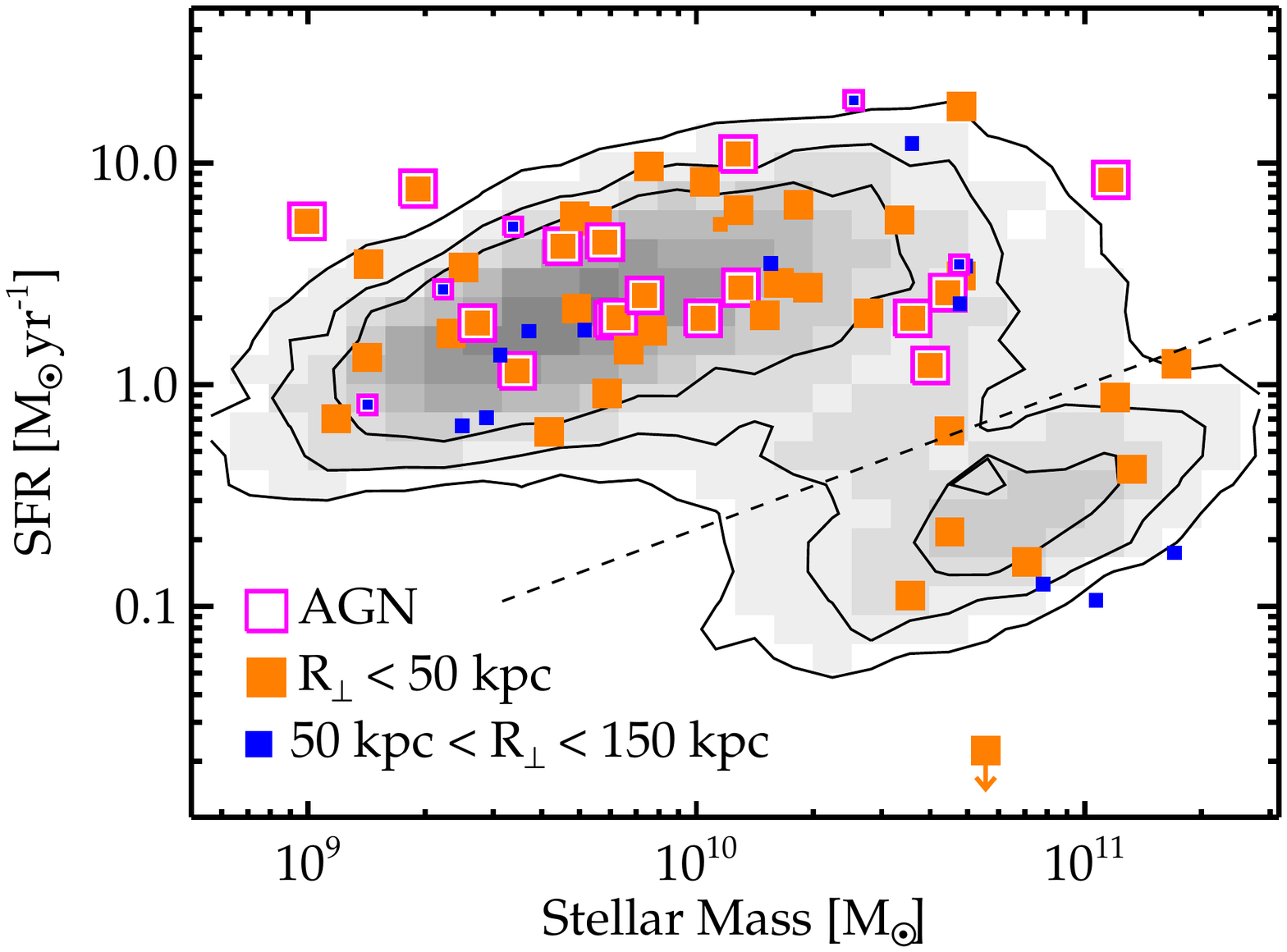}
\caption{{\it Left:} $U-B$ vs.\ $M_{\rm B}$ for the foreground
  galaxies in our pair sample.  Objects in pairs with $\mrperp \le 50$
  kpc are indicated with large orange squares, and objects in more
  widely separated pairs are marked in blue.  Contours and gray shading show the $U-B$
  vs.\ $M_{\rm B}$ distribution of all PRIMUS galaxies having $0.35 <
  z^{\rm PR} < 0.8$.  Symbols outlined in magenta indicate pairs in
  which the background galaxy hosts a bright AGN.
{\it Right:}  SFR vs.\ stellar mass for the sample foreground galaxies.  Symbol sizes and colors are consistent with those in the
left-hand panel.  Contours show the SFR-$M_*$ distribution of
PRIMUS galaxies at $0.35 <  z^{\rm PR} < 0.8$.  
The dashed line indicates the dividing line between star-forming and
quiescent galaxies adopted from a fit to the minimum of the galaxy
distribution by \citet{Berti2016} assuming $z\sim0.6$.
A handful (7) of the f/g galaxies in our sample of close pairs has
SFR $\lesssim 0.1-1~M_{\odot}~\rm yr^{-1}$ and occupy the``red sequence'' in
the color-magnitude diagram shown at left.
However, the vast majority
of our f/g galaxies 
lie along the main locus of
the star-forming sequence at $z\sim0.4-0.8$.  
\label{fig.primus_ms}}
\end{center}
\end{figure*}

Having estimated rest-frame colors, luminosities, SFRs and stellar
masses for our foreground galaxy sample as described in
Section~\ref{sec.supplementary}, we now examine these properties in
the context of the $z\sim0.5$ galaxy population.
Figure~\ref{fig.primus_ms} (left) shows 
the distribution of rest-frame $U-B$ color and absolute $B$-band
magnitude of these objects with colored points.  Symbols outlined in
magenta mark pairs in which the b/g galaxy hosts a bright AGN.  
 These latter pairs should be considered QSO-galaxy pairs,
  albeit with fainter b/g sightlines than are typically used, whereas
  the former are {\it bona fide} galaxy-galaxy pairs in which the b/g
  sightline is not dominated by a bright nuclear source.
Six pairs for which our coverage of \ion{Mg}{2} $\lambda 2796$ at $z_{\rm f/g}$ is affected
by line blending are excluded here (and from all following analysis;
these objects are indicated in the column reporting $W_{2796}$ in Table~\ref{tab.galewinfo}).
The black contours and gray shading indicate the distribution of 
all galaxies with high-quality redshift measurements in
the PRIMUS catalog at $0.35 < z^{\rm PR} < 0.8$, with the degree
of shading scaled to the density of  objects.  The SFR-stellar mass
distributions for the same galaxy samples are shown at right.  A small
minority of the f/g objects in our close pair sample lie along the
``red sequence'' in the color-magnitude diagram
\citep[e.g.,][]{Willmer2006} and exhibit low SFRs ($\lesssim0.1-1~M_{\odot}~\rm
yr^{-1}$), while the majority of the sample lies along the main
locus of the star-forming sequence at $z\sim0.4-0.8$.  
In the following analysis, we adopt the fit to the minimum of the
bimodal galaxy distribution in the parent PRIMUS survey reported by \citet{Berti2016} as the
dividing line between the star-forming and quiescent objects:
\begin{equation}\label{eq.sfcrit}
\log {\rm SFR} = -1.29 + 0.65 (\log
\frac{M_{*}}{M_{\odot}} - 10) + 1.33 (z - 0.1),
\end{equation}
with SFR having units of $M_{\odot}~\rm yr^{-1}$. 
Thus, the SFR dividing the star-forming sequence from the locus of quiescent galaxies at a
given $M_*$ increases with increasing redshift.  The slope and
intercept of this criterion for objects at $z\sim0.6$ is
indicated in the right-hand panel of Figure~\ref{fig.primus_ms} with a
dashed line.

\begin{figure*}
\begin{center}
\includegraphics[angle=0,width=7in,trim=0 0 0 0,clip=]{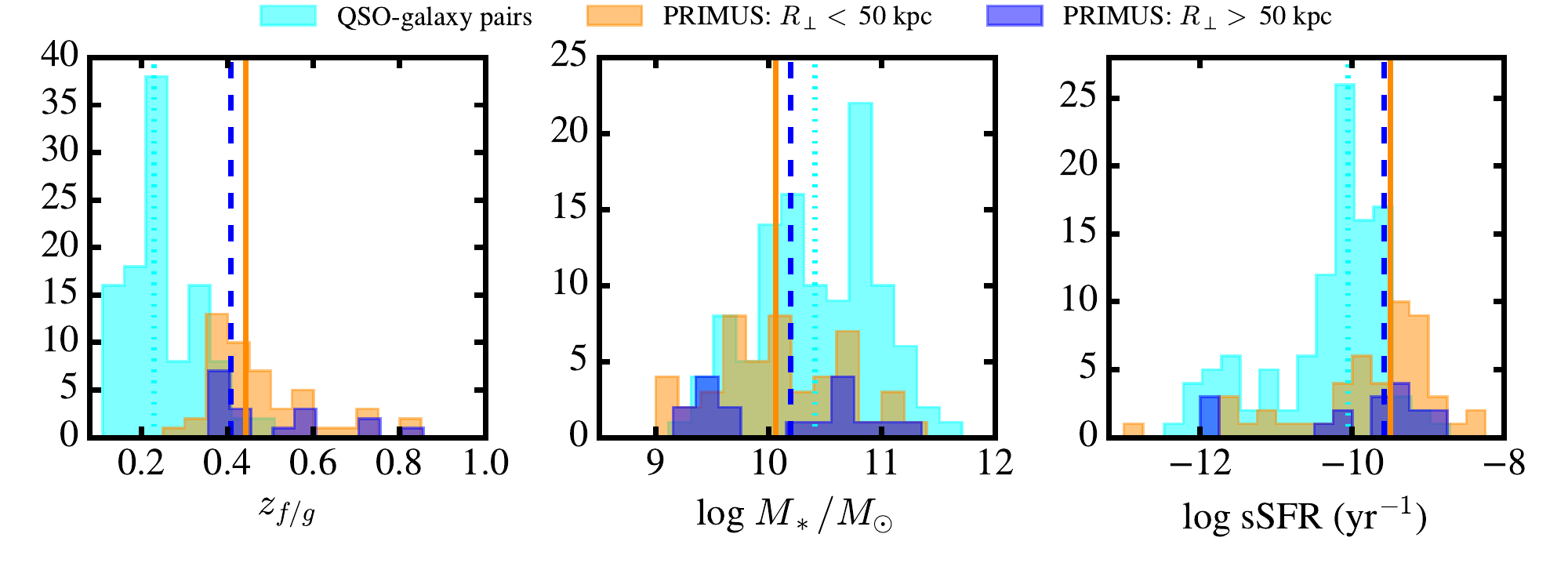}
\caption{\emph{Left:} Redshift distribution of our PRIMUS f/g galaxy
  sample.  The distribution of galaxy redshifts for pairs having
  $\mrperp < 50$ kpc is shown in orange, with the $z_{\rm f/g}$
  distribution for wider pairs shown in blue.  The redshift
  distribution of galaxies included in the QSO-galaxy pair studies of
  \citet{Chen2010} and \citet{Werk2013} is shown in cyan.  The median
  value for each subsample is shown with a vertical line of the same color.
\emph{Middle:} Stellar mass distribution of our PRIMUS f/g galaxy sample and
QSO-galaxy pair comparison samples.  The colors are chosen as in the
left-hand panel.  
\emph{Right:} Specific SFR distribution of our PRIMUS f/g galaxy sample and
QSO-galaxy pair comparison samples.  The PRIMUS pairs tend to lie at
higher redshifts than the comparison sample galaxies.  PRIMUS pairs
having $\mrperp < 50$ kpc have a
median $\log M_*/M_{\odot}$ lower by $\sim0.4$ dex and median $\log
\rm sSFR$ higher by $\sim0.6$ dex.  The few PRIMUS pairs at $\mrperp > 50$ kpc
have a median $\log M_*/M_{\odot}$ and median $\log
\rm sSFR$ 
very close to that of the $\mrperp < 50$ kpc pairs.
\label{fig.primusqso_galcomp}}
\end{center}
\end{figure*}

We show the frequency distribution of redshifts, stellar masses, and specific
SFRs for these galaxies in Figure~\ref{fig.primusqso_galcomp}.   In
preparation for the analysis to follow in
Section~\ref{sec.results_cgm}, 
we also compare these distributions
with those for the f/g galaxies included in the QSO-galaxy pair
comparison samples discussed in \S\ref{sec.sample_qsocomp} \citep{Chen2010,Werk2013}.
The f/g galaxies in these samples tend to lie at lower
redshifts than the PRIMUS f/g galaxies, with a median redshift $z_{\rm
  f/g}^{\rm QSO} = 0.23$.  However, the QSO-galaxy pair and PRIMUS
pair samples span a similar (and broad) range
in stellar mass and specific SFR, with the median $\log M_*/M_{\odot}$
and $\log \rm sSFR$ values for the PRIMUS f/g galaxies in close pairs
($\mrperp < 50$ kpc) offset by
$-0.4$ dex and $+0.6$ dex, respectively, from the median values of the
comparison sample distributions.  
Considering only those systems lying on the star-forming sequence as
defined by Eq.~\ref{eq.sfcrit}, the median $\log \rm sSFR/[yr^{-1}] =
-9.5$ and $\log M_*/M_{\odot} = 9.9$ for close
PRIMUS pairs, while the
QSO-galaxy star-forming pair sample has a median $\log \rm sSFR/[yr^{-1}] = -10.0$  and $\log M_*/M_{\odot} = 10.3$.  
This $+0.5$ dex offset in sSFR is
consistent with the ``best-fit'' relation
between galaxy SFR, stellar mass, and age of the Universe from
\citet[][Eq. 28]{Speagle2014}, adopting $\log M_*/M_{\odot} = 9.9$ at age
$t=9.1$ Gyr ($z=0.4$) and $\log M_*/M_{\odot} = 10.3$ at age
$t=11.0$ Gyr ($z=0.2$). 
This reinforces the assertions made above and in
\citet{Chen2010} and \citet{Werk2013} that these galaxies are
representative of the star-forming population at the corresponding epochs.
We note that the PRIMUS pairs at $\mrperp > 50$ kpc have a 
median $\log M_*/M_{\odot}$ within $+0.15$ dex of the PRIMUS pairs having
$\mrperp < 50$ kpc; however, the subsample of these wide pairs which
are star-forming have a distribution of $M_*$ which is overall lower,
with median $\log M_*/M_{\odot} = 9.6$ and $\log \rm sSFR/[yr^{-1}] =
-9.5$.


\section{Background Galaxy Sizes and Morphologies}\label{sec.bg}

\subsection{Half-Light Radii}

The salient characteristics of our background galaxies are those which
differentiate them from QSO sightlines: namely, their sizes and
morphologies. Most germane to our analysis is the spatial distribution
of sources contributing to the continuum emission of each galaxy at
$\lambda_{\rm obs} \sim 3650 - 5180$ \AA, or $\lambda_{\rm rest} \sim
2200 - 2700$ \AA\  -- i.e., the portion of the b/g galaxy
  continuum probing \ion{Mg}{2} at $z_{\rm f/g}$. Because such
sources in $z \sim$ $0.5 - 1$ 
galaxies cannot be resolved from the ground, an
ideal dataset for measuring this distribution would be \emph{HST} imaging
with the ACS or WFC3/UVIS cameras in filters covering the SDSS $g$ band
(e.g., F475W). 

\begin{figure}
\begin{center}
\includegraphics[angle=0,width=3.5in,trim=0 150 0 150,clip=]{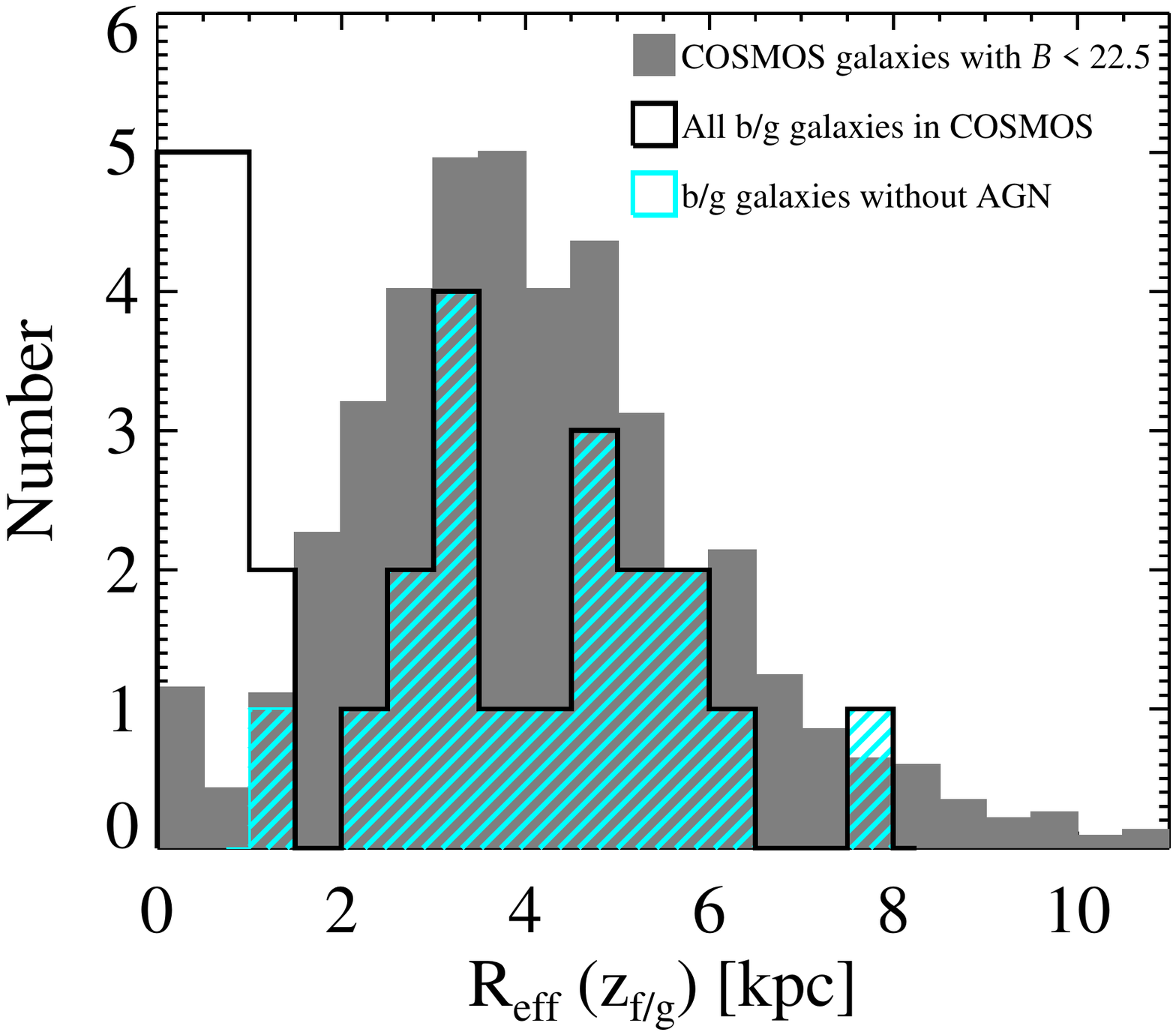}
\caption{Distribution of half-light radii measured in \emph{HST}/ACS F814W
  imaging of the 
b/g galaxies in our sample which are located in the COSMOS field
(black). 
The radii are estimated at the redshift of the corresponding f/g
galaxy. 
The distribution of sizes for those b/g galaxies without a dominant 
broad-line AGN is shown in cyan.  The distribution of half-light radii
for all galaxies in the COSMOS field having $0.4 < z^{\rm PR} <
1.0$ and $B_{\rm AB}<22.5$ is shown in gray.  These latter sizes are estimated at the
redshift of the target ($z^{\rm PR}$).  
\label{fig.primus_bgradhist}}
\end{center}
\end{figure}

This type of imaging is not currently available;
however, those pairs located in the COSMOS fields have
been deeply imaged in the ACS F814W band, sensitive to 
$\lambda_{\rm obs} \sim 7700 - 8400$ \AA.  
To assess galaxy sizes in this passband, we make use of the publicly available
COSMOS ACS $I$-band Photometry Catalog\footnote{www.irsa.ipac.caltech.edu/data/COSMOS/datasets.html}
\citep{Leauthaud2007} generated using the SExtractor photometry
detection software \citep{Bertin1996}.  From this catalog, we select
measurements of half-light radii for 30 b/g galaxies which have
redshifts at least 1000 $\mkms$ larger than the corresponding f/g
galaxy, and for which our followup spectroscopy covers \ion{Mg}{2} at
$z_{\rm f/g}$.  We then calculate the projected physical extent of
these half-light radii at $z_{\rm f/g}$, $R_{\rm eff} \rm (z_{f/g})$,
and show the distribution of these values in
Figure~\ref{fig.primus_bgradhist} in black.  We show the size
distribution of those b/g galaxies which lack a bright broad-line AGN
in cyan.  To compare the sizes of these particular galaxies to the
population from which they are selected, we also use the measurements
in the \citet{Leauthaud2007} catalog to calculate effective
radii for the $\sim1000$ PRIMUS galaxies in the COSMOS field having redshifts in
the range $0.4 < z^{\rm PR} < 1.0$ and having apparent $B$-band
magnitudes $B_{\rm AB}<22.5$.
The distribution of these sizes, normalized to an arbitrary value, is shown in Figure~\ref{fig.primus_bgradhist} in gray.

Our PRIMUS b/g galaxies have a broad range in sizes, with the smallest object extending
over only $R_{\rm eff} \rm (z_{f/g}) = 0.4$ kpc, and the largest
having $R_{\rm eff} \rm (z_{f/g}) = 7.9$ kpc.  
Indeed, comparing the black and gray histograms, we see that the b/g
galaxies include a
significantly higher fraction of very compact sources than the overall
bright galaxy population.  However, the distribution of PRIMUS b/g
galaxies without bright AGN (identified spectroscopically) is
qualitatively similar to that of the broader COSMOS population: the
median radii are $R_{\rm  eff} \rm (z_{f/g}) = 4.1$ kpc and $R_{\rm
  eff} \rm (z^{\rm PR}) = 3.9$ kpc, respectively.  Furthermore, 
 the minimum $R_{\rm eff}$  of the former (cyan) distribution is $R_{\rm
  eff} \rm (z_{f/g}) = 1.0$ kpc.
The high surface-brightness regions
of rest-frame optical emission from
these systems (i.e., the inner regions producing half of the
total emission)
are therefore subtending projected distances (or half-light diameters) of at least $\sim2$ kpc  and up to
$\gtrsim8$ kpc
across the halos of the corresponding f/g galaxies.
And because these sizes are typical of the bright ($B_{\rm AB}<22.5$) galaxy
population in the COSMOS field, we assume they are also representative of
the sizes of the remainder of our b/g galaxy sample.

\begin{figure}
\begin{center}
\includegraphics[angle=0,width=3.5in,trim=0 150 0 150,clip=]{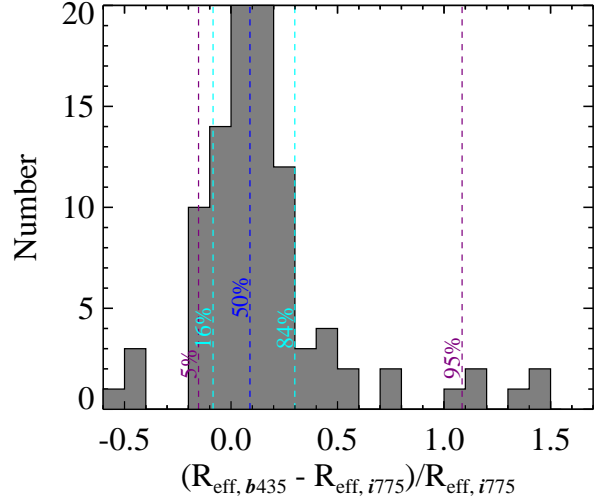}
\caption{Distribution of the offset between half-light radii
  measured in the \emph{HST}/ACS F435W and F775W filters for all galaxies
  observed in the Team Keck Redshift Survey \citep{Wirth2004} to have
  secure redshifts in the range  $0.4 < z < 1.0$ and having F435W
  magnitudes $b435 < 23$.  The 5th, 16th,
  50th, 84th, and 95th-percentile values of the distribution are
  marked with vertical dashed lines.  The $b435$-band half-light
  radius is more than $\sim12\%$ smaller than the $i775$-band
  half-light radius in only 5\% of the galaxies.
\label{fig.primus_uvoptsize}}
\end{center}
\end{figure}

\begin{figure}
\begin{center}
\includegraphics[angle=0,width=3.5in,trim=0 20 0 0,clip=]{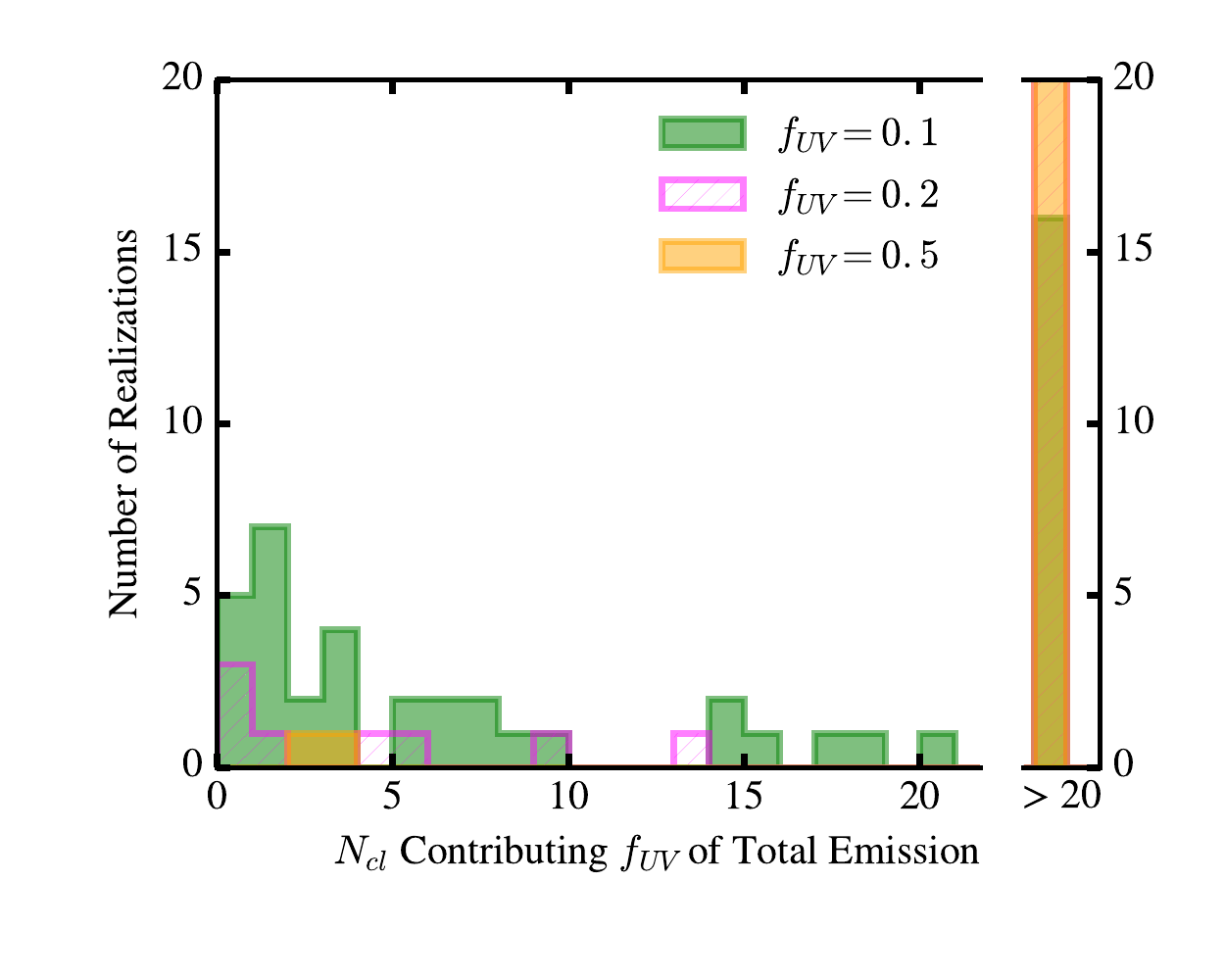}
\caption{The distribution of the number of clusters ($N_{\rm cl}$)
  which emit up to a set fraction ($f_{\rm UV}$) of the total $L^{\rm
    F225W}_{\rm tot}$ of a given SLUG
  galaxy realization.  Realizations for which the number
  of clusters producing the assigned $f_{\rm UV}$ exceeds 20 are
  indicated in the right-most bin.  For $f_{\rm UV} = 0.2$ (magenta)
  and $0.5$ (orange), the number of clusters producing $f_{\rm UV} \times L^{\rm F225W}_{\rm tot}$
  exceeds 20 in nearly all realizations, such that the height of the
  right-most bin exceeds the range of the y-axis.  It is somewhat more
  common for just a handful of clusters to contribute only a fraction
  $f_{\rm UV} = 0.1$ of the total $L^{\rm F225W}_{\rm tot}$ (green histogram).
\label{fig.primus_slug}}
\end{center}
\end{figure}


The rest-frame ultraviolet emission from these systems, however, may
have a differing morphology.  In particular, while emission at
$\lambda_{\rm rest} \sim 5300$ \AA\  includes contributions from
A stars and later spectral types, $\lambda_{\rm rest}\sim2200 - 2700$
\AA\ emission is produced exclusively by O and B stars.  
To assess the distribution of offsets between half-light radii measured
in the rest-frame UV vs.\ the rest-frame optical in a galaxy sample
similar to our own, we turn to the Team
Keck Redshift Survey \citep[TKRS;][]{Wirth2004} of the GOODS-N field
\citep{Giavalisco2004}.  This field has the advantage of deep imaging
in both the {\it HST}/ACS F435W and F775W passbands and
publicly-available photometry catalogs for each \citep{Giavalisco2004}.
From the TKRS
galaxy sample (magnitude-limited to $R_{\rm AB} < 24.4$), we select
objects having $0.4 < z < 1.0$ and apparent magnitude in the F435W
passband $b435 < 23.0$.  We then calculate the relative offset in the
half-light radii measured for each galaxy in the F435W and F775W
bands, $(R_{\mathrm{eff}, b435} - R_{\mathrm{eff},
  i775})/R_{\mathrm{eff}, i775}$.  The distribution of this quantity
is shown in Figure~\ref{fig.primus_uvoptsize}.  The median value of
this distribution is $\sim +0.09$, indicating that $R_{\mathrm{eff},  b435}$ is 
typically $\sim10\%$ larger than $R_{\mathrm{eff},  i775}$. Indeed,
$R_{\mathrm{eff},  b435}$ is $\sim15\%$ {\it smaller} than
$R_{\mathrm{eff},  i775}$ only below the 5th-percentile value of the
distribution.  This suggests that the b/g galaxy sizes we measure in
the F814W {\it HST}/ACS passband (Figure~\ref{fig.primus_bgradhist})
are similar to the sizes of the beams produced by their $\lambda_{\rm rest}\sim2200 - 2700$
\AA\ emission.

\subsection{The Detailed Distribution of Rest-Frame UV Continuum Emission}

A significant fraction of the young O and B
stars which are producing continuum emission in the UV are known
to form in embedded clusters within molecular clouds
\citep{LadaLada2003}.  Stellar winds and radiation from these
clusters are  expected to disrupt and destroy the remainder of their nascent clouds within a
few Myr \citep{Murray2010}.  
Because these young, bright clusters are likely to survive disruption by, e.g.,
stellar mass loss or tidal disturbances for at least $\gtrsim 10^7 - 10^8$ yr \citep{Fall2009}, 
the integrated (but instantaneous) UV emission from a star-forming galaxy may therefore be dominated by
light from massive star clusters.
In the context of intervening absorption studies,
because young clusters have radii of only $r_{cl} \sim 0.1-10$ pc \citep{LadaLada2003,Murray2010},
b/g galaxies 
may be viewed as a closely-spaced set of numerous pencil-beam
sightlines.
 A recent survey of the young cluster population of local star-forming
galaxies indicates that massive spirals may host several
hundred or more than $1000$
individual clusters \citep{Krumholz2015b,Grasha2015};
however, 
for our purposes, it is of interest to consider in particular the total
number of such clusters which make a dominant contribution to the
rest-frame UV continuum.  For instance, in the extreme case that a galaxy's UV light
is dominated by only a single massive young cluster, the b/g beam of
that galaxy would have a morphology similar to that of a QSO.
While this scenario is unlikely in view of our finding that the
half-light radii of $z\sim0.4-1.0$ galaxies is similar at both
$\lambda_{\rm obs}\sim 4400$ \AA\ and $\lambda_{\rm obs}\sim 7750$
\AA, if it were to arise it would 
weaken our experimental leverage on the sizes of f/g absorbers.

To estimate the
number and luminosity of young clusters in a ``typical'' b/g galaxy at
the epoch of observation, we make use of the SLUG stellar population
synthesis
code\footnote{http://slug2.readthedocs.io/en/latest/intro.html} \citep{daSilva2012,Krumholz2015}.
SLUG predicts the spectrum of a given stellar population with an
explicit accounting for the stochastic nature of star and star cluster
formation.  Rather than adopting initial mass functions, cluster mass
functions, star formation histories, etc., which are constant in time,
SLUG assigns each of these relations a probability distribution
function.  To simulate the growth of a galaxy, the user must first select
the fraction ($f_c$) of stars which are expected to form in clusters
(as opposed to the field).  SLUG then calculates the total mass in
stars which must be formed at a given time step as set by the star
formation history.  The code draws cluster masses from the input
cluster mass function until it has formed the appropriate amount of
mass in clusters.  It then populates each cluster with stars by drawing from
the initial mass function probability distribution.  The stars are allowed to age over
time, and the clusters are also disrupted (and join the field population) on a time scale drawn from
the specified cluster lifetime function.

At each timestep, SLUG computes the composite spectrum of all stars in
the simulation, as well as the spectrum of each individual star
cluster.  In addition, the code package includes throughput curves for
numerous filters, allowing the user to calculate the total luminosity
of the system as well as the luminosity of individual clusters in several
passbands in common use.  We simulate a galaxy with a continuous star
formation rate of $0.1~M_{\odot}~\rm yr^{-1}$ and with the fraction
$f_c$ set to 1 (i.e., such that all stars form in clusters) for a
total of 200 Myr.  We adopt
the default settings specified in SLUG for the remaining simulation
inputs, including a Chabrier IMF \citep{Chabrier2005}, a cluster mass
function $dN_{cl}/dM_{cl} \propto M_{cl}^{-2}$, and a cluster lifetime function
$dN_{cl}/dt \propto t^{-1.9}$.  We then generate 48 realizations of this
simulation, recording the luminosity of each galaxy and each
individual cluster in the F225W filter available with the
\emph{HST}/WFC3 UVIS channel in the galaxy's rest frame.  This filter
has an effective wavelength $\lambda_{\rm eff} \sim 2359$ \AA\ and a
width of 467 \AA, and thus samples the spectral window of interest.

At the final time step of each realization, we rank order
the clusters by their F225W luminosity ($L^{\rm F225W}_{\rm cl}$).  We
then calculate the cumulative luminosity of the clusters at each rank
position, and divide this luminosity by the total integral F225W
luminosity of the system ($L^{\rm F225W}_{\rm tot}$).  Using these
cumulative distributions, we then count the number of clusters which
emit some fraction, $f_{\rm UV}$, of the total $L^{\rm F225W}_{\rm
  tot}$ in each realization.
Figure~\ref{fig.primus_slug} shows the distribution of these cluster
counts for three values of $f_{\rm UV} = 0.1, 0.2,$ and 0.5 (in green,
magenta, and orange, respectively).
In about half of our galaxy realizations, at least $10\%$ of the total 
$L^{\rm F225W}_{\rm tot}$ is produced by fewer than 10 clusters (see
the green histogram).
Indeed, in five realizations, the brightest cluster produces
\emph{more} than 10\% of $L^{\rm F225W}_{\rm tot}$.  However, it is
unusual for fewer than 10 clusters to produce more than $20\%$ of the 
total UV emission from each simulation (magenta histogram), and  fewer
than 10 clusters produce more than half of $L^{\rm F225W}_{\rm tot}$
in only two realizations (orange histogram).  This demonstrates that 
the UV continuum emission from such systems is not dominated by only a
handful of bright sources, but instead is generated in approximately
equal measure by many tens or
hundreds of star clusters.  

 We have used SLUG to verify that the number of these realizations that are
  dominated by a few bright clusters decreases with increasing SFR.
  This strongly
  suggests that our bright, blue b/g galaxy sample 
is less likely to be dominated by only a few bright clusters than
implied by
  Figure~\ref{fig.primus_slug}.  Furthermore, if we relax our
  assumption of a constant SFR, we find that we produce model galaxies
which are indeed dominated by a handful of bright star clusters
only if they are observed very close to the onset of a burst of star
formation.  For instance, for models with an
exponentially-decaying starburst
with a decay time $= 10$ Myr (and which produce a mean SFR of
$0.1~M_{\odot}~\rm yr^{-1}$ over 100 Myr), approximately two-thirds of
the model realizations yield UV continuum emission which is dominated
(at the $>50\%$ level)
by fewer than 10 clusters if they are observed within 20 Myr of the burst
onset.  At later times (i.e., within $\ge 40$ Myr), these clusters
have aged or have been disrupted, such that only $< 10\%$ of realizations remain
dominated by individual clusters to this extent.
Given this very short timescale, it is
unlikely that our b/g galaxy sample is composed primarily of objects in such a
cluster-dominated phase.

In what follows, we will use this analysis
to bolster our assumption that the UV continuum beams provided by our b/g galaxy
sample are made up of numerous point sources with a similar spatial
extent as is measured in the F814W band (i.e., in Figure~\ref{fig.primus_bgradhist}).
This will inform our
interpretation of our absorption line analysis as discussed in
\S\ref{sec.results_cgm} and in \refPaperII.

\section{The \ion{Mg}{2}-Absorbing CGM as Probed by PRIMUS Galaxies}\label{sec.results_cgm}

\subsection{The $W_{2796}$-$\mrperp$ Relation}\label{sec.results_wrperp}

\begin{figure*}
\begin{center}
\includegraphics[angle=0,width=\columnwidth,trim=0 0 0
-10,clip=]{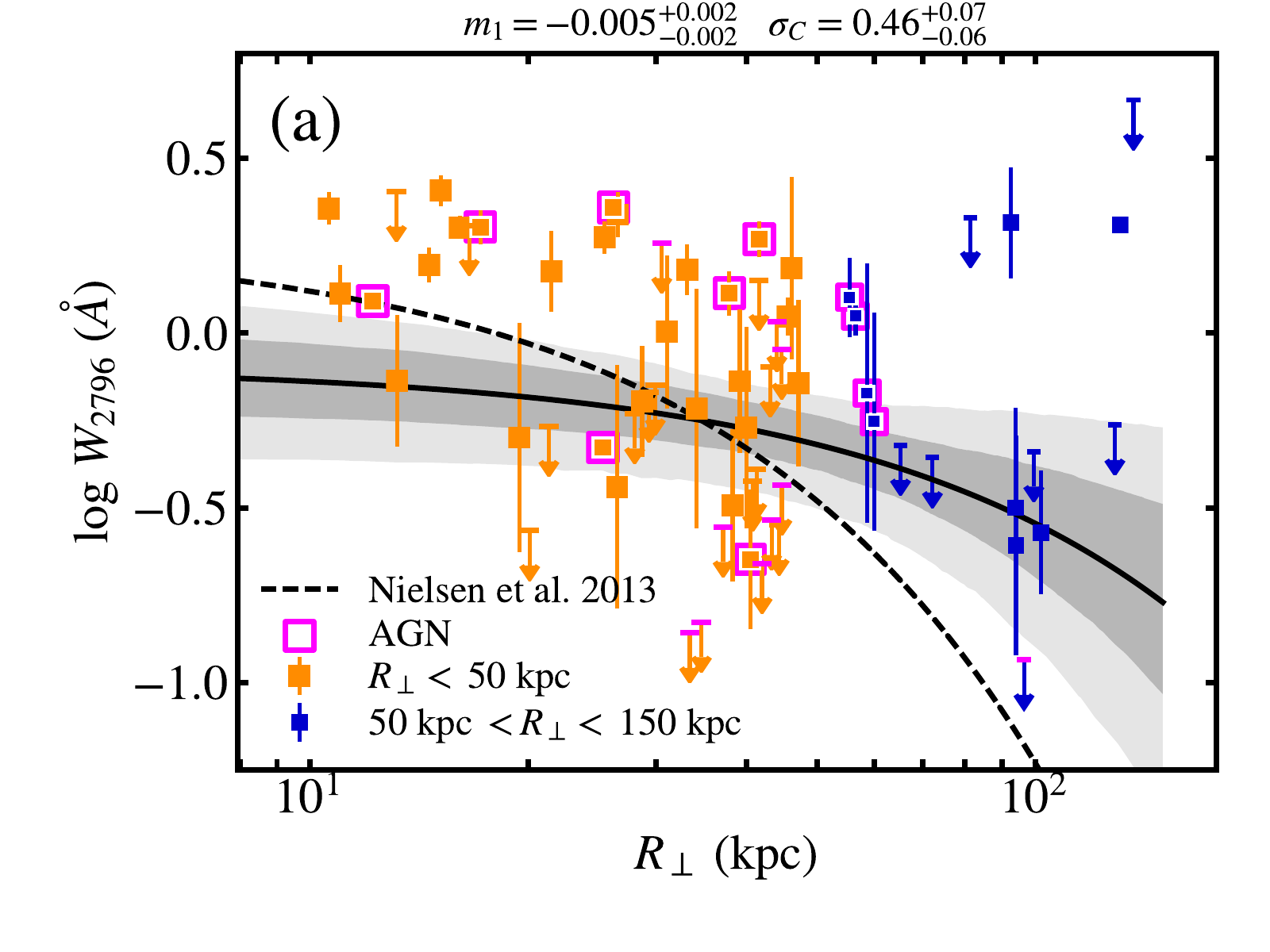}
\includegraphics[angle=0,width=\columnwidth,trim=0 0 0
-10,clip=]{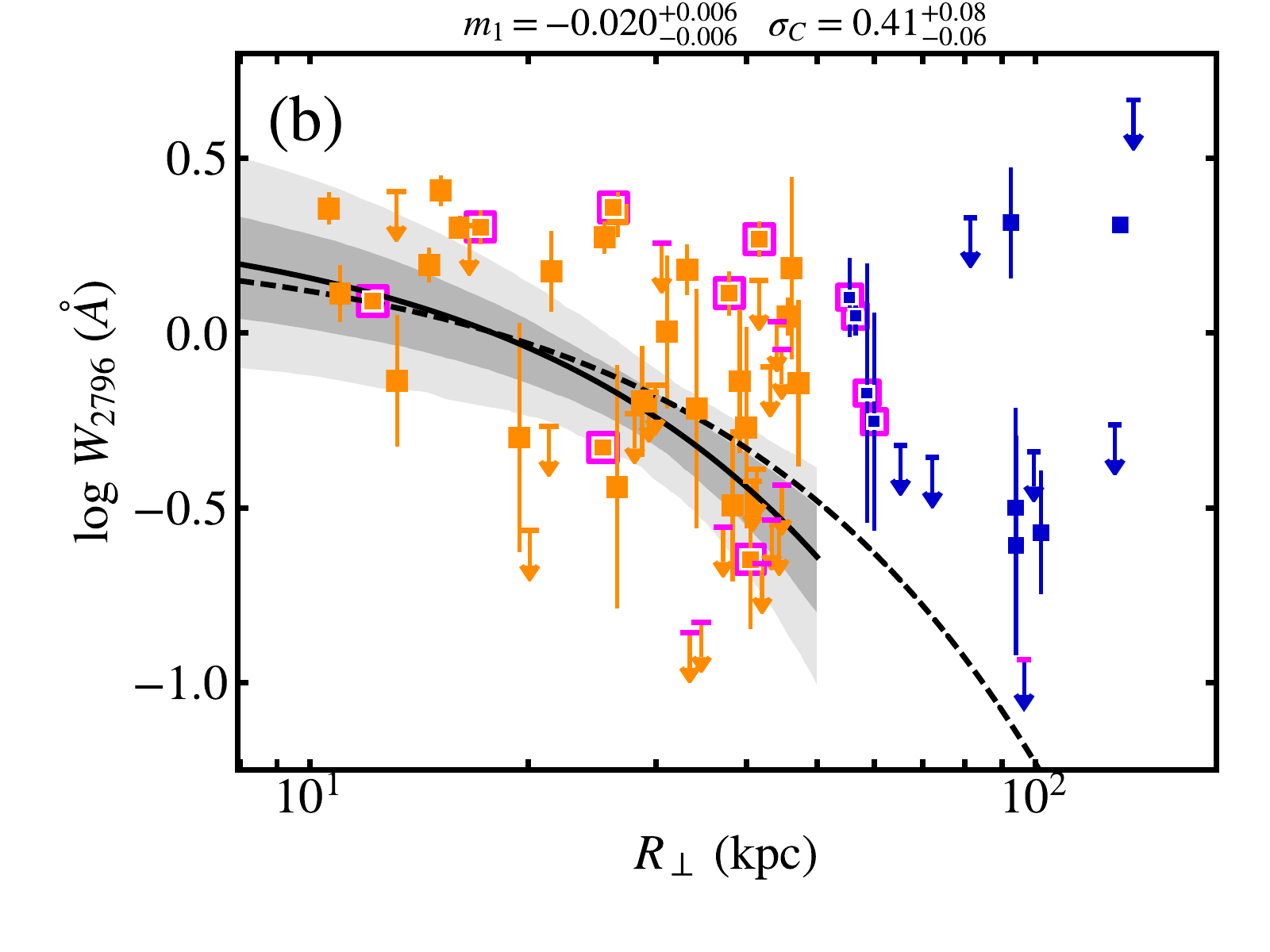}
\caption{\emph{(a)} $W_{2796}$ measured along PRIMUS b/g galaxy
  sightlines vs.\ 
  projected distance $\mrperp$ from the associated f/g galaxies.  The
  symbol colors and sizes match those used in
  Figures~\ref{fig.primus_s2n} and \ref{fig.primus_ms}.  
  Symbols outlined in magenta indicate pairs in
  which the background galaxy hosts a bright AGN.  The solid black line
  shows the ``best-fit'' linear relation between $\log W_{2796}$ and
  $\mrperp$ determined as described in
  Section~\ref{sec.results_wrperp} for the full dataset.  
The dark and light gray contours indicate the inner $\pm34\%$ and
$\pm47.5\%$ of the locus of fits determined from random draws
from the PPDF of this linear model.
The  best-fit values of the slope ($m_1$) and intrinsic scatter
  ($\sigma_C$) and their uncertainty intervals are printed above the plot.  The dashed black curve shows the log-linear fit to the
  QSO-galaxy pair Magiicat dataset from \citet{Nielsen2013}. 
  \emph{(b)}  The $W_{2796}$ measurements and \citet{Nielsen2013}
  relation are the same as shown in panel \emph{(a)}.  Here, the solid
  black line and dark
  and light gray contours show the best-fit relation between $\log
  W_{2796}$ and $\mrperp$ and the analogous uncertainty
  intervals
  for measurements with $\mrperp <
  50$ kpc.  
\label{fig.primus_ewrho}}
\end{center}
\end{figure*}

Figure~\ref{fig.primus_ewrho} shows our constraints on $W_{2796}$
measured around each of the sample f/g galaxies as a function of the
pair projected separation ($\mrperp$).  Pairs with $\mrperp < 50$ kpc
are indicated with large orange squares, and pairs with larger impact
parameters are shown with dark blue squares.  Symbols for pairs in which the b/g
galaxy is host to a bright QSO or broad-line AGN are outlined in
magenta.   It is those pairs without this indication that have b/g
sightlines which are not dominated by a bright nuclear source (see Figure~\ref{fig.primus_bgradhist}), and
hence which may be considered to offer a truly novel (i.e., spatially-extended) probe of f/g absorption.
$W_{2796}$ measurements for which the $\pm1\sigma$ uncertainty
interval extends to $<0.05$ \AA\ are shown as $2\sigma$ upper limits.  
Fifteen of our securely-detected absorbers have 
$W_{2796} >1.0$ \AA; $\sim6$ of these absorbers exhibit $W_{2796} > 2$ \AA.
Furthermore, we are approximately equally likely to detect such strong
absorbers toward non-AGN hosts as we are toward bright AGN.
The overall sensitivity of our survey is lower than that of
the QSO-galaxy comparison samples, such that a number of our b/g sightlines yield
quite weak upper limits on $W_{2796}$.  However, while the b/g QSOs in
our sample provide the most constraining $W_{2796}$ limits (at $\sim
0.15$ \AA), we are able to place limits as low as $\sim0.3$ \AA\ using
a few of our non-AGN host b/g objects.  

Numerous QSO-galaxy pair studies have noted a decline in $W_{2796}$
with increasing $\mrperp$ at high statistical significance
\citep[e.g.,][]{LanzettaBowen1990, Steidel1995,Kacprzak2008, Chen2010,Nielsen2013}.
To characterize the relation between $W_{2796}$ and $\mrperp$ in our
sample, we model our dataset assuming a linear dependence of $\log
W_{2796}$ on $ \mrperp$ as described by 
\citet{Nielsen2013}:
\begin{equation}\label{eq.W_Rperp}
  \log W_{2796} = b + m_1 \mrperp. 
\end{equation}
Following the methodology of \citet{Chen2010}, the likelihood
function for this model can be written
\begin{multline*}
  \mathcal{L}(\overbar{W}) = \left (\prod_{i=1}^{n} \frac{1}{\sqrt{2\pi
      s_i^2}} \exp \left
      \{-\frac{1}{2} \left [\frac{W_i - \overbar{W}}{s_i} \right
      ]^2 \right \} \right )\\
 \times \left (\prod_{i=1}^{m}
    \int_{-\infty}^{W_i} \frac{dW'}{\sqrt{2\pi
      s_i^2}} \exp \left
      \{-\frac{1}{2} \left [\frac{W' - \overbar{W}}{s_i} \right
      ]^2 \right \} \right ),  
\label{eq.likelihood}
\end{multline*}
with $W_i$ representing the $\log W_{2796}$ value for each measurement
$i$, and $\overbar{W}$ equal to the value of $\log W_{2796}$ given by the
model at each $R_{\perp,i}$.  
The first product includes all $n$
systems which yield a direct measurement of $\log W_{2796}$, and the
second includes the $m$ systems for which our constraint on $\log
W_{2796}$ is an upper limit.  We assume that the 
Gaussian variance in this expression has two components:
\begin{equation}
  s_i^2 = \sigma_i^2 + \sigma_C^2, 
\end{equation}
with $\sigma_i$ representing the measurement uncertainty in $W_{2796, i}$, and $\sigma_C$ an
additional factor which accounts for intrinsic scatter in the
relation.\footnote{This follows the recommendations for model fitting
  offered in the online documentation for the \texttt{emcee} software package at http://dan.iel.fm/emcee/current/user/line/.}
Hence, because we are making the assumption that the dispersion in the
quantity $\log W_{2796}$ is Gaussian, we are equivalently assuming
that the scatter in $W_{2796}$  is
lognormal.  
While this assumption is not unreasonable, we caution that the number of measurements
in both our PRIMUS pair and QSO-comparison datasets is insufficient to
perform a test with the power to rule out this claim (i.e., to
constrain the shape of the $\log W_{2796}$ distribution over a narrow
range in $\mrperp$).  

We sample the posterior probability density function (PPDF) for this
model using the Markov Chain Monte Carlo technique as implemented in
the Python software package \texttt{emcee}, an open source code 
which uses an affine-invariant ensemble sampler \citep{DFM2013}.  We
adopt uniform probability densities over the intervals $-5.0 < m_1 <
5.0$, $-10.0 < b < 10.0$, and $-10.0 < \ln \sigma_C < 10.0$ as priors.  We
find that Markov chains generated by 100 ``walkers'' each taking 6000 steps
(and discarding the first 1000 steps)
provide a thorough sampling of the PPDF in each parameter dimension.
The code outputs the parameter values with maximum likelihood, as well
as marginalized PPDFs.  In the following, we adopt the median and
$\pm34$th-percentiles of these PPDFs as the ``best''
value of each parameter and its uncertainty interval.

The results of this procedure are shown in
Figure~\ref{fig.primus_ewrho}.  
Panel \emph{(a)} shows 
the best-fit relation between
$\mrperp$ and $\log W_{2796}$ for 
all the datapoints plotted (i.e., for 0 kpc $< \mrperp <$ 150 kpc).  
The dark and light gray contours 
were obtained by first 
selecting 1000 sets of parameter values from
the PPDF at random, determining the corresponding $\overbar{W}$ values
for each parameter set, and then filling in the region of the
figure containing the inner $\pm34\%$ and $\pm47.5\%$ of $\overbar{W}$ values at each point along the x-axis.
This relation is quite flat, with a slope of only $m_1 =
-0.005\pm0.002$, an intercept $b = -0.09_{-0.12}^{+0.12}$, and an
intrinsic scatter $\sigma_C = 0.46_{-0.06}^{+0.07}$.  Such a flat slope
is in fact inconsistent with the log-linear fit
to the Magiicat dataset over a similar range in $\mrperp$ (5 kpc $<
\mrperp < $ 200 kpc) presented in \citet[][with $m_1 = -0.015 \pm
0.002$ and $b = 0.27 \pm 0.11$; dashed line in Figure~\ref{fig.primus_ewrho}]{Nielsen2013}.

A by-eye comparison of the distribution of the points in the
figure and the locus of the \citet{Nielsen2013} relation suggests
that the PRIMUS pair dataset may be offset to higher $W_{2796}$ at
high impact parameters in particular; and furthermore, that it is
these high $W_{2796}$ values at large $\mrperp$ which tend to flatten
the best-fit slope.  
Restricting our fitting
procedure to the measurements within $\mrperp <50$ kpc (i.e., to the
region of parameter space over which our sampling is most thorough),
we show the resulting best fit  and
corresponding uncertainty
intervals with a solid black line and gray contours in Figure~\ref{fig.primus_ewrho}b.  This yields
a somewhat steeper relation, consistent with that of \citet{Nielsen2013}, with  $m_1 =
-0.020\pm0.006$, an intercept $b = 0.36\pm0.19$, and $\sigma_C = 0.41_{-0.06}^{+0.08}$.
We  note that whereas \citet{Nielsen2013} limited their
f/g galaxy sample to include only ``isolated'' objects (i.e.,
objects without a neighbor within 100 kpc and having a velocity
separation of $<500\mkms$), the PRIMUS pair sample is not 
restricted based on environment.  Indeed, \citet{Chen2010} found that
$W_{2796}$ exhibits no significant trend with increasing $\mrperp$
around galaxies within group environments, suggesting that a
simultaneous fit to $\log W_{2796}$ vs.\ $\mrperp$ including both
isolated and group f/g galaxies would yield a  flatter relation
than that reported by \citet{Nielsen2013}.
At the same time, however, the strongest absorber among the eight sightlines probing group
environments in the \citet{Chen2010} sample has $W_{2796} =
0.79\pm0.03$ \AA\ --
i.e., well below several of the PRIMUS f/g absorbers at $\mrperp > 50$ kpc.




\subsection{The Relation Between $W_{2796}$ and Intrinsic Host Galaxy
  Properties}\label{sec.results_wrintrinsic}

We now test our sample for additional correlations between $W_{2796}$ and
intrinsic host galaxy properties (e.g., $M_*$, SFR, and sSFR) at fixed
values of $\mrperp$.  Such correlations have been noted in several
studies, beginning with \citet{Chen2010b}.   These authors found that the scatter
in the relation between $\log W_{2796}$ and $\log \mrperp$ is reduced when additional terms which scale
linearly with $\log M_*$
and $\log~\rm sSFR$ are included.  More recently, using a larger
QSO-galaxy pair sample (including both absorption-selected galaxies
and pairs selected without prior knowledge of the presence of halo
\ion{Mg}{2} absorption), \citet{Nielsen2013} presented strong evidence for an
increase in $W_{2796}$ with increasing host galaxy $B$- and $K$-band
luminosity at fixed $\mrperp$, and further reported a weak
dependence of  $W_{2796}$ on galaxy color of marginal statistical significance.
\citet{Churchill2013}, in their discussion of the same dataset,
interpreted these results as indicative of a positive scaling between 
 $W_{2796}$ and the virial mass of the host halo at fixed
$\mrperp$.  \citet{Lan2014} additionally reported that relative to all
galaxies within $\mrperp < 50$ kpc of \ion{Mg}{2} absorbers, those
with higher
SFR and sSFR are associated with increasingly enhanced excess
$W_{2796}$.  They further drew a distinction between star-forming
and quiescent host galaxies, finding that while this excess $W_{2796}$
tends to increase with the $M_*$ of star-forming galaxies,
quiescent galaxies do not give rise to a significant $W_{2796}$ excess
regardless of their $M_*$ (within $\mrperp < 50$ kpc).  And at higher
redshifts ($z\sim2$), discovery of the high
incidence and absorption strength in low-ionization metal transitions 
measured in the massive halos of QSO host galaxies has likewise
pointed to a positive correlation between cool gas absorption strength and 
halo mass, at least among the active star-forming and QSO hosts which
have been studied at such early epochs \citep{QPQ5,QPQ7}.

\begin{figure}
\begin{center}
\includegraphics[angle=0,width=\columnwidth,trim=0 0 0
0,clip=]{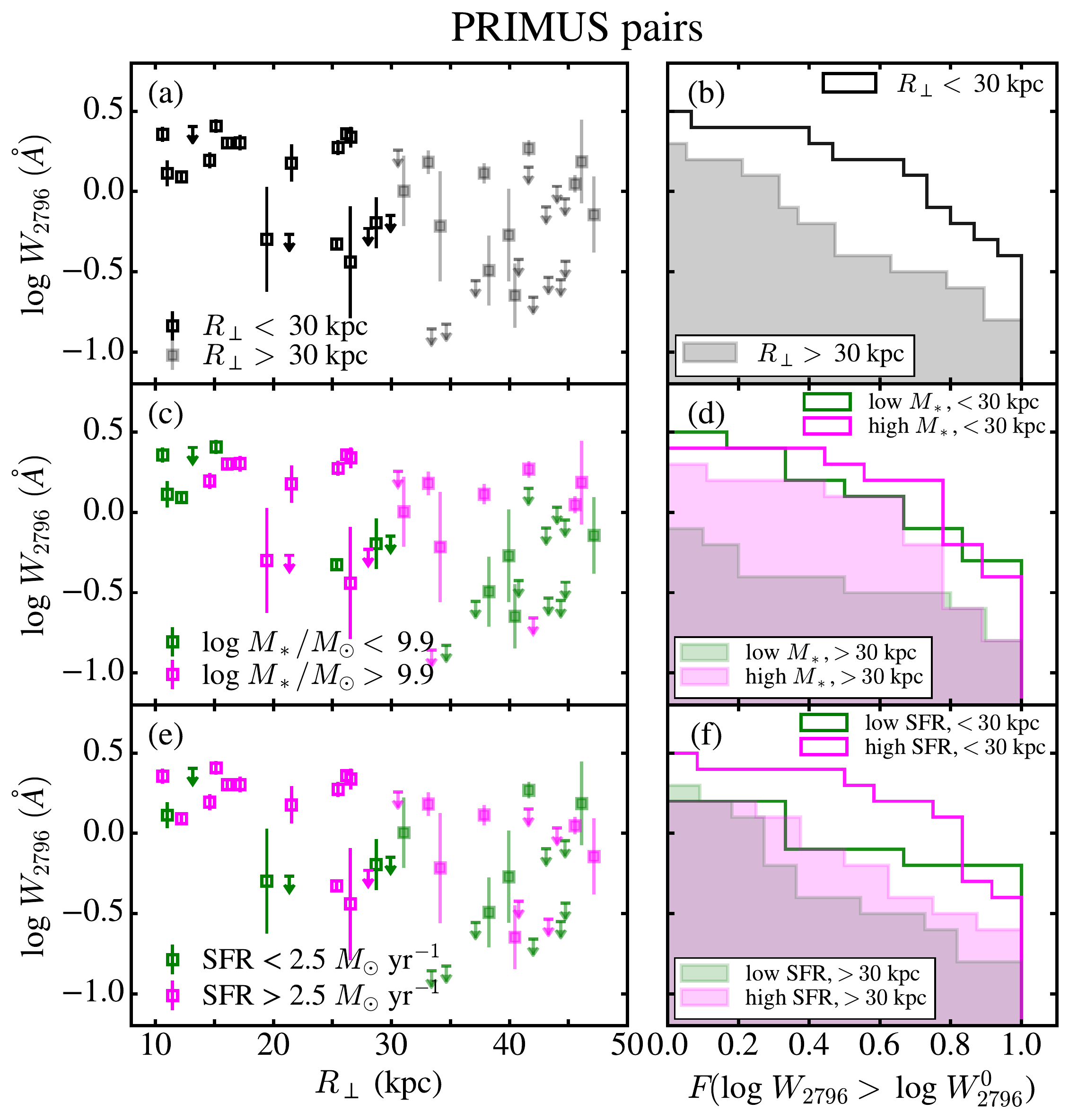}
\caption{\emph{(a)} $\log W_{2796}$ vs.\ projected distance
  $\mrperp$ measured along PRIMUS b/g sightlines
  probing star-forming f/g halos.  Sightlines within $\mrperp < 30$ kpc and at 30 kpc $< \mrperp < $
  50 kpc are indicated with open black and filled gray symbols,
  respectively.  \emph{(b)} Cumulative distribution of $\log W_{2796}$ in
  each subsample shown in panel \emph{(a)}.  The x-axis value
  indicates the fraction of sightlines having $\log W_{2796}$
  greater than the y-axis value within $\mrperp < 30$ kpc (open black histogram) and with 30 kpc $< \mrperp < $
  50 kpc (filled gray histogram).  
  \emph{(c)} The identical $\log W_{2796}$ vs.\ 
  $\mrperp$ distribution shown in panel \emph{(a)}, with the point
  color indicating f/g galaxies with $\log M_{*}/M_{\odot} < 9.9$
  (green) and $\log M_{*}/M_{\odot} > 9.9$ (magenta).  Panel
  \emph{(d)} shows the cumulative distributions of $\log W_{2796}$ for these low- and high-$M_*$ subsamples
  at $\mrperp < 30$ kpc (open green and magenta histograms) and at 30
  kpc $< \mrperp < $ 50 kpc (filled green and magenta histograms).  
  Panel \emph{(e)} shows the same measurements once again,
  here with the datapoints color-coded according to the SFR of the f/g
  galaxy as indicated in the legend.  \emph{(f)} Cumulative
  distributions of $\log W_{2796}$ for these low- and high-SFR subsamples
  as described above.
Upper limits on $W_{2796}$ are included in all cumulative distributions
  at their $2\sigma$ values if they are $< 0.5$ \AA.  All other limits
  are excluded.  Sightlines passing close to
  higher-$M_*$ galaxies tend to yield higher $W_{2796}$, particularly
  at 30 kpc $< \mrperp <$ 50 kpc.  Larger $W_{2796}$ values also tend
  to arise around host galaxies with higher SFR.
\label{fig.primus_ewcumdist}}
\end{center}
\end{figure}

We approach a test for such correlations by comparing 
the cumulative distributions of $W_{2796}$
values among subsamples of sightlines in our survey.  First, we
isolate the handful of sightlines ($\sim7$) which probe quiescent galaxy
halos, defined 
as described by Eq.~\ref{eq.sfcrit}.
To control for the possible effect of galaxy quiescence, and because
there are relatively few of these sightlines in our sample, we exclude
them from all cumulative distributions described below.  We then
subdivide our sample into two bins with $\mrperp < 30$ kpc and 30 kpc $<
\mrperp <$ 50 kpc as shown in Figure~\ref{fig.primus_ewcumdist}a with
black open and gray filled points, respectively.  In Figure~\ref{fig.primus_ewcumdist}b, we show the 
cumulative distribution ($F(\log W_{2796} > \log W_{2796}^0)$) of $\log W_{2796}$ (i.e., the fraction of systems having
$\log W_{2796}$ greater than a given value $\log W_{2796}^{0}$) for sightlines having
$\mrperp < 30$ kpc (black open histogram) and for those at 30 kpc $<
\mrperp <$ 50 kpc (gray filled histogram).  
All upper limits on
$W_{2796}$ are included in these distributions at their
$2\sigma$ values if they are $< 0.5$ \AA; weaker limits are excluded,
as they may have the effect of inflating $F(\log W_{2796} > \log W_{2796}^0)$ at large
$W_{2796}$ relative to the $F(\log W_{2796} > \log W_{2796}^0)$ measured in a more sensitive
spectroscopic survey.  

\begin{figure}
\begin{center}
\includegraphics[angle=0,width=\columnwidth,trim=0 0 0
0,clip=]{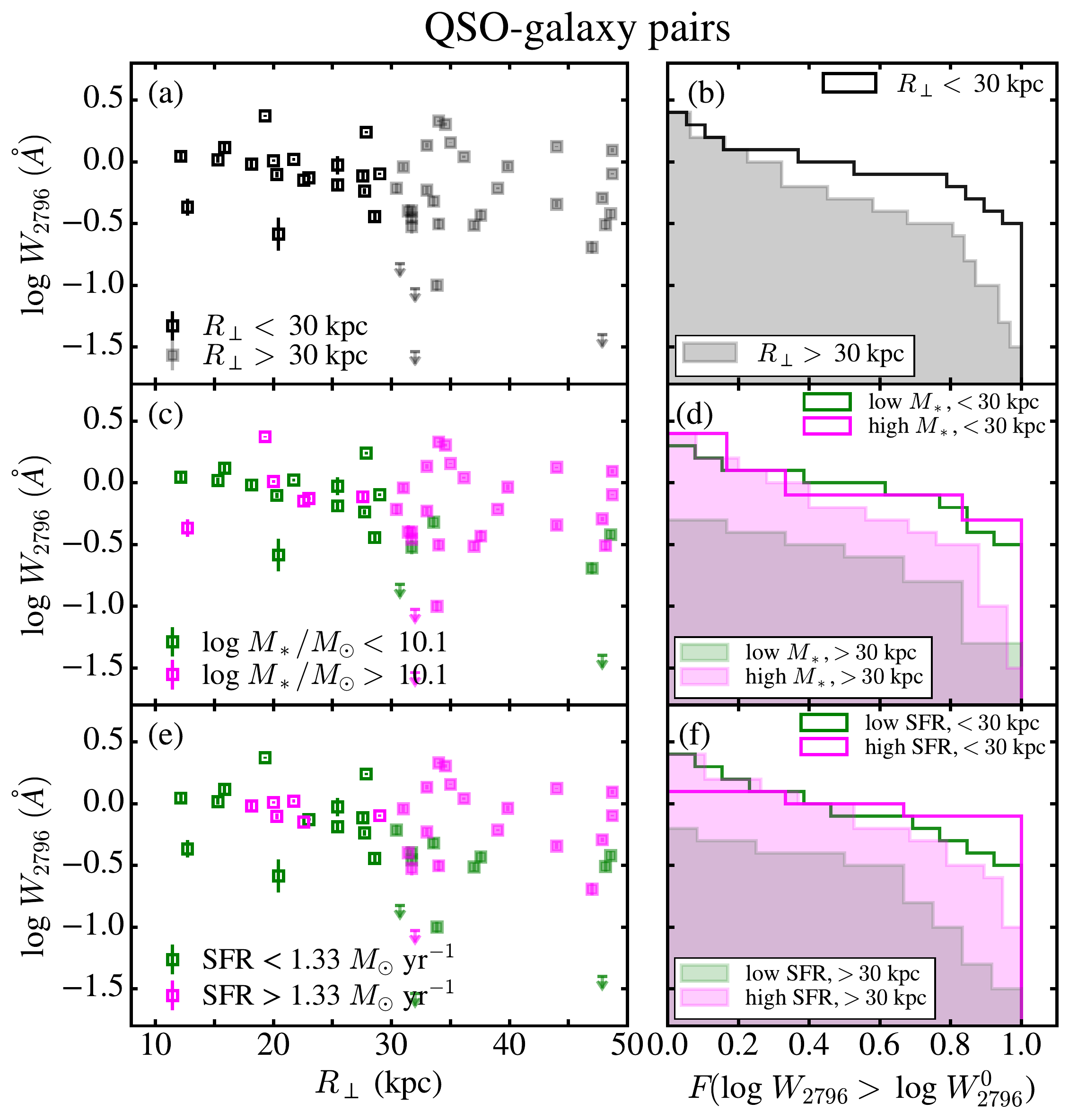}
\caption{Same as Figure~\ref{fig.primus_ewcumdist}, for our
  QSO-galaxy comparison dataset adopted from \citet{Chen2010} and \citet{Werk2013}.  Here we use slightly different
  values of $\log M_{*}/M_{\odot}$ and SFR to subdivide the sample in
  panels \emph{(c)}, \emph{(d)}, \emph{(e)}, and \emph{(f)} as
  indicated in the legends.  
\label{fig.primus_qsoewcumdist}}
\end{center}
\end{figure}

It is evident from Figure~\ref{fig.primus_ewcumdist}b that the cumulative distribution within $\mrperp < 30$ kpc includes a
higher frequency of large $W_{2796}$ values relative to the $F(\log
W_{2796} > \log W_{2796}^0)$ at larger impact
parameters.  We test the statistical significance of this offset using
\texttt{ASURV} Rev 1.2 \citep{Lavalley1992}, a software package designed for statistical analysis of
censored data invoking methods presented in \citet{FeigelsonNelson1985}.
Using \texttt{ASURV}, we perform a Gehan's generalized Wilcoxon test of the
probability that these two $W_{2796}$ distributions are drawn from the
same parent population.  Here, we do not exclude any upper limits on
$W_{2796}$, regardless of their value, as \texttt{ASURV} properly accounts for
the weighting of censored data.  The results of this test, along with 
the number of sightlines  and the median $W_{2796}$ 
in each subsample, are included in the first column of data in Table~\ref{tab.galpropsplit}.
The median $W_{2796}$ value is nearly 1 \AA\ higher for sightlines within 30 kpc
relative to sightlines at 30 kpc $< \mrperp <$ 50 kpc, and the
probability that these distributions are drawn from the same parent
population is only $P=0.003$.  This result is unsurprising, and is
fully consistent with the findings of \S\ref{sec.results_wrperp}.  

We next further subdivide each of the samples described above
by the $M_*$, SFR, and sSFR of the associated f/g galaxies, again
excluding any quiescent systems.  
Figure~\ref{fig.primus_ewcumdist}c
includes the same measurements
plotted in panel \emph{(a)}, here with green symbols indicating 
sightlines probing the halos of f/g galaxies having $\log M_* /
M_{\odot} < 9.9$, and with magenta symbols indicating sightlines
probing higher-$M_*$ systems.  Figure~\ref{fig.primus_ewcumdist}d 
shows 
$F(\log W_{2796} > \log W_{2796}^0)$ for each of these subsamples:
those with 
$\mrperp < 30$ kpc  are shown with green and magenta open histograms,  
and the subsamples 
at 30 kpc $< \mrperp <$ 50 kpc are shown with green and magenta filled histograms.
The cumulative distributions for the two
subsamples at small impact parameters appear similar, 
 and our statistical test ($P=0.892$) fails to rule
out the null hypothesis that they are drawn from the same parent population
(Table~\ref{tab.galpropsplit}).  
At 30 kpc $< \mrperp <$ 50
kpc, however, there is a very low probability ($P=0.006$) that the subsamples
divided by $M_*$ are drawn from the same distribution, with the
high-$M_*$ subsample exhibiting a median $W_{2796}$ $\sim0.9$ \AA\
above that of the low-$M_*$ subsample.  
We then repeat this analysis, instead subdividing the samples at SFR $=
2.5~M_{\odot}~\rm yr^{-1}$,  
and show the resulting scatterplot and cumulative distributions in 
Figure~\ref{fig.primus_ewcumdist} panels \emph{(e)} and \emph{(f)}.  We find that in general,
subsamples probing higher-SFR f/g galaxy halos tend to exhibit higher
$W_{2796}$, but that the probability of the low- and high-SFR
subsamples originating from the same distribution rules out the null
hypothesis only within $\mrperp < 30$ kpc ($P=0.049$).  
Finally, we subdivide the $\mrperp < 30$ kpc and 30 kpc $< \mrperp < $
50 kpc 
subsamples by sSFR at $\log \rm sSFR~ [yr^{-1}] = -9.46$,
finding that the corresponding cumulative distributions 
appear similar in both impact parameter bins, 
 and finding no evidence suggesting their parent populations are distinct
(see Table~\ref{tab.galpropsplit}).

Broadly speaking, we see evidence for
larger-$W_{2796}$ systems associated with host galaxies with higher
$M_*$ or SFR.  This trend is not statistically significant in every
$\mrperp$ bin tested; however, we consider our results to be
qualitatively similar to those reported in the studies described
above, as
our relatively small sample size and
low-S/N spectroscopy cannot rule out the persistence of these relationships
over a wider range in $\mrperp$.  To more directly test for
consistency with previous work, however, we now
perform the same analysis laid out above on the QSO-galaxy comparison
sample described in \S\ref{sec.fg}.  In 
Figure~\ref{fig.primus_qsoewcumdist}, we show the $W_{2796}$ vs.\ $\mrperp$
distributions and cumulative
distribution of $W_{2796}$ values in this comparison sample in
subsamples divided by $\mrperp$ (panels \emph{(a)} and \emph{(b)}),
$M_*$ (panels \emph{(c)} and \emph{(d)}), and SFR (panels \emph{(e)} and \emph{(f)}).   Here, we have
excluded sightlines probing quiescent galaxies 
defined as in Eq.~\ref{eq.sfcrit}.
We treat upper limits on $W_{2796}$ as
described above; in practice, all upper limits are included in these
distributions, as the QSO spectroscopy tends to be much more sensitive
than our b/g galaxy spectroscopy.  
We also adopt slightly different values for
the $M_*$ and SFR at which we split these subsamples,
adjusting our divisions 
so that there are at least 5 objects in each.  
The specific values
chosen are indicated in the legends in
Figure~\ref{fig.primus_qsoewcumdist}. 
For completeness, we also subdivide
this sample by sSFR, and show the results of our tests
for consistency among the corresponding cumulative distributions in Table~\ref{tab.galpropsplit}.

We do not rule out consistency between
the distributions of $W_{2796}$
in these QSO sightlines at impact parameters $< 30$ kpc when dividing the
sample by $M_*$ or SFR.
  At 30 kpc $<
\mrperp <$ 50 kpc, however, the median $W_{2796}$ of the high-$M_*$
and high-SFR subsamples are $\sim0.4$ \AA\ higher than in the
low-$M_*$ and SFR subsamples, and there is a very low probability that
these subsets are drawn from the same parent population ($P<0.02$).
We find no evidence for significant differences between the subsamples divided
by sSFR.  

 These
  findings are fully consistent with those reported above for the
  PRIMUS subsamples divided both by $M_*$ and sSFR.  We note that the PRIMUS
  subsamples divided by SFR yield somewhat different results: we fail
  to rule out the null hypothesis for low- vs.\ high-SFR PRIMUS f/g
  galaxies at 30 kpc $<\mrperp <$ 50 kpc, and rule it out at
  $\sim2\sigma$ significance at $\mrperp < 30$ kpc.  It is possible
  that this discrepancy is due to the differences in background beam sizes used to
probe the PRIMUS vs.\ QSO-galaxy f/g samples; however, given the overall
agreement between all other two-sample test results, we argue that this
discrepancy is more likely due to
differences in the SFR distributions of the two samples and/or to the
large uncertainties in our SFR estimates (i.e., relative to the
uncertainties associated with our estimates of $M_*$.)

All together, the foregoing analysis of both the PRIMUS and
QSO-galaxy pair samples
points to an increase in $W_{2796}$ with both $M_*$ and SFR.   Although this
dependence is not found to be statistically significant in both impact parameter
bins studied, we consider our findings to further demonstrate a qualitative consistency
between the trends in the $W_{2796}$ distributions observed toward
b/g QSOs and b/g galaxies.

\subsection{Comparison between Galaxy-Galaxy and QSO-Galaxy Samples}\label{sec.results_comparison}

\begin{figure*}
\begin{center}
\includegraphics[angle=0,width=7in,trim=0 0 0
0,clip=]{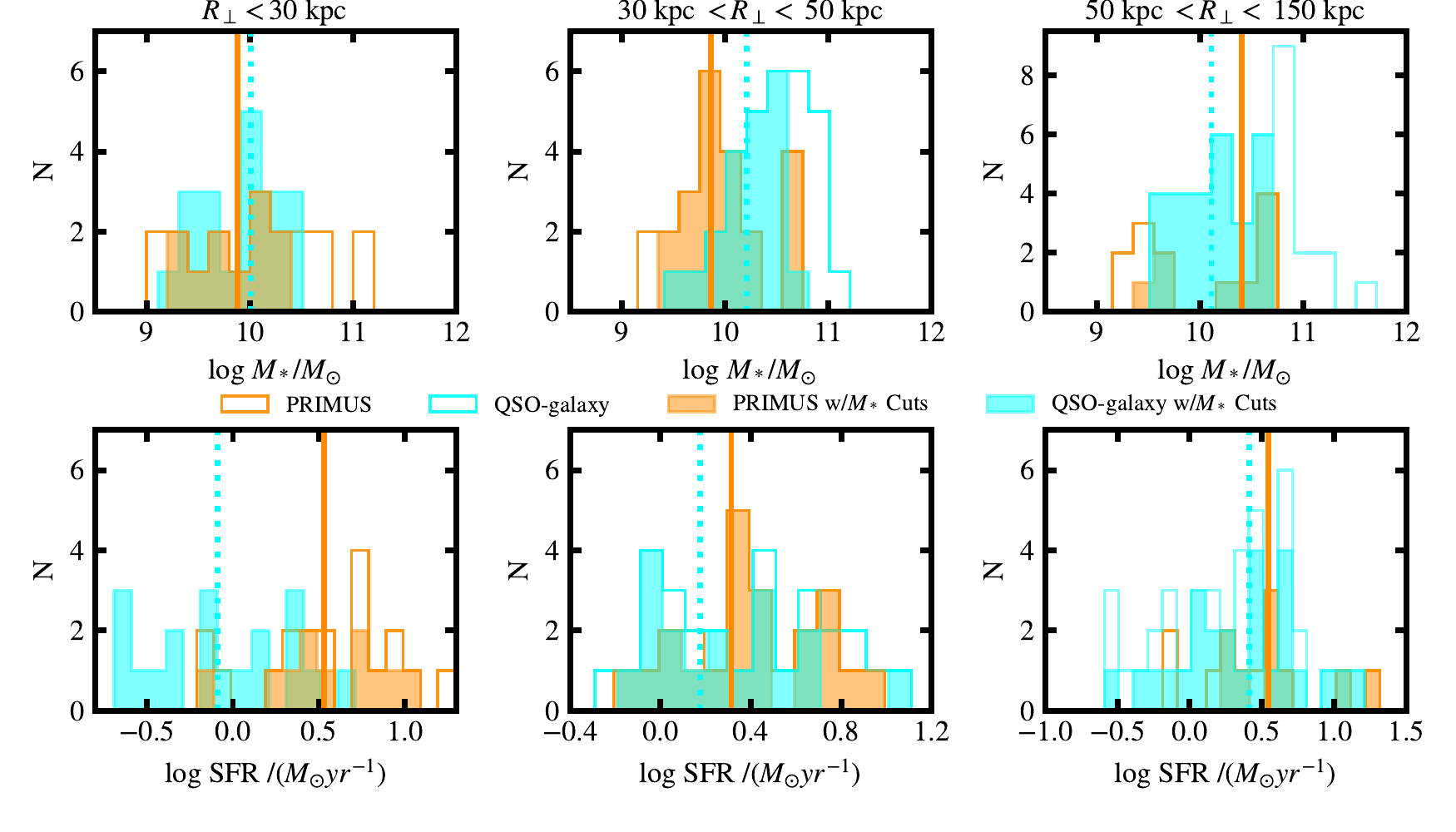}
\caption{\emph{Top Row:} The distribution of $M_*$ for all
  star-forming f/g galaxies in PRIMUS (orange open and filled histograms) and in the QSO-galaxy
  comparison sample (cyan open and filled histograms) within $\mrperp
  < 30$ kpc (left), with
  30 kpc $< \mrperp <$ 50 kpc (middle), and with  50 kpc $< \mrperp <$
  150 kpc (right).  The filled portion of each
  histogram shows systems we select for comparison between the two samples
  (such that they span the same ranges in $M_*$ in each impact
  parameter bin).  The median value of each filled histogram is shown
  with a vertical curve of the same color.  \emph{Bottom Row:}  The
  distribution of SFR for each subsample shown in the top row.
  The median values are indicated with vertical curves as above.  The
  SFRs of the f/g galaxies in the QSO-galaxy comparison sample tend to
  be much lower ($\sim0.6$ dex in the median) than in our PRIMUS
  sample at $\mrperp < 30$ kpc.  At 30 kpc $< \mrperp <$ 50 kpc
  these SFRs have similar distributions.  
\label{fig.primus_compare_to_stacks}}
\end{center}
\end{figure*}

We also note, however, that the PRIMUS sightline subsample
within $\mrperp < 30$ kpc exhibits a median
$W_{2796}=1.3$ \AA, $\sim0.5$ \AA\ higher than the analogous
QSO-galaxy comparison
sightline sample.  To investigate the possible origin of this offset,
in Figure~\ref{fig.primus_compare_to_stacks} we examine the distributions
of $M_*$ (top row) and SFR (bottom row) in the $\mrperp < 30$ kpc
(left column) and
30 kpc $< \mrperp <$ 50 kpc (middle column) subsamples discussed
above.
The same distributions for sightlines at 50 kpc $< \mrperp <$ 150 kpc
are shown for completeness in the right-most column.
The combined open/filled orange histograms show these distributions
for all star-forming f/g galaxies in PRIMUS, while the combined
open/filled cyan histograms include all star-forming f/g galaxies in
the QSO-galaxy comparison sample.  Comparing these distributions in
the left-hand column, we see that the PRIMUS f/g galaxy sample
includes a few objects with $\log M_*/M_{\odot} > 10.5$, while the
stellar masses in the 
QSO-galaxy sample do not exceed this limit.  The
QSO-galaxy sample also has SFRs which are lower overall by $\gtrsim
0.5$ dex.  The evidence presented in
\S\ref{sec.results_wrintrinsic} for larger $W_{2796}$ around f/g
galaxies of higher $M_*$ and/or SFR suggests that the higher median
$W_{2796}$ measured around the PRIMUS f/g galaxies may indeed be
due to the larger stellar masses or SFRs of this subsample.  
Alternatively, these offsets could in principle also arise from 
the differences in the sizes of the b/g
beams used in the two studies.  Here, we test the former hypothesis;
however, we will return to the potential effects of b/g beam size on
the $W_{2796}$ distribution of f/g absorbers in \refPaperII.


At 30 kpc $< \mrperp <$ 50 kpc (middle column), the PRIMUS
f/g galaxies tend to have slightly lower $M_*$ values than their
counterparts in the QSO-galaxy 
sample, but occupy a comparable
range in SFR.  Indeed, their median $W_{2796}$ values are  similar (0.27
\AA\ vs.\ 0.43 \AA; see Table~\ref{tab.galpropsplit}).  However, to test for full consistency between
these samples, we now select subsets of the distributions shown in
Figure~\ref{fig.primus_compare_to_stacks} for a more detailed
comparison.  At small impact parameters ($\mrperp < 30$ kpc), we
exclude galaxies from the PRIMUS sample which have $M_*$ outside 
the range spanned by the QSO-galaxy comparison sample;
i.e., we exclude galaxies with $\log M_*/M_{\odot} < 9.1$ or $\log
M_*/M_{\odot} > 10.4$.  Those galaxies which remain are included in
the filled orange histograms.  At larger impact parameters (30 kpc $<
\mrperp <$ 50 kpc), we exclude PRIMUS galaxies with stellar masses
lower than the least massive galaxy in the QSO-galaxy comparison
sample ($\log M_*/M_{\odot} < 9.4$; filled orange histograms).
We also remove QSO-galaxy  pairs with f/g masses higher than the most massive PRIMUS galaxy
in this bin ($\log M_*/M_{\odot} >10.7$; filled cyan histograms).  
Finally, for a complete comparison, we select a subsample of PRIMUS sightlines at 
50 kpc $<\mrperp <$ 150 kpc with f/g galaxies within the stellar mass
range spanned by the QSO-galaxy comparison sample ($9.5 < \log
M_*/M_{\odot} < 11.6$), and refine this QSO-galaxy subsample by
removing objects with $\log M_*/M_{\odot} > 10.7$.

The median values of each of these trimmed sample distributions are indicated with
vertical lines.  The median stellar masses of the modified samples are
very similar, and the median SFRs are within $<0.2$ dex at 30 kpc $<
\mrperp <$ 50 kpc and 50 kpc $<
\mrperp <$ 150 kpc.  The SFRs of the trimmed samples at $\mrperp < 30$
kpc remain offset by $\sim0.6$ dex, however, and we must consider this
caveat as we proceed with our comparison of the $W_{2796}$
distributions of these subsamples.

We now wish to compare the median and dispersion in $W_{2796}$
in each of these trimmed samples.  However, because many of
our PRIMUS sightlines yield only upper limits on $W_{2796}$, it is not
straightforward to calculate these statistics from analysis of the
measurements of $W_{2796}$ in individual spectra.  Instead, we 
coadd our spectroscopy of these sightlines to obtain the median 
normalized flux value as a function of wavelength, and use a
bootstrapping analysis to estimate the dispersion in these flux
values.  We then perform absorption line analysis on the resulting
coadds to assess the median and dispersion in $W_{2796}$.  

\begin{figure}
\begin{center}
\includegraphics[angle=0,width=3.5in,trim=30 170 180 160,clip=]{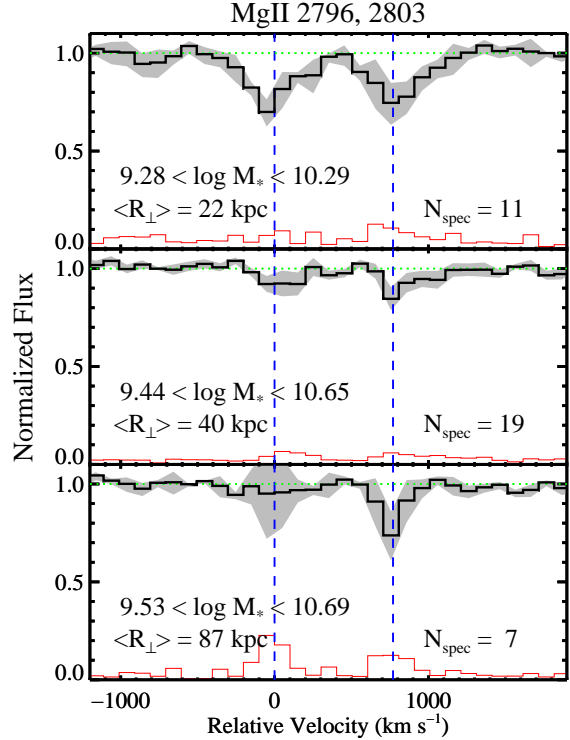}
\caption{Coadded spectra (black) of our PRIMUS b/g galaxy sightlines in the
  region around the \ion{Mg}{2} 2796, 2803 transition at $z_{\rm f/g}$.  The spectra in
  the top, middle, and bottom panels
  include all pairs in the orange filled histograms in the
  left-, middle-, and right-hand columns of
  Figure~\ref{fig.primus_compare_to_stacks}, respectively, except for
  those pairs with b/g spectra having $\rm
  S/N($\ion{Mg}{2}$) < 4$ \AA$^{-1}$.  
  The mean $\mrperp$ of all sightlines in each
  coadd is noted at the lower left of each panel, and the number of
  sightlines is noted at lower right.  The filled gray curves
  show the $\pm34$th-percentile interval for the fluxes in our
  bootstrap sample in each pixel.  The red histogram shows this same
  $1\sigma$ error array.  The vertical blue dashed lines mark the rest
  velocity of each transition in the \ion{Mg}{2} doublet.  
\label{fig.primus_qsocompstacks}}
\end{center}
\end{figure}

In detail, we first expunge those spectra having particularly low $\rm
S/N$ (i.e., $\rm S/N
($\ion{Mg}{2}$) < 4$ \AA$^{-1}$), as we have found that the poor
continuum normalization of these sightlines can introduce spurious
features into the final coadds (see, e.g., the \ion{Mg}{2} profile for
object 1611 in Appendix Figure~\ref{fig.primus_allmgii_p2}).  In
practice, this eliminates only two sightlines from the subsamples at 
30 kpc $< \mrperp <$ 50 kpc and 50 kpc $< \mrperp <$ 150 kpc, and does
not eliminate any sightlines within 30 kpc.
Then, 
using the method described in Section 3 of
\citet{QPQ7}, we linearly interpolate the continuum-normalized flux in
each remaining spectrum onto $100\mkms$-wide pixels.  We determine the median
flux in each pixel to construct the final median spectrum.  We also
generate 100 bootstrap samples of the spectra, calculating the median of
each in the same manner.  
 Finally, we determine the 
 continuum level of the resulting coadds via
  a linear fit in the velocity windows
 $-3000\mkms < \delta v < -385\mkms$ and
  $1155\mkms < \delta v < 3000\mkms$ (with $\delta v = 0\mkms$ at
  $\lambda=2796.35$ \AA),  
   renormalizing each coadd to ensure it has a continuum level $\approx 1$.
We measure the equivalent width of the
  \ion{Mg}{2} $\lambda 2796$ feature 
in the coadds in the relative velocity
  window $-385\mkms < \delta v < 385\mkms$, such that the red edge of
  this range falls at the midpoint between the $\lambda2796.35$ and
  $\lambda2803.53$ transitions.  The median coadds of the spectra in
  each of the trimmed PRIMUS subsamples described above are shown in 
  Figure~\ref{fig.primus_qsocompstacks}.  

\begin{figure}
\begin{center}
\includegraphics[angle=0,width=3.5in,trim=10 470 180 0,clip=]{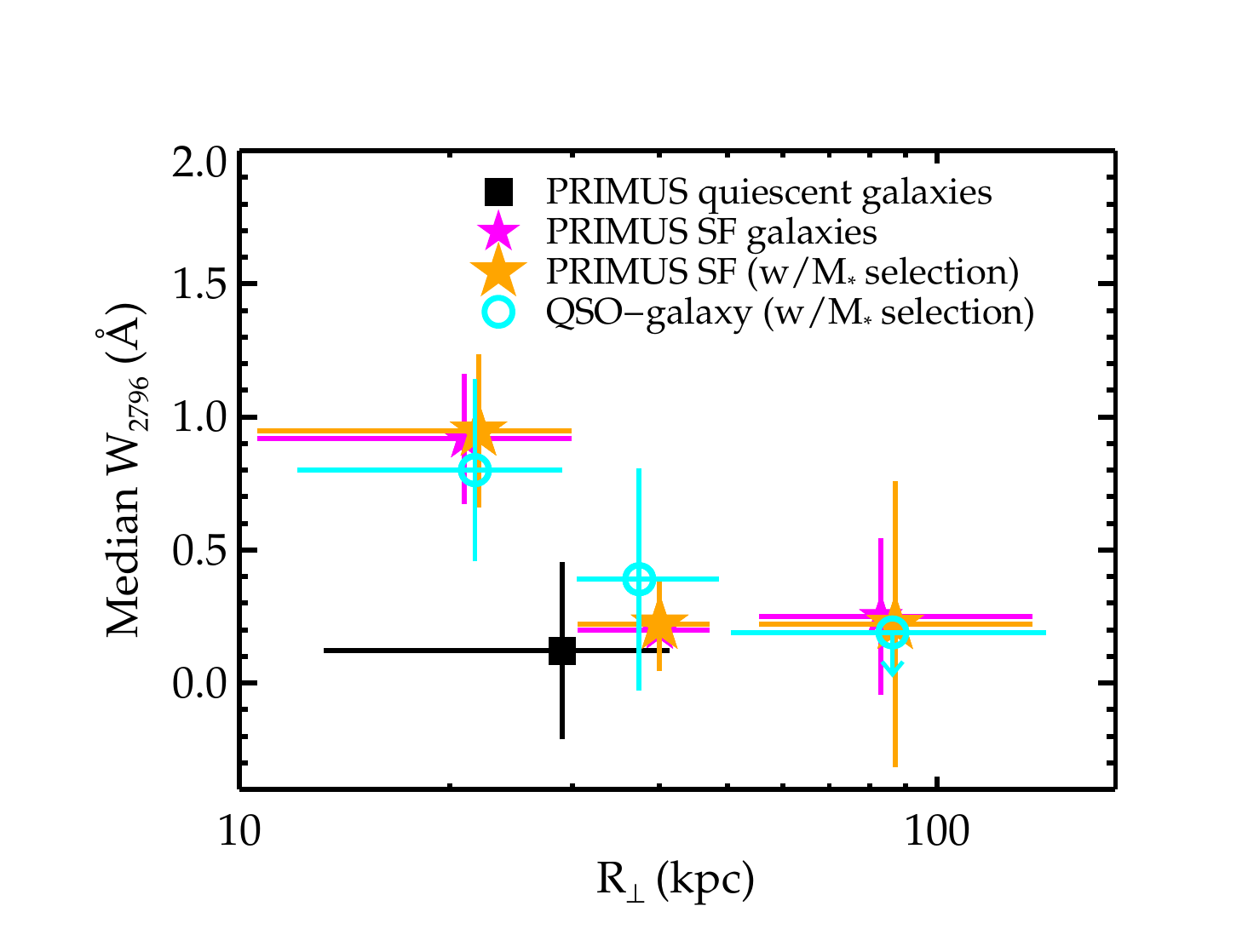}
\caption{Median $W_{2796}$ in coadded spectra of the subsamples of
  PRIMUS sightlines described in \S\ref{sec.results_comparison} and
  shown in Figure~\ref{fig.primus_qsocompstacks} (filled orange
  stars).  The filled magenta stars show the median $W_{2796}$ in the
  coadded spectra of all PRIMUS sightlines probing f/g star-forming
  galaxy halos and having $\rm
  S/N($\ion{Mg}{2}$) > 4$ \AA$^{-1}$ at $\mrperp < 30$ kpc, 30 kpc $<
  \mrperp <$ 50 kpc, and $\mrperp > 50$ kpc (i.e., the galaxies
  shown with the open orange histograms in Figure~\ref{fig.primus_compare_to_stacks}).  The filled black square shows 
$W_{2796}$ measured in the coadded spectrum of PRIMUS sightlines
probing quiescent f/g galaxy halos (shown in Figure~\ref{fig.primus_redseqstacks}). 
The vertical error bars  indicate the
  dispersion in $W_{2796}$ measured from median coadds of 100
  bootstrap realizations of each subsample.
The open cyan circles at $\mrperp < 50$ kpc show the median $W_{2796}$ and dispersion in the
$W_{2796}$ values in the individual QSO-galaxy comparison 
sightlines included in each trimmed subsample described in
\S\ref{sec.results_comparison}.  The open cyan circle at $\mrperp =
86$ kpc is placed at the upper 25th-percentile value of the cumulative
distribution function of $W_{2796}$ values in the QSO-galaxy
comparison dataset with $\mrperp > 50$ kpc, and is calculated using
the \texttt{ASURV} software package.
The $x$-axis locations of the points are set by the mean $\mrperp$ of the
sightlines in each coadd,  and the horizontal error bars
show the full range in these values.
The subsamples represented by the orange
and cyan points have been selected to cover the same ranges in $M_*$ in each
$\mrperp$ bin, and have similar median $M_*$ values. 
\label{fig.primus_medewstacks}}
\end{center}
\end{figure}

  The $W_{2796}$ measured from each of these coadds are shown in
  orange  in Figure~\ref{fig.primus_medewstacks}.  The vertical error
  bars on these points indicate the $\pm1\sigma$ dispersion in
  $W_{2796}$ measured from the 100 bootstrap realizations of the
  median coadds described above.  We measure a large median $W_{2796}
  = 0.95 \pm 0.29$ \AA\ at $\mrperp < 30$ kpc.  
At 30 kpc $< \mrperp <$ 50 kpc, the
  sample absorption strength is significantly weaker, with a median
  $W_{2796} = 0.22 \pm 0.16$ \AA.  Finally, at $\mrperp > 50$ kpc, our
  sightlines exhibit a median $W_{2796}$  similar to that of the 30 kpc $<
    \mrperp <$ 50 kpc sample, albeit with
  a large scatter.  Indeed, the 
 comparably
large $W_{2796}$ values at these large impact
  parameters likely give rise to the flat best-fit $\log W_{2796} -
  \mrperp$ relation discussed in \S\ref{sec.results_wrperp}.
  Again, we note that in QSO-galaxy pair studies targeting
  the CGM of relatively isolated f/g galaxies, the average/median $W_{2796}$ has been
  ubiquitously observed to {\it decline} with increasing $\mrperp$.
  These differences are suggestive of a physical effect which enhances
  $W_{2796}$ in the extended environments of magnitude-selected
  galaxies. 
 We require a larger and higher-$\rm S/N$ sample at $\mrperp > 50$ kpc to confirm this overall trend.  

We also calculate the median and dispersion in $W_{2796}$ in each of
the QSO-galaxy comparison subsamples within $\mrperp < 50$ kpc defined in
Figure~\ref{fig.primus_compare_to_stacks}, and show the results in
cyan in Figure~\ref{fig.primus_medewstacks}.  
 Working with the $W_{2796}$ measurements themselves (rather than
  the QSO spectra), we estimate the
dispersion by using an iterative sigma-clipping algorithm
to identify outliers defined
to lie $>3\sigma$ from the central value (as in our estimate of the
dispersion in the bootstrap samples above).  We iteratively mask these
outliers, compute the central value of the distribution, and reassess the
dispersion in $W_{2796}$ taking this new central value into account.
Among the 38 QSO
sightlines considered here, only four do not yield significant
detections of \ion{Mg}{2} absorption, such that our estimates of the
sample dispersion will not be significantly biased by the inclusion of censored
measurements.  As the majority of QSO sightlines at $\mrperp > 50$ kpc
yield upper limits, we again turn to the \texttt{ASURV} survival
analysis software to quantify the $W_{2796}$ distribution at these
large impact parameters, placing the cyan upper limit in Figure~\ref{fig.primus_medewstacks}
at the upper 25th-percentile of
the 
 Kaplan-Meier estimator for the
distribution function of $W_{2796}$ values.
This figure demonstrates a
striking similarity between the $W_{2796}$ in the PRIMUS and
QSO-galaxy comparison samples, particularly at $\mrperp < 50$ kpc.  We
note that the dispersion in the QSO-galaxy subsample is somewhat
larger at 30 kpc $< \mrperp < $ 50 kpc, and will discuss
potential reasons for this discrepancy in \refPaperII.
Overall, we interpret these measurements as suggestive that the CGM of
galaxies having a comparable range in $M_*$ give rise to similar
\ion{Mg}{2} absorption strength toward both our b/g galaxy sightlines
at $z_{\rm f/g} \sim 0.4$ and toward b/g QSO sightlines at $z_{\rm f/g} \sim
0.2$.  

\begin{figure}
\begin{center}
\includegraphics[angle=0,width=3.5in,trim=30 170 180 400,clip=]{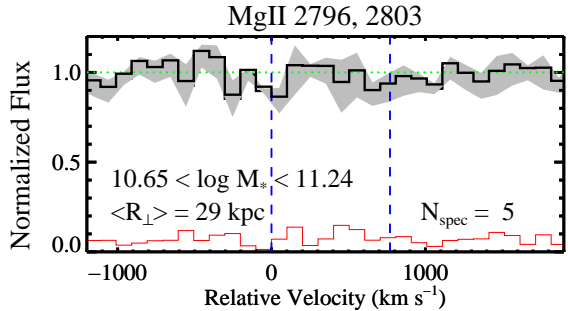}
\caption{Coadded spectra (black) of our PRIMUS b/g galaxy sightlines 
probing quiescent f/g galaxy halos
in the  region around the \ion{Mg}{2} 2796, 2803 transition.  All
spectra probe within $\mrperp < 50$ kpc and have $\rm
  S/N($\ion{Mg}{2}$) > 4$ \AA$^{-1}$.  The filled gray curves
  show the $\pm34$th-percentile interval for the fluxes in our
  bootstrap sample in each pixel.  The red histogram shows this same
  $1\sigma$ error array.  Plot labels and horizontal and
  vertical dashed lines are as described in
  the caption of Figure~\ref{fig.primus_qsocompstacks}.  
\label{fig.primus_redseqstacks}}
\end{center}
\end{figure}

Finally, to assess the strength of \ion{Mg}{2} absorption surrounding
the quiescent galaxies in our sample, we coadd spectroscopy of five
sightlines passing within $\mrperp < 50$ kpc of objects 
sitting below the threshold for star-forming systems discussed in \S\ref{sec.fg}
 and show the resulting stack in
Figure~\ref{fig.primus_redseqstacks}.  The coadd does not exhibit
detectable absorption, yielding a median $W_{2796} =
0.12\pm0.33$ \AA.  This measurement is
consistent with the median $W_{2796}$ measured for each of the PRIMUS
star-forming f/g galaxy subsamples at $\mrperp > 30$ kpc, and is
discrepant by only $1.9\sigma$ with the median $W_{2796}$ for
star-forming galaxies within $\mrperp < 30$ kpc.
We also note that f/g \ion{Mg}{2} absorption is indeed securely
detected toward two of these five sightlines individually
 (having galaxy pair IDs 405 and 611, and $W_{2796} =
0.73\pm0.34$ \AA\ and $0.73\pm0.32$ \AA), suggestive of a large dispersion
in absorption strength in quiescent galaxy environments.
A few previous studies have presented evidence for weaker \ion{Mg}{2}
absorption around host galaxies which are redder in color: e.g.,
 \citet{GauthierChen2011} reported
 a lower incidence
of strong \ion{Mg}{2} absorbers around quiescent Luminous Red Galaxies (LRGs)
than around $\lesssim L^*$ galaxies at $z\sim0.2-0.3$; and
\citet{Bordoloi2011} measured weaker \ion{Mg}{2} absorption in stacked
spectra probing red vs.\ blue f/g hosts.  While the central value of
our $W_{2796}$ measurement around star-forming hosts at $\mrperp < 30$
kpc is indeed higher
than the same measurement around quiescent systems, 
we lack the $\rm
S/N$ required to confirm the detection of this trend.  We present a
more detailed comparison to the results of \citet{Bordoloi2011} in the
following subsection.



\subsection{Comparison with the CGM Probed by Stacked Background
  Sightline Samples}\label{sec.results_bordoloi}

As a final test of the consistency of our measurements with the
literature, here we compare our results to those of additional studies
which have constrained halo \ion{Mg}{2} absorption properties via
the coaddition of numerous low-$\rm S/N$ spectra of background sightlines.  
The primary comparison study, \citet{Bordoloi2011}, mined the zCOSMOS galaxy
redshift survey
\citep{Lilly2007} for projected galaxy pairs within $\mrperp < 200$
kpc.  The $\sim4000$ f/g galaxies satisfying their selection criteria span a redshift
range $0.5 < z < 0.9$.  The zCOSMOS spectra of the b/g galaxies have a
low spectral resolution ($R\sim200$) and hence cannot resolve the two
transitions in the \ion{Mg}{2} doublet.  Therefore, this work reported
the median equivalent width of the blended doublet  ($W_{\rm MgII}$) detected
in coadded spectra of subsamples of $\sim75-150$ b/g objects.  

\citet{Bordoloi2011} investigated the dependence of $W_{\rm MgII}$ on
several f/g host galaxy properties, including color, stellar mass,
environment, and orientation.  Given the galaxy characteristics already at hand for
the PRIMUS sample, we choose to compare to the \citet{Bordoloi2011}
subsamples selected by a combination of stellar mass and color.  To
differentiate between blue and red galaxies, these authors invoked a
division in $(u-B)$ color just slightly bluer than that inspired by \citet{Peng2010}:
\begin{equation}
(u - B)_{\rm AB} = 0.98 + 0.075 \log \frac{M_*}{10^{10}~M_{\odot}} - 0.18z.
\end{equation}\label{eq.colordiv}
We note that the locus of this cut sits blueward of the minimum of the bimodal
galaxy distribution, such that some star-forming galaxies fall into
the ``red'' subsample.

\begin{figure}
\begin{center}
\includegraphics[angle=0,width=3.5in,trim=30 170 180 0,clip=]{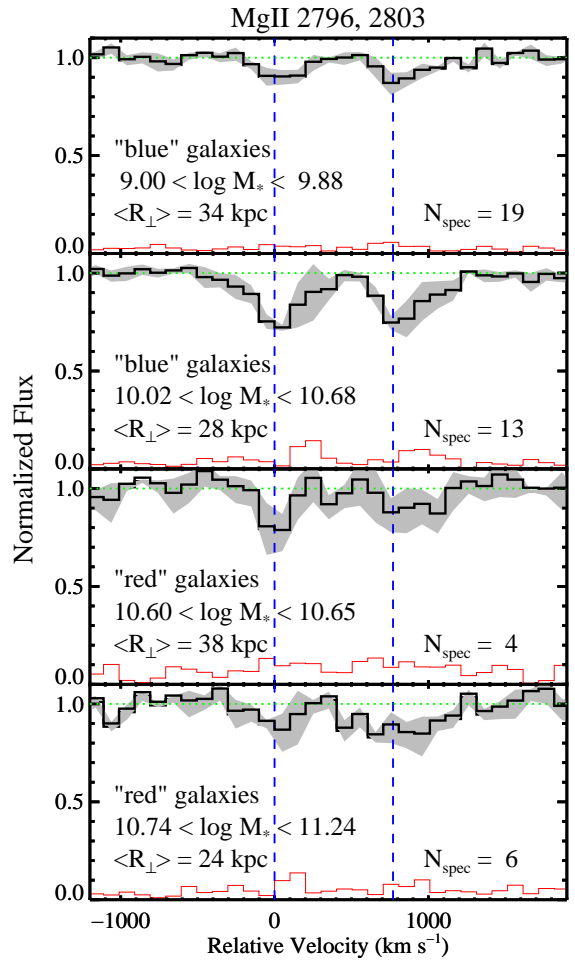}
\caption{Coadded spectra (black) of PRIMUS b/g galaxy sightlines
  probing blue, low-$M_*$ galaxies (top), blue, high-$M_*$ galaxies
  (top middle), red, low-$M_*$ galaxies (bottom middle), and red, high-$M_*$ galaxies (bottom) in the region around the
  \ion{Mg}{2} 2796, 2803 transitions.  The specific color and $M_*$
  selection criteria used are described in
  \S\ref{sec.results_bordoloi} (see Eq.~\ref{eq.colordiv}) and are chosen to match the sample
  selection of \citet{Bordoloi2011}.  All spectra probe within
  $\mrperp < 50$ kpc and have $\rm
  S/N($\ion{Mg}{2}$) > 4$ \AA$^{-1}$.  The filled gray curves
  show the $\pm34$th-percentile interval for the fluxes in our
  bootstrap sample in each pixel.  The red histogram shows this same
  $1\sigma$ error array.  Plot labels and horizontal and
  vertical dashed lines are as described in
  the caption of Figure~\ref{fig.primus_qsocompstacks}. 
\label{fig.primus_bordcompstacks}}
\end{center}
\end{figure}

Though the passbands used to calculate this color are not explicitly
specified, we assume that the quantity $(u-B)$ is similar to the $(U-B)$
color we use in Figure~\ref{fig.primus_ms}a and apply this cut without
adjustment to our f/g galaxy sample.  \citet{Bordoloi2011} 
further subdivided these blue and red samples by stellar mass,
separating the blue galaxies into bins above and below $\log
M_*/M_{\odot} = 9.88$ and the red galaxies at $\log M_*/M_{\odot} =
10.68$.  We adopt the same subdivisions for the portion of our pair sample having
$\mrperp < 50$ kpc, finding that
the resulting blue galaxy subsamples each contain $>10$ sightlines,
while the red galaxy subsamples contain just handfuls of objects (4-6).
We show the
coadded spectra for these four subsamples in
Figure~\ref{fig.primus_bordcompstacks}, with the top two panels
showing the coadds of the low- and high-$M_*$ blue subsamples, and the
bottom panels showing the coadds of our red subsamples.  
Strong absorption is evident in the blue, high-$M_*$ panel,
while the blue, low-$M_*$ and red subsamples each exhibit a  weak absorption signal.

\begin{figure}
\begin{center}
\includegraphics[angle=0,width=3.5in,trim=10 480 180 0,clip=]{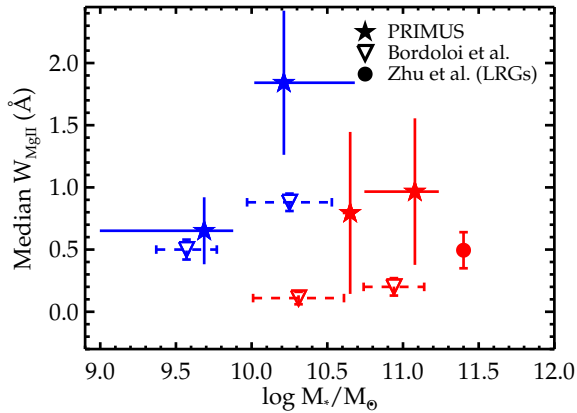}
\caption{Median $W_{\rm MgII}$ in coadded spectra of subsamples of
  PRIMUS sightlines described in \S\ref{sec.results_bordoloi} and
  shown in Figure~\ref{fig.primus_bordcompstacks}.  The filled red
  stars show the median $W_{\rm MgII}$ in the red PRIMUS subsamples,
  while the filled blue stars show the same measurement for the blue
  PRIMUS subsamples.  The x-axis locations are set by the median $\log
  M_*/M_{\odot}$ value for the subsamples, and the error bars in the
  $x$-direction show the full range in these values.  The error bars in
  the $y$-direction indicate the dispersion in $W_{\rm MgII}$ measured
  from median coadds of 100 bootstrap realizations of each subsample.
The open triangles show measurements of median $W_{\rm MgII}$ around
subsamples of blue and red galaxies from \citet{Bordoloi2011}.  The
$x$-axis values and uncertainty intervals of these points indicate the
mean and $\pm1\sigma$ dispersion in $\log M_{*}/M_{\odot}$ values in
each subsample.  The filled circle shows the mean $W_{\rm MgII}$
measured in coadded QSO sightlines passing within 45 kpc of $z\sim0.5$
LRGs reported in \citet{Zhu2014}.
\label{fig.primus_medewbordcomp}}
\end{center}
\end{figure}

To determine the total absorption from both doublet transitions
(for consistency with Bordoloi et al.), we measure the
equivalent width in these spectra over a velocity window
$-385\mkms<\delta v < 1155\mkms$, which spans an interval from $385\mkms$
blueward of the $\lambda2796$ transition to $\approx385\mkms$ redward of the
$\lambda2803$ transition.  The resulting median $W_{\rm MgII}$ values are
plotted with filled stars in Figure~\ref{fig.primus_medewbordcomp},
along with the analogous measurements from \citet[][shown with open
triangles]{Bordoloi2011}\footnote{\citet{Bordoloi2011} estimated uncertainty
  in $W_{\rm MgII}$ using a bootstrapping approach, generating 1000
  coadds from sets of random draws from each subsample.  The
  dispersion in $W_{\rm MgII}$ among this sample of 1000 was adopted
  as the error in this quantity.}.  
The filled red circle shows a measurement of the
mean $W_{\rm MgII}$ within $\mrperp < 45$ kpc of a sample of 35 $z\sim0.5$ LRGs 
assessed using coadded SDSS QSO spectra by \citet{Zhu2014}.
The x-axis value of this point is  the median stellar mass of the full sample of
LRGs used in this study ($\log M_*/M_{\odot} = 11.4$), which includes
galaxies in projected LRG-QSO pairs with separations as large as
$\mrperp = 18$ Mpc.

The large dispersion in the $W_{\rm MgII}$ values measured from the PRIMUS
subsamples (and their small sizes) again limits our ability to draw firm conclusions from this
analysis.  However, we note the close consistency between the $W_{\rm MgII}$
measured around the low-$M_*$ blue galaxy subsamples
drawn from both PRIMUS and \citet{Bordoloi2011}.  The $W_{\rm MgII}$ measurements
for the remaining PRIMUS subsamples tend to have higher central values
than those from zCOSMOS, with the high-$M_*$ blue subsamples having an
offset of the greatest significance ($1.7\sigma$).  
And while the median $W_{\rm MgII}$ values for the red PRIMUS samples
are indeed lower than that measured for the high-$M_*$ blue subsample,
overall we fail to reproduce the highly significant offset between
$W_{\rm MgII}$ around blue vs.\ red galaxies reported in the zCOSMOS
analysis.
The results shown in Figure~\ref{fig.primus_medewstacks} are
  similarly suggestive of lower $W_{2796}$ around quiescent vs.\
  star-forming galaxies, but again, these measurements are discrepant
  at $\lesssim 2\sigma$.
Given that these \citet{Bordoloi2011} subsamples all include at least 99
pairs, this discrepancy may
arise simply due to stochasticity in our sparse sampling of the red galaxy
population.  
However, we also note that the trend of lower $W_{\rm MgII}$ around redder
host galaxies is not reproduced in higher-$\rm S/N$ studies of
individual projected QSO-galaxy pairs. For instance,
\citet{Nielsen2013} found no significant difference in the incidence
of \ion{Mg}{2} absorbers around blue vs.\ red galaxies within $\mrperp
< 50$ kpc, even after testing several limiting $W_{2796}$ values (of 0.1, 0.3, 0.6, and 1.0 \AA).
\citet{Chen2010} likewise reported no strong dependence of $W_{2796}$ on
f/g galaxy color.  This apparent disagreement in the literature may
be resolved with larger samples of sightlines (surpassing those of
Bordoloi et al.\ in size) 
probing each of these galaxy
subpopulations. 

\section{Summary}\label{sec.summary}

We have presented spectroscopy of 72 projected pairs of galaxies
selected from the PRIsm MUlti-object Survey redshift catalog
\citep{Coil2011primus} with  impact parameters $\mrperp < 150$ kpc.
Fifty-one of these pairs have $\mrperp < 50$
kpc, and hence probe circumgalactic environments known to yield strong intervening
absorption traced by the \ion{Mg}{2} transition.  The sample
foreground (f/g) galaxies span a range in redshifts $0.35 < z_{\rm f/g} < 0.8$ and
include both star-forming and quiescent systems to a stellar mass limit
$M_* \gtrsim10^9~M_{\odot}$.  The background (b/g) galaxies, selected to a
magnitude limit $B_{\rm AB} \lesssim 22.3$, are distributed in
redshift over the range
$0.4 < z_{\rm b/g} < 2.0$.  While a third of these galaxies host bright AGN
which give rise to broad emission line features, the
remaining b/g sightlines exhibit no signs of broad-line AGN in their
optical spectroscopy.  The
$\sim20$ of these latter systems which have been imaged by {\it HST}/ACS have
half-light radii in the range $2~\mathrm{kpc} \lesssim R_{\rm eff}
\lesssim 8~\mathrm{kpc}$ estimated at the redshift of the
corresponding f/g system.

Our spectroscopy covers the
\ion{Mg}{2} $\lambda \lambda 2796, 2803$ doublet in the rest-frame of
the foreground galaxy at high $\rm S/N$,
constraining the \ion{Mg}{2} equivalent width to a typical limit
$W_{2796}\gtrsim 0.5$ \AA\ in 
individual b/g galaxy sightlines.  
We make secure ($>2\sigma$ significant) detections of the $\lambda2796$ transition in 20 sightlines
passing within $\mrperp < 50$ kpc, and place a $2\sigma$ upper
limit on $W_{2796}$ of $<0.5$ \AA\ in an additional 11 close sightlines.
This is the first work presenting a
sample of more than $\sim1-2$ individual b/g galaxy sightlines with
securely-detected absorbers arising from intervening systems.

We have 
 shown that the $W_{2796}$ associated with the f/g galaxy halos
in this sample declines with increasing $\mrperp$ of the b/g galaxy
within $\mrperp < 50$ kpc,
consistent with the findings of numerous works probing the
circumgalactic medium (CGM) of f/g galaxies with similar properties using
b/g QSO sightlines.  Our analysis additionally constrains the intrinsic
scatter in the relationship between $\mrperp$ and $W_{2796}$
(\S\ref{sec.results_wrperp}).  
 We have demonstrated that $W_{2796}$ is
higher around galaxies with higher SFR and/or 
stellar mass ($M_*$), and have
shown that these trends are exhibited with
statistical significance within the ``inner'' CGM
 (at $30~\mathrm{kpc} < \mrperp < 50~\mathrm{kpc}$ in the case of
 $M_*$ and at $\mrperp < 30~\mathrm{kpc}$ in the case of SFR; \S\ref{sec.results_wrintrinsic}).
Qualitatively similar trends are
likewise exhibited in
projected QSO-galaxy
pair studies of circumgalactic \ion{Mg}{2} absorption at $z<1$.
Finally, we have shown that the median $W_{2796}$ 
observed toward both b/g galaxies and a sample of b/g QSOs taken from
the literature which probe the environments of 
 f/g galaxies with a similar range in $M_*$ at similar
impact parameters are statistically consistent
(\S\ref{sec.results_comparison}).


All together, these findings point to a broad-brush conformity in the
mean properties of the CGM as observed toward both extended ($>$ kpc) galaxy b/g
beams and the pencil-beam ($<10^{-2}$ pc) sightlines offered by b/g QSOs.
In \refPaperII, we will compare this sample
of \ion{Mg}{2} absorbers and those observed toward b/g QSOs in more detail,
focusing
in particular on the dispersion in their respective equivalent
width distributions.
With the adoption of a few simplifying assumptions,
we will demonstrate that together, these two datasets set a 
lower limit on the coherence scale of \ion{Mg}{2}-absorbing
circumgalactic material.  Such a limit is unique in the context of CGM
studies, and has important implications for the mass, survival time,
and origin of the cool, photoionized gas which pervades galaxy environments.




\acknowledgements

K.H.R.R acknowledges support for this project from the Clay
Postdoctoral Fellowship, and from the Alexander von
Humboldt foundation in the form of the Humboldt Postdoctoral
Fellowship.  The Humboldt foundation is funded by the German Federal
Ministry for Education and Research.
A.L.C. acknowledges support from NSF CAREER award AST-1055081.
Funding for PRIMUS is provided by NSF (AST-0607701, AST-0908246, 
AST-0908442, AST-0908354) and NASA (Spitzer-1356708, 08-ADP08-0019,
NNX09AC95G).

These findings are in part based on observations collected at the
European Organisation for Astronomical Research in the Southern
Hemisphere under ESO programs 088.A-0529(A) and 090.A-0485(A).
In addition, a significant portion 
of the data presented herein were obtained at the W. M. Keck
Observatory, 
which is operated as a scientific partnership among the 
California Institute of Technology, the University of California, and
the National Aeronautics and Space Administration. The Observatory was
made possible by the 
generous financial support of the W. M. Keck Foundation.

The authors wish to thank the CFHTLS, COSMOS, and SWIRE teams for
making their data and catalogs public.  
We acknowledge Robert da Silva for his guidance in the use of the SLUG
software package.  
We thank James Bullock, Joe
Hennawi, and Brice M{\'e}nard for insightful comments on this
analysis, and 
we gratefully acknowledge Xavier
Prochaska for sharing code to generate spectroscopic coadds, and for
numerous valuable discussions of this work and his reading of
an earlier version of this manuscript.

Finally, the authors wish to recognize and acknowledge the very
significant cultural role and reverence that the summit of Mauna Kea
has always had within the indigenous Hawaiian community. We are most
fortunate to have the opportunity to conduct observations from this
mountain.


\begin{thebibliography}{}
\expandafter\ifx\csname natexlab\endcsname\relax\def\natexlab#1{#1}\fi

\bibitem[{{Adelberger} {et~al.}(2005){Adelberger}, {Shapley}, {Steidel},
  {Pettini}, {Erb}, \& {Reddy}}]{Adelberger2005}
{Adelberger}, K.~L., {Shapley}, A.~E., {Steidel}, C.~C., {et~al.} 2005, \apj,
  629, 636

\bibitem[{{Appenzeller} {et~al.}(1998){Appenzeller}, {Fricke}, {F{\"u}rtig},
  {G{\"a}ssler}, {H{\"a}fner}, {Harke}, {Hess}, {Hummel}, {J{\"u}rgens},
  {Kudritzki}, {Mantel}, {Meisl}, {Muschielok}, {Nicklas}, {Rupprecht},
  {Seifert}, {Stahl}, {Szeifert}, \& {Tarantik}}]{Appenzeller1998}
{Appenzeller}, I., {Fricke}, K., {F{\"u}rtig}, W., {et~al.} 1998, The
  Messenger, 94, 1

\bibitem[{{Barger} {et~al.}(2008){Barger}, {Cowie}, \& {Wang}}]{Barger2008}
{Barger}, A.~J., {Cowie}, L.~L., \& {Wang}, W.-H. 2008, \apj, 689, 687

\bibitem[{{Bergeron} \& {Stasi{\'n}ska}(1986)}]{BergeronStasinska1986}
{Bergeron}, J., \& {Stasi{\'n}ska}, G. 1986, \aap, 169, 1

\bibitem[{{Berti} {et~al.}(2016){Berti}, {Coil}, {Behroozi}, {Eisenstein},
  {Bray}, {Cool}, \& {Moustakas}}]{Berti2016}
{Berti}, A.~M., {Coil}, A.~L., {Behroozi}, P.~S., {et~al.} 2016, ArXiv
  e-prints, arXiv:1608.05084

\bibitem[{{Bertin} \& {Arnouts}(1996)}]{Bertin1996}
{Bertin}, E., \& {Arnouts}, S. 1996, \aaps, 117, 393

\bibitem[{{Bigelow} \& {Dressler}(2003)}]{BigelowDressler2003}
{Bigelow}, B.~C., \& {Dressler}, A.~M. 2003, in Society of Photo-Optical
  Instrumentation Engineers (SPIE) Conference Series, Vol. 4841, Instrument
  Design and Performance for Optical/Infrared Ground-based Telescopes, ed.
  M.~{Iye} \& A.~F.~M. {Moorwood}, 1727--1738

\bibitem[{{Binney} {et~al.}(2009){Binney}, {Nipoti}, \&
  {Fraternali}}]{Binney2009}
{Binney}, J., {Nipoti}, C., \& {Fraternali}, F. 2009, MNRAS, accepted
  (arXiv:0902.4525), arXiv:0902.4525

\bibitem[{{Bond} {et~al.}(2001){Bond}, {Churchill}, {Charlton}, \&
  {Vogt}}]{Bond2001}
{Bond}, N.~A., {Churchill}, C.~W., {Charlton}, J.~C., \& {Vogt}, S.~S. 2001,
  \apj, 562, 641

\bibitem[{{Bordoloi} {et~al.}(2011){Bordoloi}, {Lilly}, {Knobel}, {Bolzonella},
  {Kampczyk}, {Carollo}, {Iovino}, {Zucca}, {Contini}, {Kneib}, {Le Fevre},
  {Mainieri}, {Renzini}, {Scodeggio}, {Zamorani}, {Balestra}, {Bardelli},
  {Bongiorno}, {Caputi}, {Cucciati}, {de la Torre}, {de Ravel}, {Garilli},
  {Kova{\v c}}, {Lamareille}, {Le Borgne}, {Le Brun}, {Maier}, {Mignoli},
  {Pello}, {Peng}, {Perez Montero}, {Presotto}, {Scarlata}, {Silverman},
  {Tanaka}, {Tasca}, {Tresse}, {Vergani}, {Barnes}, {Cappi}, {Cimatti},
  {Coppa}, {Diener}, {Franzetti}, {Koekemoer}, {L{\'o}pez-Sanjuan},
  {McCracken}, {Moresco}, {Nair}, {Oesch}, {Pozzetti}, \&
  {Welikala}}]{Bordoloi2011}
{Bordoloi}, R., {Lilly}, S.~J., {Knobel}, C., {et~al.} 2011, \apj, 743, 10

\bibitem[{{Bordoloi} {et~al.}(2014){Bordoloi}, {Tumlinson}, {Werk},
  {Oppenheimer}, {Peeples}, {Prochaska}, {Tripp}, {Katz}, {Dav{\'e}}, {Fox},
  {Thom}, {Ford}, {Weinberg}, {Burchett}, \& {Kollmeier}}]{Bordoloi2014}
{Bordoloi}, R., {Tumlinson}, J., {Werk}, J.~K., {et~al.} 2014, \apj, 796, 136

\bibitem[{{Bouch{\'e}} {et~al.}(2012){Bouch{\'e}}, {Hohensee}, {Vargas},
  {Kacprzak}, {Martin}, {Cooke}, \& {Churchill}}]{Bouche2012}
{Bouch{\'e}}, N., {Hohensee}, W., {Vargas}, R., {et~al.} 2012, \mnras, 3207

\bibitem[{{Bouch{\'e}} {et~al.}(2013){Bouch{\'e}}, {Murphy}, {Kacprzak},
  {P{\'e}roux}, {Contini}, {Martin}, \& {Dessauges-Zavadsky}}]{Bouche2013}
{Bouch{\'e}}, N., {Murphy}, M.~T., {Kacprzak}, G.~G., {et~al.} 2013, Science,
  341, 50

\bibitem[{{Burchett} {et~al.}(2015){Burchett}, {Tripp}, {Prochaska}, {Werk},
  {Tumlinson}, {O'Meara}, {Bordoloi}, {Katz}, \& {Willmer}}]{Burchett2015}
{Burchett}, J.~N., {Tripp}, T.~M., {Prochaska}, J.~X., {et~al.} 2015, \apj,
  815, 91

\bibitem[{{Chabrier}(2003)}]{Chabrier2003}
{Chabrier}, G. 2003, \pasp, 115, 763

\bibitem[{{Chabrier}(2005)}]{Chabrier2005}
{Chabrier}, G. 2005, in Astrophysics and Space Science Library, Vol. 327, The
  Initial Mass Function 50 Years Later, ed. E.~{Corbelli}, F.~{Palla}, \&
  H.~{Zinnecker}, 41

\bibitem[{{Chen} {et~al.}(2010{\natexlab{a}}){Chen}, {Helsby}, {Gauthier},
  {Shectman}, {Thompson}, \& {Tinker}}]{Chen2010}
{Chen}, H., {Helsby}, J.~E., {Gauthier}, J., {et~al.} 2010{\natexlab{a}}, \apj,
  714, 1521

\bibitem[{{Chen} {et~al.}(2001){Chen}, {Lanzetta}, \& {Webb}}]{Chen2001}
{Chen}, H.-W., {Lanzetta}, K.~M., \& {Webb}, J.~K. 2001, \apj, 556, 158

\bibitem[{{Chen} {et~al.}(2010{\natexlab{b}}){Chen}, {Wild}, {Tinker},
  {Gauthier}, {Helsby}, {Shectman}, \& {Thompson}}]{Chen2010b}
{Chen}, H.-W., {Wild}, V., {Tinker}, J.~L., {et~al.} 2010{\natexlab{b}}, \apjl,
  724, L176

\bibitem[{{Churchill} {et~al.}(2000){Churchill}, {Mellon}, {Charlton},
  {Jannuzi}, {Kirhakos}, {Steidel}, \& {Schneider}}]{Churchill2000}
{Churchill}, C.~W., {Mellon}, R.~R., {Charlton}, J.~C., {et~al.} 2000, \apjs,
  130, 91

\bibitem[{{Churchill} {et~al.}(2013){Churchill}, {Trujillo-Gomez}, {Nielsen},
  \& {Kacprzak}}]{Churchill2013}
{Churchill}, C.~W., {Trujillo-Gomez}, S., {Nielsen}, N.~M., \& {Kacprzak},
  G.~G. 2013, \apj, 779, 87

\bibitem[{{Cohen} {et~al.}(1994){Cohen}, {Cromer}, \& {Southard}}]{Cohen1994}
{Cohen}, J.~G., {Cromer}, J., \& {Southard}, Jr., S. 1994, in Astronomical
  Society of the Pacific Conference Series, Vol.~61, Astronomical Data Analysis
  Software and Systems III, ed. D.~R. {Crabtree}, R.~J. {Hanisch}, \&
  J.~{Barnes}, 469--+

\bibitem[{{Coil} {et~al.}(2011{\natexlab{a}}){Coil}, {Weiner}, {Holz},
  {Cooper}, {Yan}, \& {Aird}}]{Coil2011winds}
{Coil}, A.~L., {Weiner}, B.~J., {Holz}, D.~E., {et~al.} 2011{\natexlab{a}},
  \apj, 743, 46

\bibitem[{{Coil} {et~al.}(2011{\natexlab{b}}){Coil}, {Blanton}, {Burles},
  {Cool}, {Eisenstein}, {Moustakas}, {Wong}, {Zhu}, {Aird}, {Bernstein},
  {Bolton}, \& {Hogg}}]{Coil2011primus}
{Coil}, A.~L., {Blanton}, M.~R., {Burles}, S.~M., {et~al.} 2011{\natexlab{b}},
  \apj, 741, 8

\bibitem[{{Conroy} \& {Gunn}(2010)}]{ConroyGunn2010}
{Conroy}, C., \& {Gunn}, J.~E. 2010, \apj, 712, 833

\bibitem[{{Cooke} \& {O'Meara}(2015)}]{CookeOMeara2015}
{Cooke}, J., \& {O'Meara}, J.~M. 2015, \apjl, 812, L27

\bibitem[{{Cool} {et~al.}(2013){Cool}, {Moustakas}, {Blanton}, {Burles},
  {Coil}, {Eisenstein}, {Wong}, {Zhu}, {Aird}, {Bernstein}, {Bolton}, {Hogg},
  \& {Mendez}}]{Cool2013}
{Cool}, R.~J., {Moustakas}, J., {Blanton}, M.~R., {et~al.} 2013, \apj, 767, 118

\bibitem[{{Crighton} {et~al.}(2013){Crighton}, {Hennawi}, \&
  {Prochaska}}]{Crighton2013}
{Crighton}, N.~H.~M., {Hennawi}, J.~F., \& {Prochaska}, J.~X. 2013, \apjl, 776,
  L18

\bibitem[{{Crighton} {et~al.}(2014){Crighton}, {Hennawi}, {Simcoe}, {Cooksey},
  {Murphy}, {Fumagalli}, {Prochaska}, \& {Shanks}}]{Crighton2014}
{Crighton}, N.~H.~M., {Hennawi}, J.~F., {Simcoe}, R.~A., {et~al.} 2014, ArXiv
  e-prints, arXiv:1406.4239

\bibitem[{{da Silva} {et~al.}(2012){da Silva}, {Fumagalli}, \&
  {Krumholz}}]{daSilva2012}
{da Silva}, R.~L., {Fumagalli}, M., \& {Krumholz}, M. 2012, \apj, 745, 145

\bibitem[{{Danforth} \& {Shull}(2008)}]{DanforthShull2008}
{Danforth}, C.~W., \& {Shull}, J.~M. 2008, \apj, 679, 194

\bibitem[{{Davis} {et~al.}(2003){Davis}, {Faber}, {Newman}, {Phillips},
  {Ellis}, {Steidel}, {Conselice}, {Coil}, {Finkbeiner}, {Koo}, {Guhathakurta},
  {Weiner}, {Schiavon}, {Willmer}, {Kaiser}, {Luppino}, {Wirth}, {Connolly},
  {Eisenhardt}, {Cooper}, \& {Gerke}}]{Davis2003}
{Davis}, M., {Faber}, S.~M., {Newman}, J., {et~al.} 2003, in Society of
  Photo-Optical Instrumentation Engineers (SPIE) Conference Series, Vol. 4834,
  Society of Photo-Optical Instrumentation Engineers (SPIE) Conference Series,
  ed. P.~{Guhathakurta}, 161--172

\bibitem[{{Diamond-Stanic} {et~al.}(2015){Diamond-Stanic}, {Coil}, {Moustakas},
  {Tremonti}, {Sell}, {Mendez}, {Hickox}, \& {Rudnick}}]{ADS2015}
{Diamond-Stanic}, A.~M., {Coil}, A.~L., {Moustakas}, J., {et~al.} 2015, ArXiv
  e-prints, arXiv:1507.01945

\bibitem[{{Fall} {et~al.}(2009){Fall}, {Chandar}, \& {Whitmore}}]{Fall2009}
{Fall}, S.~M., {Chandar}, R., \& {Whitmore}, B.~C. 2009, \apj, 704, 453

\bibitem[{{Farina} {et~al.}(2014){Farina}, {Falomo}, {Scarpa}, {Decarli},
  {Treves}, \& {Kotilainen}}]{Farina2014}
{Farina}, E.~P., {Falomo}, R., {Scarpa}, R., {et~al.} 2014, \mnras, 441, 886

\bibitem[{{Feigelson} \& {Nelson}(1985)}]{FeigelsonNelson1985}
{Feigelson}, E.~D., \& {Nelson}, P.~I. 1985, \apj, 293, 192

\bibitem[{{Foreman-Mackey} {et~al.}(2013){Foreman-Mackey}, {Hogg}, {Lang}, \&
  {Goodman}}]{DFM2013}
{Foreman-Mackey}, D., {Hogg}, D.~W., {Lang}, D., \& {Goodman}, J. 2013, \pasp,
  125, 306

\bibitem[{{Frank} {et~al.}(2007){Frank}, {Bentz}, {Stanek}, {Mathur},
  {Dietrich}, {Peterson}, \& {Atlee}}]{Frank2007}
{Frank}, S., {Bentz}, M.~C., {Stanek}, K.~Z., {et~al.} 2007, \apss, 312, 325

\bibitem[{{Gauthier} \& {Chen}(2011)}]{GauthierChen2011}
{Gauthier}, J.-R., \& {Chen}, H.-W. 2011, \mnras, 418, 2730

\bibitem[{{Giacconi} {et~al.}(2001){Giacconi}, {Rosati}, {Tozzi}, {Nonino},
  {Hasinger}, {Norman}, {Bergeron}, {Borgani}, {Gilli}, {Gilmozzi}, \&
  {Zheng}}]{Giacconi2001}
{Giacconi}, R., {Rosati}, P., {Tozzi}, P., {et~al.} 2001, \apj, 551, 624

\bibitem[{{Giavalisco} {et~al.}(2004){Giavalisco}, {Ferguson}, {Koekemoer},
  {Dickinson}, {Alexander}, {Bauer}, {Bergeron}, {Biagetti}, {Brandt},
  {Casertano}, {Cesarsky}, {Chatzichristou}, {Conselice}, {Cristiani}, {Da
  Costa}, {Dahlen}, {de Mello}, {Eisenhardt}, {Erben}, {Fall}, {Fassnacht},
  {Fosbury}, {Fruchter}, {Gardner}, {Grogin}, {Hook}, {Hornschemeier}, {Idzi},
  {Jogee}, {Kretchmer}, {Laidler}, {Lee}, {Livio}, {Lucas}, {Madau},
  {Mobasher}, {Moustakas}, {Nonino}, {Padovani}, {Papovich}, {Park},
  {Ravindranath}, {Renzini}, {Richardson}, {Riess}, {Rosati}, {Schirmer},
  {Schreier}, {Somerville}, {Spinrad}, {Stern}, {Stiavelli}, {Strolger},
  {Urry}, {Vandame}, {Williams}, \& {Wolf}}]{Giavalisco2004}
{Giavalisco}, M., {Ferguson}, H.~C., {Koekemoer}, A.~M., {et~al.} 2004, \apjl,
  600, L93

\bibitem[{{Grasha} {et~al.}(2015){Grasha}, {Calzetti}, {Adamo}, {Kim},
  {Elmegreen}, {Gouliermis}, {Aloisi}, {Bright}, {Christian}, {Cignoni},
  {Dale}, {Dobbs}, {Elmegreen}, {Fumagalli}, {Gallagher}, {Grebel}, {Johnson},
  {Lee}, {Messa}, {Smith}, {Ryon}, {Thilker}, {Ubeda}, \&
  {Wofford}}]{Grasha2015}
{Grasha}, K., {Calzetti}, D., {Adamo}, A., {et~al.} 2015, \apj, 815, 93

\bibitem[{{Ilbert} {et~al.}(2009){Ilbert}, {Capak}, {Salvato}, {Aussel},
  {McCracken}, {Sanders}, {Scoville}, {Kartaltepe}, {Arnouts}, {Le Floc'h},
  {Mobasher}, {Taniguchi}, {Lamareille}, {Leauthaud}, {Sasaki}, {Thompson},
  {Zamojski}, {Zamorani}, {Bardelli}, {Bolzonella}, {Bongiorno}, {Brusa},
  {Caputi}, {Carollo}, {Contini}, {Cook}, {Coppa}, {Cucciati}, {de la Torre},
  {de Ravel}, {Franzetti}, {Garilli}, {Hasinger}, {Iovino}, {Kampczyk},
  {Kneib}, {Knobel}, {Kovac}, {Le Borgne}, {Le Brun}, {F{\`e}vre}, {Lilly},
  {Looper}, {Maier}, {Mainieri}, {Mellier}, {Mignoli}, {Murayama}, {Pell{\`o}},
  {Peng}, {P{\'e}rez-Montero}, {Renzini}, {Ricciardelli}, {Schiminovich},
  {Scodeggio}, {Shioya}, {Silverman}, {Surace}, {Tanaka}, {Tasca}, {Tresse},
  {Vergani}, \& {Zucca}}]{Ilbert2009}
{Ilbert}, O., {Capak}, P., {Salvato}, M., {et~al.} 2009, \apj, 690, 1236

\bibitem[{{Johnson} {et~al.}(2015){Johnson}, {Chen}, \&
  {Mulchaey}}]{Johnson2015}
{Johnson}, S.~D., {Chen}, H.-W., \& {Mulchaey}, J.~S. 2015, \mnras, 452, 2553

\bibitem[{{Kacprzak} {et~al.}(2012){Kacprzak}, {Churchill}, \&
  {Nielsen}}]{Kacprzak2012}
{Kacprzak}, G.~G., {Churchill}, C.~W., \& {Nielsen}, N.~M. 2012, ArXiv
  e-prints, arXiv:1205.0245

\bibitem[{{Kacprzak} {et~al.}(2008){Kacprzak}, {Churchill}, {Steidel}, \&
  {Murphy}}]{Kacprzak2008}
{Kacprzak}, G.~G., {Churchill}, C.~W., {Steidel}, C.~C., \& {Murphy}, M.~T.
  2008, \aj, 135, 922

\bibitem[{{Kere{\v s}} {et~al.}(2005){Kere{\v s}}, {Katz}, {Weinberg}, \&
  {Dav{\'e}}}]{Keres2005}
{Kere{\v s}}, D., {Katz}, N., {Weinberg}, D.~H., \& {Dav{\'e}}, R. 2005,
  \mnras, 363, 2

\bibitem[{{Krumholz} {et~al.}(2015{\natexlab{a}}){Krumholz}, {Fumagalli}, {da
  Silva}, {Rendahl}, \& {Parra}}]{Krumholz2015}
{Krumholz}, M.~R., {Fumagalli}, M., {da Silva}, R.~L., {Rendahl}, T., \&
  {Parra}, J. 2015{\natexlab{a}}, \mnras, 452, 1447

\bibitem[{{Krumholz} {et~al.}(2015{\natexlab{b}}){Krumholz}, {Adamo},
  {Fumagalli}, {Wofford}, {Calzetti}, {Lee}, {Whitmore}, {Bright}, {Grasha},
  {Gouliermis}, {Kim}, {Nair}, {Ryon}, {Smith}, {Thilker}, {Ubeda}, \&
  {Zackrisson}}]{Krumholz2015b}
{Krumholz}, M.~R., {Adamo}, A., {Fumagalli}, M., {et~al.} 2015{\natexlab{b}},
  \apj, 812, 147

\bibitem[{{Lada} \& {Lada}(2003)}]{LadaLada2003}
{Lada}, C.~J., \& {Lada}, E.~A. 2003, \araa, 41, 57

\bibitem[{{Lan} {et~al.}(2014){Lan}, {M{\'e}nard}, \& {Zhu}}]{Lan2014}
{Lan}, T.-W., {M{\'e}nard}, B., \& {Zhu}, G. 2014, \apj, 795, 31

\bibitem[{{Lanzetta} \& {Bowen}(1990)}]{LanzettaBowen1990}
{Lanzetta}, K.~M., \& {Bowen}, D. 1990, \apj, 357, 321

\bibitem[{{Lavalley} {et~al.}(1992){Lavalley}, {Isobe}, \&
  {Feigelson}}]{Lavalley1992}
{Lavalley}, M., {Isobe}, T., \& {Feigelson}, E. 1992, in Astronomical Society
  of the Pacific Conference Series, Vol.~25, Astronomical Data Analysis
  Software and Systems I, ed. D.~M. {Worrall}, C.~{Biemesderfer}, \&
  J.~{Barnes}, 245

\bibitem[{{Leauthaud} {et~al.}(2007){Leauthaud}, {Massey}, {Kneib}, {Rhodes},
  {Johnston}, {Capak}, {Heymans}, {Ellis}, {Koekemoer}, {Le F{\`e}vre},
  {Mellier}, {R{\'e}fr{\'e}gier}, {Robin}, {Scoville}, {Tasca}, {Taylor}, \&
  {Van Waerbeke}}]{Leauthaud2007}
{Leauthaud}, A., {Massey}, R., {Kneib}, J.-P., {et~al.} 2007, \apjs, 172, 219

\bibitem[{{Lee} {et~al.}(2014){Lee}, {Hennawi}, {Stark}, {Prochaska}, {White},
  {Schlegel}, {Eilers}, {Arinyo-i-Prats}, {Suzuki}, {Croft}, {Caputi},
  {Cassata}, {Ilbert}, {Garilli}, {Koekemoer}, {Le Brun}, {Le F{\`e}vre},
  {Maccagni}, {Nugent}, {Taniguchi}, {Tasca}, {Tresse}, {Zamorani}, \&
  {Zucca}}]{Lee2014}
{Lee}, K.-G., {Hennawi}, J.~F., {Stark}, C., {et~al.} 2014, \apjl, 795, L12

\bibitem[{{Lee} {et~al.}(2016){Lee}, {Hennawi}, {White}, {Prochaska},
  {Font-Ribera}, {Schlegel}, {Rich}, {Suzuki}, {Stark}, {Le F{\`e}vre},
  {Nugent}, {Salvato}, \& {Zamorani}}]{Lee2016}
{Lee}, K.-G., {Hennawi}, J.~F., {White}, M., {et~al.} 2016, \apj, 817, 160

\bibitem[{{Lehner} {et~al.}(2007){Lehner}, {Savage}, {Richter}, {Sembach},
  {Tripp}, \& {Wakker}}]{Lehner2007}
{Lehner}, N., {Savage}, B.~D., {Richter}, P., {et~al.} 2007, \apj, 658, 680

\bibitem[{{Lehner} {et~al.}(2013){Lehner}, {Howk}, {Tripp}, {Tumlinson},
  {Prochaska}, {O'Meara}, {Thom}, {Werk}, {Fox}, \& {Ribaudo}}]{Lehner2013}
{Lehner}, N., {Howk}, J.~C., {Tripp}, T.~M., {et~al.} 2013, \apj, 770, 138

\bibitem[{{Lilly} {et~al.}(2007){Lilly}, {Le F{\`e}vre}, {Renzini}, {Zamorani},
  {Scodeggio}, {Contini}, {Carollo}, {Hasinger}, {Kneib}, {Iovino}, {Le Brun},
  {Maier}, {Mainieri}, {Mignoli}, {Silverman}, {Tasca}, {Bolzonella},
  {Bongiorno}, {Bottini}, {Capak}, {Caputi}, {Cimatti}, {Cucciati}, {Daddi},
  {Feldmann}, {Franzetti}, {Garilli}, {Guzzo}, {Ilbert}, {Kampczyk}, {Kovac},
  {Lamareille}, {Leauthaud}, {Borgne}, {McCracken}, {Marinoni}, {Pello},
  {Ricciardelli}, {Scarlata}, {Vergani}, {Sanders}, {Schinnerer}, {Scoville},
  {Taniguchi}, {Arnouts}, {Aussel}, {Bardelli}, {Brusa}, {Cappi}, {Ciliegi},
  {Finoguenov}, {Foucaud}, {Franceschini}, {Halliday}, {Impey}, {Knobel},
  {Koekemoer}, {Kurk}, {Maccagni}, {Maddox}, {Marano}, {Marconi}, {Meneux},
  {Mobasher}, {Moreau}, {Peacock}, {Porciani}, {Pozzetti}, {Scaramella},
  {Schiminovich}, {Shopbell}, {Smail}, {Thompson}, {Tresse}, {Vettolani},
  {Zanichelli}, \& {Zucca}}]{Lilly2007}
{Lilly}, S.~J., {Le F{\`e}vre}, O., {Renzini}, A., {et~al.} 2007, \apjs, 172,
  70

\bibitem[{{Maller} \& {Bullock}(2004)}]{MB2004}
{Maller}, A.~H., \& {Bullock}, J.~S. 2004, \mnras, 355, 694

\bibitem[{{Martin} {et~al.}(2012){Martin}, {Shapley}, {Coil}, {Kornei},
  {Bundy}, {Weiner}, {Noeske}, \& {Schiminovich}}]{Martin2012}
{Martin}, C.~L., {Shapley}, A.~E., {Coil}, A.~L., {et~al.} 2012, ArXiv
  e-prints, arXiv:1206.5552

\bibitem[{{McCourt} {et~al.}(2015){McCourt}, {O'Leary}, {Madigan}, \&
  {Quataert}}]{McCourt2015}
{McCourt}, M., {O'Leary}, R.~M., {Madigan}, A.-M., \& {Quataert}, E. 2015,
  \mnras, 449, 2

\bibitem[Mendez et al.(2016)]{Mendez2016} Mendez, A.~J., Coil, A.~L., Aird, J., et al.\ 2016, \apj, 821, 55 

\bibitem[{{Moustakas} {et~al.}(2013){Moustakas}, {Coil}, {Aird}, {Blanton},
  {Cool}, {Eisenstein}, {Mendez}, {Wong}, {Zhu}, \& {Arnouts}}]{Moustakas2013}
{Moustakas}, J., {Coil}, A.~L., {Aird}, J., {et~al.} 2013, \apj, 767, 50

\bibitem[{{Murray} {et~al.}(2010){Murray}, {Quataert}, \&
  {Thompson}}]{Murray2010}
{Murray}, N., {Quataert}, E., \& {Thompson}, T.~A. 2010, \apj, 709, 191

\bibitem[{{Nielsen} {et~al.}(2013){Nielsen}, {Churchill}, \&
  {Kacprzak}}]{Nielsen2013}
{Nielsen}, N.~M., {Churchill}, C.~W., \& {Kacprzak}, G.~G. 2013, \apj, 776, 115

\bibitem[{{Nielsen} {et~al.}(2015){Nielsen}, {Churchill}, {Kacprzak}, {Murphy},
  \& {Evans}}]{Nielsen2015}
{Nielsen}, N.~M., {Churchill}, C.~W., {Kacprzak}, G.~G., {Murphy}, M.~T., \&
  {Evans}, J.~L. 2015, \apj, 812, 83

\bibitem[{{Oliver} {et~al.}(2000){Oliver}, {Rowan-Robinson}, {Alexander},
  {Almaini}, {Balcells}, {Baker}, {Barcons}, {Barden}, {Bellas-Velidis},
  {Cabrera-Guerra}, {Carballo}, {Cesarsky}, {Ciliegi}, {Clements}, {Crockett},
  {Danese}, {Dapergolas}, {Drolias}, {Eaton}, {Efstathiou}, {Egami}, {Elbaz},
  {Fadda}, {Fox}, {Franceschini}, {Genzel}, {Goldschmidt}, {Graham},
  {Gonzalez-Serrano}, {Gonzalez-Solares}, {Granato}, {Gruppioni},
  {Herbstmeier}, {H{\'e}raudeau}, {Joshi}, {Kontizas}, {Kontizas},
  {Kotilainen}, {Kunze}, {La Franca}, {Lari}, {Lawrence}, {Lemke},
  {Linden-V{\o}rnle}, {Mann}, {M{\'a}rquez}, {Masegosa}, {Mattila}, {McMahon},
  {Miley}, {Missoulis}, {Mobasher}, {Morel}, {N{\o}rgaard-Nielsen}, {Omont},
  {Papadopoulos}, {Perez-Fournon}, {Puget}, {Rigopoulou}, {Rocca-Volmerange},
  {Serjeant}, {Silva}, {Sumner}, {Surace}, {Vaisanen}, {van der Werf}, {Verma},
  {Vigroux}, {Villar-Martin}, \& {Willott}}]{Oliver2000}
{Oliver}, S., {Rowan-Robinson}, M., {Alexander}, D.~M., {et~al.} 2000, \mnras,
  316, 749

\bibitem[{{Peng} {et~al.}(2010){Peng}, {Lilly}, {Kova{\v c}}, {Bolzonella},
  {Pozzetti}, {Renzini}, {Zamorani}, {Ilbert}, {Knobel}, {Iovino}, {Maier},
  {Cucciati}, {Tasca}, {Carollo}, {Silverman}, {Kampczyk}, {de Ravel},
  {Sanders}, {Scoville}, {Contini}, {Mainieri}, {Scodeggio}, {Kneib}, {Le
  F{\`e}vre}, {Bardelli}, {Bongiorno}, {Caputi}, {Coppa}, {de la Torre},
  {Franzetti}, {Garilli}, {Lamareille}, {Le Borgne}, {Le Brun}, {Mignoli},
  {Perez Montero}, {Pello}, {Ricciardelli}, {Tanaka}, {Tresse}, {Vergani},
  {Welikala}, {Zucca}, {Oesch}, {Abbas}, {Barnes}, {Bordoloi}, {Bottini},
  {Cappi}, {Cassata}, {Cimatti}, {Fumana}, {Hasinger}, {Koekemoer},
  {Leauthaud}, {Maccagni}, {Marinoni}, {McCracken}, {Memeo}, {Meneux}, {Nair},
  {Porciani}, {Presotto}, \& {Scaramella}}]{Peng2010}
{Peng}, Y.-j., {Lilly}, S.~J., {Kova{\v c}}, K., {et~al.} 2010, \apj, 721, 193

\bibitem[{{Pierre} {et~al.}(2004){Pierre}, {Valtchanov}, {Altieri}, {Andreon},
  {Bolzonella}, {Bremer}, {Disseau}, {Dos Santos}, {Gandhi}, {Jean}, {Pacaud},
  {Read}, {Refregier}, {Willis}, {Adami}, {Alloin}, {Birkinshaw}, {Chiappetti},
  {Cohen}, {Detal}, {Duc}, {Gosset}, {Hjorth}, {Jones}, {Le F{\`e}vre},
  {Lonsdale}, {Maccagni}, {Mazure}, {McBreen}, {McCracken}, {Mellier},
  {Ponman}, {Quintana}, {Rottgering}, {Smette}, {Surdej}, {Starck}, {Vigroux},
  \& {White}}]{Pierre2004}
{Pierre}, M., {Valtchanov}, I., {Altieri}, B., {et~al.} 2004, \jcap, 9, 11

\bibitem[{{Prochaska} {et~al.}(2013{\natexlab{a}}){Prochaska}, {Hennawi}, \&
  {Simcoe}}]{QPQ5}
{Prochaska}, J.~X., {Hennawi}, J.~F., \& {Simcoe}, R.~A. 2013{\natexlab{a}},
  \apjl, 762, L19

\bibitem[{{Prochaska} {et~al.}(2011){Prochaska}, {Kasen}, \&
  {Rubin}}]{Prochaska2011}
{Prochaska}, J.~X., {Kasen}, D., \& {Rubin}, K. 2011, ArXiv e-prints,
  arXiv:1102.3444

\bibitem[{{Prochaska} {et~al.}(2014){Prochaska}, {Wingyee Lau}, \&
  {Hennawi}}]{QPQ7}
{Prochaska}, J.~X., {Wingyee Lau}, M., \& {Hennawi}, J.~F. 2014, ArXiv
  e-prints, arXiv:1409.6344

\bibitem[{{Prochaska} {et~al.}(2013{\natexlab{b}}){Prochaska}, {Hennawi},
  {Lee}, {Cantalupo}, {Bovy}, {Djorgovski}, {Ellison}, {Lau}, {Martin},
  {Myers}, {Rubin}, \& {Simcoe}}]{QPQ6}
{Prochaska}, J.~X., {Hennawi}, J.~F., {Lee}, K.-G., {et~al.}
  2013{\natexlab{b}}, \apj, 776, 136

\bibitem[{{Rees} \& {Ostriker}(1977)}]{ReesOstriker1977}
{Rees}, M.~J., \& {Ostriker}, J.~P. 1977, \mnras, 179, 541

\bibitem[{{Rubin} {et~al.}(2015){Rubin}, {Hennawi}, {Prochaska}, {Simcoe},
  {Myers}, \& {Lau}}]{Rubin2015}
{Rubin}, K.~H.~R., {Hennawi}, J.~F., {Prochaska}, J.~X., {et~al.} 2015, \apj,
  808, 38

\bibitem[{{Rubin} {et~al.}(2014){Rubin}, {Prochaska}, {Koo}, {Phillips},
  {Martin}, \& {Winstrom}}]{Rubin2013}
{Rubin}, K.~H.~R., {Prochaska}, J.~X., {Koo}, D.~C., {et~al.} 2014, \apj, 794,
  156

\bibitem[{{Rubin} {et~al.}(2010{\natexlab{a}}){Rubin}, {Prochaska}, {Koo},
  {Phillips}, \& {Weiner}}]{Rubin2010}
{Rubin}, K.~H.~R., {Prochaska}, J.~X., {Koo}, D.~C., {Phillips}, A.~C., \&
  {Weiner}, B.~J. 2010{\natexlab{a}}, \apj, 712, 574

\bibitem[{{Rubin} {et~al.}(2010{\natexlab{b}}){Rubin}, {Weiner}, {Koo},
  {Martin}, {Prochaska}, {Coil}, \& {Newman}}]{RubinTKRS2009}
{Rubin}, K.~H.~R., {Weiner}, B.~J., {Koo}, D.~C., {et~al.} 2010{\natexlab{b}},
  \apj, 719, 1503

\bibitem[{{Schaye} {et~al.}(2007){Schaye}, {Carswell}, \& {Kim}}]{Schaye2007}
{Schaye}, J., {Carswell}, R.~F., \& {Kim}, T.-S. 2007, \mnras, 379, 1169

\bibitem[{{Scoville} {et~al.}(2007){Scoville}, {Abraham}, {Aussel}, {Barnes},
  {Benson}, {Blain}, {Calzetti}, {Comastri}, {Capak}, {Carilli}, {Carlstrom},
  {Carollo}, {Colbert}, {Daddi}, {Ellis}, {Elvis}, {Ewald}, {Fall},
  {Franceschini}, {Giavalisco}, {Green}, {Griffiths}, {Guzzo}, {Hasinger},
  {Impey}, {Kneib}, {Koda}, {Koekemoer}, {Lefevre}, {Lilly}, {Liu},
  {McCracken}, {Massey}, {Mellier}, {Miyazaki}, {Mobasher}, {Mould}, {Norman},
  {Refregier}, {Renzini}, {Rhodes}, {Rich}, {Sanders}, {Schiminovich},
  {Schinnerer}, {Scodeggio}, {Sheth}, {Shopbell}, {Taniguchi}, {Tyson}, {Urry},
  {Van Waerbeke}, {Vettolani}, {White}, \& {Yan}}]{Scoville2007}
{Scoville}, N., {Abraham}, R.~G., {Aussel}, H., {et~al.} 2007, \apjs, 172, 38

\bibitem[{{Shakura} \& {Sunyaev}(1973)}]{SS1973}
{Shakura}, N.~I., \& {Sunyaev}, R.~A. 1973, \aap, 24, 337

\bibitem[{{Simcoe} {et~al.}(2004){Simcoe}, {Sargent}, \& {Rauch}}]{Simcoe2004}
{Simcoe}, R.~A., {Sargent}, W.~L.~W., \& {Rauch}, M. 2004, \apj, 606, 92

\bibitem[{{Speagle} {et~al.}(2014){Speagle}, {Steinhardt}, {Capak}, \&
  {Silverman}}]{Speagle2014}
{Speagle}, J.~S., {Steinhardt}, C.~L., {Capak}, P.~L., \& {Silverman}, J.~D.
  2014, \apjs, 214, 15

\bibitem[{{Steidel}(1995)}]{Steidel1995}
{Steidel}, C.~C. 1995, in QSO Absorption Lines, ed. G.~{Meylan}, 139--+

\bibitem[{{Steidel} {et~al.}(2010){Steidel}, {Erb}, {Shapley}, {Pettini},
  {Reddy}, {Bogosavljevi{\'c}}, {Rudie}, \& {Rakic}}]{Steidel2010}
{Steidel}, C.~C., {Erb}, D.~K., {Shapley}, A.~E., {et~al.} 2010, \apj, 717, 289

\bibitem[{{Tripp} {et~al.}(2011){Tripp}, {Meiring}, {Prochaska}, {Willmer},
  {Howk}, {Werk}, {Jenkins}, {Bowen}, {Lehner}, {Sembach}, {Thom}, \&
  {Tumlinson}}]{Tripp2011}
{Tripp}, T.~M., {Meiring}, J.~D., {Prochaska}, J.~X., {et~al.} 2011, Science,
  334, 952

\bibitem[{{Weiner} {et~al.}(2009){Weiner}, {Coil}, {Prochaska}, {Newman},
  {Cooper}, {Bundy}, {Conselice}, {Dutton}, {Faber}, {Koo}, {Lotz}, {Rieke}, \&
  {Rubin}}]{Weiner2009}
{Weiner}, B.~J., {Coil}, A.~L., {Prochaska}, J.~X., {et~al.} 2009, \apj, 692,
  187

\bibitem[{{Werk} {et~al.}(2013){Werk}, {Prochaska}, {Thom}, {Tumlinson},
  {Tripp}, {O'Meara}, \& {Peeples}}]{Werk2013}
{Werk}, J.~K., {Prochaska}, J.~X., {Thom}, C., {et~al.} 2013, \apjs, 204, 17

\bibitem[{{Werk} {et~al.}(2014){Werk}, {Prochaska}, {Tumlinson}, {Peeples},
  {Tripp}, {Fox}, {Lehner}, {Thom}, {O'Meara}, {Ford}, {Bordoloi}, {Katz},
  {Tejos}, {Oppenheimer}, {Dav{\'e}}, \& {Weinberg}}]{Werk2014}
{Werk}, J.~K., {Prochaska}, J.~X., {Tumlinson}, J., {et~al.} 2014, ArXiv
  e-prints, arXiv:1403.0947

\bibitem[{{Willmer} {et~al.}(2006){Willmer}, {Faber}, {Koo}, {Weiner},
  {Newman}, {Coil}, {Connolly}, {Conroy}, {Cooper}, {Davis}, {Finkbeiner},
  {Gerke}, {Guhathakurta}, {Harker}, {Kaiser}, {Kassin}, {Konidaris}, {Lin},
  {Luppino}, {Madgwick}, {Noeske}, {Phillips}, \& {Yan}}]{Willmer2006}
{Willmer}, C.~N.~A., {Faber}, S.~M., {Koo}, D.~C., {et~al.} 2006, \apj, 647,
  853

\bibitem[{{Wirth} {et~al.}(2004){Wirth}, {Willmer}, {Amico}, {Chaffee},
  {Goodrich}, {Kwok}, {Lyke}, {Mader}, {Tran}, {Barger}, {Cowie}, {Capak},
  {Coil}, {Cooper}, {Conrad}, {Davis}, {Faber}, {Hu}, {Koo}, {Le Mignant},
  {Newman}, \& {Songaila}}]{Wirth2004}
{Wirth}, G.~D., {Willmer}, C.~N.~A., {Amico}, P., {et~al.} 2004, \aj, 127, 3121

\bibitem[{{Wittman} {et~al.}(2002){Wittman}, {Tyson}, {Dell'Antonio}, {Becker},
  {Margoniner}, {Cohen}, {Norman}, {Loomba}, {Squires}, {Wilson}, {Stubbs},
  {Hennawi}, {Spergel}, {Boeshaar}, {Clocchiatti}, {Hamuy}, {Bernstein},
  {Gonzalez}, {Guhathakurta}, {Hu}, {Seljak}, \& {Zaritsky}}]{Wittman2002}
{Wittman}, D.~M., {Tyson}, J.~A., {Dell'Antonio}, I.~P., {et~al.} 2002, in
  Society of Photo-Optical Instrumentation Engineers (SPIE) Conference Series,
  Vol. 4836, Survey and Other Telescope Technologies and Discoveries, ed. J.~A.
  {Tyson} \& S.~{Wolff}, 73--82

\bibitem[{{Zhu} {et~al.}(2014){Zhu}, {M{\'e}nard}, {Bizyaev}, {Brewington},
  {Ebelke}, {Ho}, {Kinemuchi}, {Malanushenko}, {Malanushenko}, {Marchante},
  {More}, {Oravetz}, {Pan}, {Petitjean}, \& {Simmons}}]{Zhu2014}
{Zhu}, G., {M{\'e}nard}, B., {Bizyaev}, D., {et~al.} 2014, \mnras, 439, 3139

\end{thebibliography}

\startlongtable
\tabletypesize{\scriptsize}
\begin{deluxetable*}{lccccccccccc}
\tablecolumns{12}
\tablecaption{Galaxy Pair Targets \label{tab.pairinfo}}
\tablewidth{0pt}
\tablehead{\colhead{Pair ID}  & \multicolumn{2}{c}{Background Galaxy} & \multicolumn{2}{c}{Foreground Galaxy} & \colhead {$z^{\rm PR}_{\rm b/g}$\tablenotemark{a}} & \colhead {$z^{\rm PR}_{\rm f/g}$\tablenotemark{a}} & \colhead {$z_{\rm b/g}\tablenotemark{b}$} & \colhead {$z_{\rm f/g}\tablenotemark{b}$} & \colhead{$B_{\rm AB}(\rm b/g)$} & \colhead{$B_{\rm AB}(\rm f/g)$} & \colhead{Angular Separation} \\
 & R.A. & Decl. & R.A. & Decl. &  &  &  &  &  &  & arcsec  }
\startdata
 101 & 23:26:33.73 & +00:00:04.1 & 23:26:33.89 & -00:00:03.0 &  0.85 &  0.49 &    0.84640 &    0.49274 &  21.9 &  21.9 &   7.52\\
 102 & 23:27:29.20 & +00:20:27.6 & 23:27:29.04 & +00:20:29.7 &  1.35 &  1.24 &    1.34876 &    1.24400 &  22.2 &  23.6 &   3.19\\
 103* & 23:30:55.20 & +00:05:57.5 & 23:30:54.90 & +00:05:54.9 &  0.69 &  0.48 &    0.69275 &    0.47989 &  21.4 &  22.7 &   5.12\\
 104* & 23:30:50.56 & +00:15:55.8 & 23:30:50.63 & +00:16:02.0 &  1.94 &  0.41 &    1.94227 &    0.40986 &  19.9 &  23.1 &   6.35\\
 201 & 02:16:17.89 & -04:32:04.6 & 02:16:17.42 & -04:32:01.8 &  0.50 &  0.44 &    0.49836 &    0.44695 &  22.0 &  22.7 &   7.51\\
 202 & 02:16:15.62 & -04:32:59.8 & 02:16:15.79 & -04:33:02.1 &  0.67 &  0.53 &    0.66387 &    0.66415 &  21.7 &  22.7 &   3.49\\
 203 & 02:15:33.15 & -04:26:11.5 & 02:15:32.80 & -04:26:08.7 &  0.50 &  0.35 &    0.50979 &    0.35138 &  22.0 &  21.5 &   5.92\\
 204 & 02:17:50.82 & -04:16:00.3 & 02:17:50.70 & -04:15:57.0 &  0.61 &  0.45 &    0.60910 &    0.44265 &  21.8 &  22.2 &   3.74\\
 207* & 02:21:08.80 & -03:59:44.0 & 02:21:08.38 & -03:59:40.0 &  0.68 &  0.43 &    0.68732 &    0.43092 &  21.5 &  22.7 &   7.48\\
 208* & 02:20:49.50 & -04:30:31.2 & 02:20:49.54 & -04:30:28.7 &  1.81 &  0.60 &    1.81307 &    0.59729 &  22.0 &  22.2 &   2.57\\
 209 & 02:22:12.28 & -05:07:24.0 & 02:22:12.14 & -05:07:24.8 &  0.62 &  0.35 &    0.61682 &    0.35012 &  21.7 &  22.5 &   2.23\\
 210 & 02:21:09.56 & -04:55:25.5 & 02:21:09.64 & -04:55:26.2 &  0.88 &  0.79 &    0.87675 &    0.80620 &  22.3 &  22.9 &   1.41\\
 211 & 02:22:24.78 & -05:02:29.0 & 02:22:25.05 & -05:02:31.3 &  0.58 &  0.41 &    0.57994 &    0.40878 &  21.7 &  21.5 &   4.68\\
 212* & 02:20:05.32 & -05:19:14.9 & 02:20:05.51 & -05:19:21.3 &  1.91 &  0.55 &    1.91296 &    0.55047 &  21.0 &  23.3 &   6.97\\
 213 & 02:20:41.35 & -05:35:30.2 & 02:20:41.33 & -05:35:32.7 &  0.62 &  0.47 &    0.61605 &    0.46911 &  21.6 &  22.3 &   2.57\\
 216* & 02:23:07.94 & -04:59:09.0 & 02:23:07.79 & -04:59:09.2 &  1.33 &  0.36 &    1.32355 &    ---    &  20.5 &  23.1 &   2.31\\
 217 & 02:23:20.72 & -05:32:08.7 & 02:23:20.27 & -05:32:08.8 &  0.49 &  0.36 &    0.49148 &    0.35417 &  21.1 &  21.2 &   6.86\\
 219 & 02:23:10.22 & -05:21:28.8 & 02:23:10.46 & -05:21:33.1 &  0.44 &  0.36 &    0.43907 &    0.36881 &  21.4 &  22.1 &   5.62\\
 221* & 02:19:38.75 & -05:11:03.4 & 02:19:38.62 & -05:11:00.0 &  1.74 &  0.74 &    1.75261 &    star    &  21.8 &  23.1 &   3.92\\
 223* & 02:17:03.70 & -04:37:38.3 & 02:17:04.03 & -04:37:41.7 &  1.34 &  0.70 &    1.35232 &    0.74561 &  21.7 &  23.3 &   6.01\\
 301* & 02:28:42.03 & +00:45:36.4 & 02:28:41.56 & +00:45:40.6 &  1.67 &  0.36 &    1.67360 &    0.36276 &  20.3 &  21.6 &   8.24\\
 302 & 02:32:17.94 & +00:50:02.0 & 02:32:17.92 & +00:50:05.9 &  0.46 &  0.35 &    0.47785 &    0.35051 &  21.5 &  22.3 &   3.93\\
 402* & 03:31:07.94 & -28:33:58.7 & 03:31:07.62 & -28:33:55.4 &  0.68 &  0.56 &    0.68496 &    0.56820 &  21.7 &  23.0 &   5.43\\
 403* & 03:31:07.94 & -28:33:58.7 & 03:31:08.10 & -28:34:05.3 &  0.68 &  0.58 &    0.68496 &    0.57017 &  21.7 &  23.4 &   6.85\\
 404* & 03:32:14.41 & -29:17:05.8 & 03:32:14.37 & -29:17:11.7 &  0.64 &  0.43 &    0.63664 &    0.43280 &  21.4 &  23.3 &   5.93\\
 405 & 03:31:56.65 & -29:13:14.3 & 03:31:57.04 & -29:13:19.7 &  0.69 &  0.38 &    0.69596 &    0.38218 &  21.9 &  22.8 &   7.49\\
 408 & 03:35:24.53 & -28:56:33.6 & 03:35:24.48 & -28:56:31.2 &  0.78 &  0.59 &    0.78855 &    0.59745 &  22.1 &  22.9 &   2.48\\
 409 & 03:37:04.75 & -28:49:14.8 & 03:37:04.60 & -28:49:23.8 &  0.58 &  0.36 &    0.58391 &    0.35703 &  21.7 &  21.5 &   9.22\\
 410 & 03:36:24.72 & -28:42:26.2 & 03:36:24.36 & -28:42:30.9 &  0.84 &  0.52 &    0.86234 &    0.51464 &  22.0 &  22.5 &   6.71\\
 411 & 03:36:37.56 & -28:43:02.6 & 03:36:38.13 & -28:43:02.4 &  0.41 &  0.36 &    0.41642 &    0.36680 &  20.8 &  21.9 &   7.52\\
 412 & 03:35:28.25 & -28:47:22.4 & 03:35:28.02 & -28:47:24.5 &  0.78 &  0.70 &    0.78699 &    0.73092 &  22.1 &  22.4 &   3.65\\
 413 & 03:35:42.97 & -28:21:36.6 & 03:35:42.64 & -28:21:42.6 &  0.57 &  0.43 &    0.56530 &    0.42752 &  21.6 &  22.6 &   7.39\\
 414 & 03:36:27.84 & -28:26:14.5 & 03:36:28.29 & -28:26:10.6 &  0.58 &  0.44 &    0.58390 &    0.43821 &  21.9 &  22.4 &   7.19\\
 417 & 03:28:06.08 & -28:30:57.7 & 03:28:06.22 & -28:31:01.6 &  0.77 &  0.57 &    0.78619 &    0.56884 &  22.1 &  23.1 &   4.30\\
 419 & 03:27:58.78 & -29:06:23.9 & 03:27:58.89 & -29:06:17.3 &  0.62 &  0.49 &         ---    &       ---    &  22.0 &  22.5 &   6.73\\
 601 & 10:01:07.45 & +02:26:26.1 & 10:01:07.11 & +02:26:20.9 &  1.00 &  0.49 &    0.98752 &    0.49290 &  21.9 &  23.3 &   7.32\\
 602* & 10:01:18.58 & +02:27:39.3 & 10:01:18.22 & +02:27:42.8 &  1.05 &  0.53 &    1.04212 &    0.51684 &  21.0 &  23.0 &   6.51\\
 603 & 10:01:30.53 & +02:19:00.2 & 10:01:30.50 & +02:19:03.0 &  0.96 &  0.69 &    0.98001 &    0.69700 &  22.3 &  22.8 &   2.81\\
 604 & 09:59:31.46 & +02:19:03.2 & 09:59:31.27 & +02:19:05.0 &  0.72 &  0.66 &    0.73213 &    0.25032 &  22.1 &  22.8 &   3.36\\
 605* & 09:59:05.12 & +02:15:30.0 & 09:59:04.89 & +02:15:25.7 &  2.23 &  1.11 &    2.20732 &    ---    &  21.4 &  22.9 &   5.60\\
 606 & 09:59:59.81 & +02:28:27.9 & 09:59:59.94 & +02:28:35.8 &  0.50 &  0.35 &    0.48726 &    0.34656 &  20.9 &  22.8 &   8.14\\
 607 & 09:59:45.19 & +02:29:40.6 & 09:59:45.35 & +02:29:39.5 &  0.55 &  0.49 &    0.54752 &    0.54760 &  21.7 &  23.2 &   2.66\\
 608 & 10:00:08.41 & +02:41:55.2 & 10:00:08.30 & +02:41:56.9 &  0.66 &  0.50 &    0.64716 &    0.50361 &  22.0 &  22.0 &   2.38\\
 609* & 10:00:38.15 & +02:49:30.5 & 10:00:38.45 & +02:49:30.6 &  1.87 &  0.43 &    1.85705 &    0.43755 &  21.6 &  22.4 &   4.47\\
 610 & 10:00:40.37 & +02:49:02.0 & 10:00:40.64 & +02:48:55.9 &  0.52 &  0.49 &    0.49556 &    0.49550 &  22.1 &  22.4 &   7.28\\
 611 & 10:02:55.24 & +02:32:55.4 & 10:02:55.06 & +02:32:55.1 &  0.47 &  0.35 &    0.47130 &    0.35155 &  21.9 &  21.9 &   2.66\\
 612* & 10:01:20.26 & +02:33:41.4 & 10:01:20.24 & +02:33:43.8 &  1.83 &  0.39 &    1.84181 &    0.36079 &  20.4 &  22.5 &   2.42\\
 613* & 10:01:24.86 & +02:20:31.8 & 10:01:24.66 & +02:20:29.7 &  1.71 &  0.72 &    1.71462 &    0.74056 &  21.1 &  22.9 &   3.59\\
 614 & 10:02:56.98 & +02:17:28.4 & 10:02:56.71 & +02:17:30.5 &  0.59 &  0.36 &    0.60588 &    0.36360 &  21.7 &  21.1 &   4.50\\
 615* & 10:01:47.90 & +02:14:47.2 & 10:01:47.72 & +02:14:48.3 &  0.89 &  0.84 &    0.87988 &    0.87795 &  20.5 &  23.5 &   2.86\\
 617* & 09:59:00.21 & +02:28:11.7 & 09:59:00.58 & +02:28:14.9 &  0.66 &  0.49 &    0.65789 &    0.48812 &  21.7 &  22.2 &   6.28\\
 619 & 09:59:52.87 & +01:55:31.7 & 09:59:53.25 & +01:55:33.7 &  0.45 &  0.35 &    0.44598 &    0.35244 &  20.9 &  23.2 &   6.03\\
 620 & 10:00:22.44 & +01:56:57.2 & 10:00:22.33 & +01:56:55.2 &  0.72 &  0.44 &    star    &    0.43649 &  21.8 &  22.5 &   2.56\\
 621 & 09:59:51.78 & +02:19:37.7 & 09:59:52.41 & +02:19:38.0 &  0.38 &  0.36 &    0.37802 &    0.35741 &  21.6 &  21.5 &   9.42\\
 622 & 09:59:25.07 & +02:38:40.9 & 09:59:24.95 & +02:38:42.3 &  0.96 &  0.86 &    0.94283 &    0.94349 &  22.2 &  23.4 &   2.20\\
 623* & 10:01:47.04 & +02:02:36.6 & 10:01:47.16 & +02:02:31.2 &  1.18 &  0.84 &    1.17012 &    0.82579 &  21.8 &  23.5 &   5.71\\
 624 & 10:01:19.79 & +02:14:32.3 & 10:01:20.08 & +02:14:32.5 &  0.45 &  0.36 &    0.44751 &    0.36054 &  21.7 &  21.7 &   4.27\\
 625* & 10:01:23.01 & +02:08:51.2 & 10:01:23.21 & +02:08:46.5 &  1.27 &  0.61 &    1.25811 &    0.60352 &  21.0 &  23.4 &   5.54\\
 626 & 10:01:56.74 & +02:04:58.9 & 10:01:56.38 & +02:04:57.1 &  0.70 &  0.43 &    0.70399 &    0.42486 &  22.0 &  22.6 &   5.57\\
 1100 & 23:30:53.46 & +00:07:18.2 & 23:30:53.55 & +00:07:30.2 &  0.54 &  0.48 &    ---    &    0.48242 &  21.8 &  22.9 &  12.08\\
 1101 & 23:31:06.44 & +00:05:43.7 & 23:31:07.59 & +00:05:25.3 &  0.47 &  0.40 &    0.46690 &    0.40722 &  22.0 &  22.0 &  25.16\\
 1200* & 02:18:34.40 & -04:00:12.1 & 02:18:34.72 & -04:00:06.7 &  1.50 &  0.95 &    1.50475 &    ---    &  20.4 &  23.3 &   7.16\\
 1201* & 02:16:14.93 & -04:06:26.4 & 02:16:15.44 & -04:06:31.5 &  1.15 &  0.56 &    1.14869 &    0.56532 &  21.2 &  23.2 &   9.23\\
 1202 & 02:20:04.83 & -05:14:27.6 & 02:20:04.28 & -05:14:49.8 &  0.53 &  0.40 &    0.53765 &    0.40820 &  21.0 &  23.2 &  23.68\\
 1203 & 02:17:17.56 & -04:42:03.1 & 02:17:18.63 & -04:42:00.9 &  0.43 &  0.38 &    0.43191 &    0.37316 &  22.0 &  23.2 &  16.15\\
 1204 & 02:19:45.93 & -05:10:24.0 & 02:19:44.52 & -05:10:28.8 &  0.93 &  0.39 &    star    &    0.50097 &  21.2 &  23.2 &  21.69\\
 1205 & 02:17:10.20 & -04:36:43.0 & 02:17:10.08 & -04:36:55.6 &  0.59 &  0.37 &    0.60304 &    0.37238 &  21.2 &  22.2 &  12.69\\
 1206 & 02:16:58.45 & -04:38:47.5 & 02:16:56.90 & -04:38:52.9 &  0.45 &  0.37 &    0.07124 &    ---    &  21.5 &  23.0 &  23.89\\
 1207 & 02:15:59.08 & -04:30:46.0 & 02:15:58.34 & -04:30:31.8 &  1.16 &  0.73 &    1.27964 &    0.74230 &  21.5 &  23.4 &  17.92\\
 1208* & 02:19:34.70 & -04:41:41.0 & 02:19:35.19 & -04:41:41.8 &  2.12 &  0.89 &    2.09599 &    0.83566 &  20.8 &  22.8 &   7.42\\
 1209 & 02:17:40.60 & -04:12:31.3 & 02:17:40.04 & -04:12:47.6 &  0.43 &  0.37 &    0.43466 &    0.37102 &  22.0 &  21.6 &  18.35\\
 1210* & 02:18:09.32 & -04:27:56.9 & 02:18:09.40 & -04:28:05.1 &  1.55 &  0.58 &    1.53694 &    0.59573 &  21.6 &  23.0 &   8.32\\
 1300 & 02:32:11.66 & +00:43:34.4 & 02:32:10.94 & +00:43:23.3 &  0.82 &  0.54 &    0.81118 &    0.58506 &  21.3 &  23.2 &  15.42\\
 1400* & 03:35:01.74 & -28:53:47.5 & 03:35:02.40 & -28:53:33.5 &  2.02 &  0.37 &    2.03408 &    ---    &  19.8 &  22.6 &  16.46\\
 1401 & 03:36:59.69 & -28:55:12.1 & 03:37:00.04 & -28:54:57.5 &  0.67 &  0.46 &    star    &    0.46915 &  19.9 &  22.7 &  15.33\\
 1402 & 03:36:21.62 & -28:29:59.7 & 03:36:20.96 & -28:30:07.0 &  0.61 &  0.54 &    0.59685 &    0.54178 &  22.0 &  21.5 &  11.34\\
 1403* & 03:32:14.41 & -29:17:05.8 & 03:32:13.86 & -29:16:57.6 &  0.64 &  0.53 &    0.63664 &    0.63920 &  21.4 &  22.8 &  10.94\\
 1600 & 10:01:30.53 & +02:19:00.2 & 10:01:30.54 & +02:18:57.5 &  0.96 &  0.49 &    0.98001 &    0.49714 &  22.3  &  -99.0 &   2.64\\
 1601* & 10:02:26.11 & +02:46:10.9 & 10:02:26.25 & +02:46:19.8 &  3.03 &  0.54 &    3.02746 &    0.53593 &  21.0 &  23.1 &   9.19\\
 1602 & 10:02:03.38 & +02:02:25.1 & 10:02:02.14 & +02:02:24.4 &  0.43 &  0.36 &    0.42516 &    0.36434 &  21.1 &  21.1 &  18.54\\
 1603 & 10:02:47.94 & +02:29:28.2 & 10:02:47.17 & +02:29:17.9 &  0.60 &  0.36 &    0.60936 &    0.36689 &  22.0 &  21.5 &  15.51\\
 1604 & 10:01:30.50 & +02:19:03.0 & 10:01:30.54 & +02:18:57.5 &  0.69 &  0.49 &    0.69700 &    0.49714 &  22.8  &  -99.0 &   5.44\\
 1605* & 10:01:10.19 & +02:32:42.4 & 10:01:10.49 & +02:32:26.3 &  2.67 &  0.38 &    2.65219 &    0.37611 &  21.5 &  21.1 &  16.69\\
 1606* & 09:59:03.22 & +02:20:02.9 & 09:59:03.83 & +02:19:56.0 &  1.14 &  0.38 &    1.13109 &    0.37167 &  21.2 &  23.1 &  11.42\\
 1607 & 09:59:43.12 & +02:38:31.0 & 09:59:43.32 & +02:38:20.7 &  0.55 &  0.51 &    0.54694 &    0.29322 &  21.7 &  23.2 &  10.69\\
 1608 & 10:01:58.47 & +02:03:50.6 & 10:01:57.79 & +02:04:00.7 &  0.54 &  0.44 &    0.53148 &    0.43827 &  22.1 &  22.7 &  14.35\\
 1609 & 10:01:50.91 & +02:03:47.7 & 10:01:51.04 & +02:04:04.2 &  0.54 &  0.35 &    0.53443 &    0.35523 &  21.8 &  22.2 &  16.65\\
 1610* & 10:00:28.63 & +02:51:12.7 & 10:00:29.43 & +02:51:07.2 &  0.78 &  0.73 &    0.76735 &    0.73089 &  21.5 &  23.1 &  13.28\\
 1611 & 09:59:44.08 & +02:33:01.7 & 09:59:44.48 & +02:33:18.7 &  0.43 &  0.38 &    0.43934 &    0.37384 &  22.0 &  23.0 &  17.98\\
 1612 & 10:00:14.81 & +01:54:26.2 & 10:00:15.00 & +01:54:06.6 &  0.68 &  0.36 &    0.67049 &    0.36038 &  21.9 &  23.0 &  19.78\\
 1613 & 10:01:54.91 & +02:04:19.4 & 10:01:55.73 & +02:04:09.6 &  0.56 &  0.44 &    0.55452 &    0.43988 &  21.8 &  22.4 &  15.75\\
\enddata
\tablecomments{Pair IDs marked with an asterisk indicate pairs with a broad-line AGN in the background object.  }
\tablenotetext{a}{Redshift determined from low-dispersion PRIMUS prism spectroscopy.}
\tablenotetext{b}{Redshift determined from our Keck/LRIS or VLT/FORS2 followup spectroscopy.  We failed to obtain spectra with S/N sufficient to constrain the redshift for targets with ``---" these columns.}
\end{deluxetable*}

\tabletypesize{\footnotesize}
\begin{deluxetable*}{lllcccc}
\tablecolumns{7}
\tablecaption{Summary of Multislit Observations \label{tab.maskobs}}
\tablewidth{0pt}
\tablehead{\colhead{Mask ID}  & \colhead{R. A.} &
  \colhead{Declination} & \colhead{Pair Targets} &
  \multicolumn{2}{c}{Exposure Time (hrs)} & \colhead{Date}\\ 
 &  \colhead{J2000} & \colhead{J2000}  & & Blue & Red & }
\startdata
 \multicolumn{7}{c}{Keck/LRIS  Spectroscopy}\\ 
XMM-1                       &  02h 16m 10.135s & -04d  31m 02.341s & 201, 202, 1207& 2.06 & 1.90 & 2011 Oct 01\\
XMM-2                           &  02h   21m   03.037s & -04d   53m 54.429s  & 210 & 2.17 & 2.04 & 2012 Jan 20\\
XMM-3                           &   02h   20m  49.988s&  -04d   33m 02.711s & 208 & 1.67 & 1.57 & 2012 Jan 21\\
XMM-4                            &  02h   20m  41.674s &  -05d 32m  18.749s &213 & 0.9 & 0.78 & 2011 Oct 01\\
XMM-5                             &  02h   17m  42.879s & -04d   13m  40.708s  & 204, 1209 & 1.40 & 1.33 & 2012 Jan 21\\
XMM-6              &   02h   15m  41.290s & -04d   27m  21.499s  & 203 & 3.17 & 2.93 & 2012 Dec 14-15\\
XMM-8                            &  02h   19m  50.915s & -05d   12m 54.655s & 221, 1202, 1204 & 0.92 & 0.37 & 2012 Dec 14\\
XMM-9                            &  02h   17m  11.469s &  -04d   38m 49.408s & 223, 1203, 1205, 1206 & 1.33 & 1.25 & 2012 Dec 13\\
DEEP2 $02^{\rm h}$-1 &  02h 32m 17.631s & +00d 46m 51.016s &302, 1300 & 1.00 & 0.89 & 2011 Oct 01 \\
CDFS-1                           &   03h   32m   05.670s & -29d   15m 39.128s & 404, 405, 1403 &  1.80 & 1.67 & 2012 Dec 15\\
CDFS-2                          &  03h   37m   01.165s & -28d   52m   08.792s & 409, 1401 & 1.50 &  1.40 & 2012 Dec 15\\
COSMOS-1                 & 10h    00m  35.040s & +02d 49m  32.190s &  609, 610, 1610 & 2.08 & 1.96 & 2012 Jan 20\\
COSMOS-2                  &  10h 00m  17.445s & +01d   54m 29.869s & 620, 1612 & 1.33 & 1.26 & 2012 Jan 20\\
COSMOS-3                    &  09h   59m  49.540s & +02d 29m  46.049s & 606, 607, 1611& 1.00 & 0.94 & 2012 Jan 20\\
COSMOS-4                     &  09h   59m  40.738s & +02d   19m  37.569s & 604, 621 & 1.67 & 1.56 & 2012 Jan 20\\
COSMOS-5                      & 10h   01m  39.562s & +02d   16m 57.736s & 603, 615, 1600, 1604 & 1.86 & 1.75 & 2012 Dec 14\\
COSMOS-6                    & 10h   01m  09.715s & +02d   29m  16.189s & 601, 1605 & 1.67 & 1.58 & 2012 Jan 21\\
COSMOS-7                     & 10h    00m  11.443s & +02d   41m   08.694s & 608 & 1.43 & 1.33 & 2012 Jan 21\\
COSMOS-8                    & 10h    02m  45.330s & +02d   17m 34.108s & 614 & 0.87 & 0.79 & 2012 Jan 21\\
COSMOS-9                   & 09h   59m  32.490s  & +02d   38m  19.718s& 622, 1607  & 0.83 & 0.83 & 2012 Jan 21\\
COSMOS-10                 & 10h    01m  49.901s & +02d    02m  52.574s& 623, 1608, 1613 & 0.58 & 0.51 & 2012 Jan 21 \\
COSMOS-11                 & 10h    02m  50.485s & +02d   31m  39.362s & 611, 1603  & 2.00 & 1.87 & 2012 Dec 15\\
COSMOS-12            & 10h    01m  34.027s & +02d   14m  45.950s & 615, 624 & 1.50 & 1.34 & 2012 Dec 15\\
COSMOS-13                 &  09h   58m  53.701s &  +02d   26m  51.576s  & 617 & 1.19 & 1.14 & 2012 Dec 13\\
COSMOS-14                 & 10h    01m  48.506s  & +02d    03m 08.747s & 626, 1602, 1609, 1613 & 2.50 & 2.22 & 2012 Dec 14-15\\
DEEP2 $23^{\rm h}$-1 &  23h  31m 0.445s  & +00d 06m 45.338s & 103, 1100, 1101  & 0.78 & 0.75 & 2011 Oct 01\\
\hline\\
 \multicolumn{7}{c}{VLT/FORS2  Spectroscopy} \\ 
XMM-7                & 02h 22m 19.336s & -05d 04m 25.90s & 209, 211 & 3.00 & 0.50 & 2012 Nov 14-15\\
CDFS-1                       & 03h 36m 39.326s & -28d 41m 49.360s & 410, 411 & 1.38 & 0.25 & 2011 Nov 25\\
CDFS-3                       & 03h 31m 12.198s & -28d 35m 39.69s & 402, 403 & 0.83 & 0.50 & 2011 Nov 25\\
CDFS-4                       & 03h 35m 13.89s & -28d 54m 57.14s & 408, 1400 & 1.10 & 0.33 & 2011 Nov 25\\
CDFS-5                       & 03h 35m 26.15s & -28d 46m 43.13s  & 412& 0.92 & 0.25 & 2011 Nov 25\\
CDFS-6                       & 03h 28m 10.935s & -28d 32m 39.720s & 417 & 1.30 & 0.25 & 2012 Nov 15\\
CDFS-8                       & 03h 35m 44.255s & -28d 21m 06.330s & 413 & 0.99 & 0.25 & 2012 Nov 15\\
CDFS-9                       & 03h 27m 55.600s & -29d 08m 10.750s & 419 & 1.30 & 0.25 & 2012 Nov 15\\
CDFS-10                     & 03h 36m 22.847s & -28d 28m 35.940s & 414, 1402 & 1.50 & 0.25 & 2012 Nov 15\\
COSMOS-15          & 10h 01m 23.346s & +02d 30m 52.610s & 602, 612 & 0.67 & 0.25 & 2011 Nov 25\\
\enddata
\tablecomments{LRIS mask coordinates are given at the epoch of
  observation. FORS2 mask coordinates are given in the J2000 reference frame.}
\end{deluxetable*}

\tabletypesize{\footnotesize}
\begin{deluxetable*}{lcccc}
\tablecolumns{5}
\tablecaption{Summary of Longslit Observations \label{tab.longobs}}
\tablewidth{0pt}
\tablehead{\colhead{Pair ID}  & Field & \multicolumn{2}{c}{Exposure
    Time (hrs)} & \colhead{Date and Instrument\tablenotemark{a}}\\ 
 &  &  Blue & Red & }
\startdata
102 & DEEP2 $23^{\rm h}$ & 1.33 & 1.27 & 2012 Dec 15\\
104 & DEEP2 $23^{\rm h}$ & 0.50 & 0.48 & 2012 Dec 13\\
207  & XMM  & 0.67 & 0.26 & 2012 Nov 14 (FORS2)\\
212  & XMM  & 0.89 & 0.50 & 2012 Nov 14 (FORS2)\\
216 &  XMM  & 0.50 & 0.48 & 2012 Jan 20\\
217 &   XMM   & 0.61 & 0.53 & 2011 Oct 01\\
219 & XMM & 1.17 & 1.26 & 2012 Dec 13\\
301 & DEEP2 $02^{\rm h}$   & 0.50 & 0.48 & 2012 Jan 20\\
605 & COSMOS  & 0.50 & 0.48 & 2012 Dec 13\\
613 & COSMOS  & 0.50 & 0.48 & 2012 Dec 13\\
619 & COSMOS  & 0.67& 0.62 & 2012 Jan 21\\
625  & COSMOS   & 0.44 & 0.33 & 2012 Nov 14 (FORS2)\\
1200  & XMM & 0.5 & 0.48 & 2012 Dec 13\\ 
1201  & XMM  & 1.25 & 1.19 & 2012 Dec 13-14\\ 
1208  & XMM  & 0.5 & 0.48 & 2012 Dec 13\\  
1210  & XMM  & 1.83 & 1.74 & 2012 Dec 13-14\\ 
1601  & COSMOS  & 0.5& 0.48 & 2012 Dec 13\\ 
1606  & COSMOS  & 0.75 & 0.72 & 2012 Dec 13\\ 
\enddata
\tablenotetext{a}{The instrument used is Keck/LRIS where not explicitly indicated.}
\end{deluxetable*}

\clearpage
\startlongtable
\tabletypesize{\scriptsize}
\begin{deluxetable*}{lcccccccrrc}
\tablecolumns{11}
\tablecaption{Foreground Galaxy Properties and CGM Absorption Line Measurements \label{tab.galewinfo}}
\tablewidth{0pt}
\tablehead{\colhead{Pair ID}  & \colhead{$z_{\rm f/g}$} & \colhead{$M_B$\tablenotemark{a}} & \colhead {\it U-B\tablenotemark{a}} & \colhead {$\log M_*/M_{\odot}$} & \colhead {SFR} & \colhead {$\mrperp$} & \colhead{S/N(\ion{Mg}{2})} & \colhead{$W_{2796}$} & \colhead{$W_{2803}$} & \colhead{$\left< \delta v_{2796} \right>$}\\
 &  & (mag) & (mag) &  & ($M_{\odot}~\rm yr^{-1}$) & (kpc) & ($\rm \AA^{-1}$) & ($\rm \AA$)  & ($\rm \AA$) & ($\mkms$)  }
\startdata
 101 &     0.4927 &  -20.99 &    0.68 &   10.02 &     8.2 &    45.6 &    18.7 & $  1.116\pm  0.138$ & $  0.354\pm  0.105$ &   -23.9\\
 102 &     1.2440 &  ---  &  ---  &   10.68 &    18.1 &    26.6 &     6.0 & $  2.183\pm  0.326$ & $  2.149\pm  0.364$ &   -58.4\\
 103* &     0.4799 &  -20.34 &    0.80 &   10.11 &     2.7 &    30.5 &     3.2 & $  0.100\pm  0.904$ & $  0.117\pm  0.830$ &    -1.6\\
 104* &     0.4099 &  -19.23 &    0.70 &    9.54 &     1.2 &    34.6 &    35.8 & $ -0.093\pm  0.074$ & $  0.165\pm  0.076$ &   -73.4\\
 201 &     0.4470 &  -20.14 &    0.82 &    9.77 &     0.9 &    43.1 &     6.2 & $ -1.485\pm  0.400$ & $ -0.474\pm  0.398$ &   -92.3\\
 203 &     0.3514 &  -21.35 &    1.22 &   11.12 &     0.4 &    29.3 &     7.4 & $  0.174\pm  0.339$ & $ -0.225\pm  0.355$ &    67.1\\
 204 &     0.4426 &  -21.52 &    1.22 &   11.08 &     0.9 &    21.3 &    11.4 & $  0.012\pm  0.270$ & $  0.869\pm  0.249$ &  -181.6\\
 207* &     0.4309 &  -20.47 &    1.09 &   10.60 &     1.2 &    42.0 &    20.1 & $ -0.147\pm  0.110$ & $ -0.075\pm  0.129$ &   -33.6\\
 208* &     0.5973 &  -21.83 &    0.99 &   11.07 &     8.4 &    17.2 &    13.6 & $  2.012\pm  0.225$ & $  1.931\pm  0.261$ &   183.8\\
 209 &     0.3501 &  -19.38 &    0.55 &    9.37 &     1.7 &    11.0 &     9.2 & $  1.297\pm  0.244$ & $  1.422\pm  0.216^b$ &  -106.4\\
 210 &     0.8062 &  -20.79 &    0.52 &    9.88 &     9.6 &    10.6 &    10.7 & $  2.275\pm  0.247$ & $  1.902\pm  0.225$ &   -28.4\\
 211 &     0.4088 &  -21.22 &    0.84 &   10.68 &     3.1 &    25.5 &    10.4 & $  1.878\pm  0.204$ & $  0.888\pm  0.210$ &  -120.6\\
 212* &     0.5505 &  -19.94 &    0.70 &    9.79 &     2.0 &    44.7 &    12.6 & $  0.084\pm  0.184$ & $  0.185\pm  0.192$ &    22.6\\
 213 &     0.4691 &  -20.21 &    0.55 &    9.40 &     3.4 &    15.1 &    11.1 & $  2.558\pm  0.260$ & $  2.003\pm  0.252$ &    28.1\\
 217 &     0.3542 &  -20.87 &    0.70 &   10.26 &     6.5 &    34.1 &     5.6 & $  0.608\pm  0.482$ & $  0.050\pm  0.478$ &   -13.4\\
 219 &     0.3688 &  -20.00 &    0.70 &    9.69 &     2.2 &    28.7 &    11.7 & $  0.638\pm  0.233$ & $  0.214\pm  0.218$ &    86.5\\
 223* &     0.7456 &  -20.59 &    0.73 &    9.66 &     4.2 &    44.0 &     4.5 & $  0.203\pm  0.538$ & $  1.183\pm  0.612$ &     0.7\\
 301* &     0.3628 &  -20.83 &    0.91 &   10.56 &     2.0 &    41.6 &    13.4 & $  1.856\pm  0.221$ & $  2.128\pm  0.221$ &   187.9\\
 302 &     0.3505 &  -19.98 &    0.93 &   10.44 &     2.1 &    19.4 &     7.5 & $  0.503\pm  0.380$ & $  0.752\pm  0.368$ &   140.1\\
 402* &     0.5682 &  -19.91 &    0.54 &    9.44 &     2.7 &    35.4 &     5.2 & $  1.179\pm  0.401^b$ & $  0.371\pm  0.393$ &   -35.6\\
 403* &     0.5702 &  -20.14 &    0.72 &    9.80 &     2.0 &    44.7 &     5.0 & $  0.268\pm  0.449$ & $ -0.304\pm  0.459$ &  -138.0\\
 404* &     0.4328 &  -19.14 &    0.64 &   10.02 &     2.0 &    33.4 &    38.1 & $ -0.066\pm  0.069$ & $  0.064\pm  0.069$ &   139.4\\
 405 &     0.3822 &  -20.31 &    1.23 &   10.65 &     0.2 &    39.1 &     7.9 & $  0.729\pm  0.345$ & $  0.031\pm  0.305$ &   -95.2\\
 408 &     0.5974 &  -20.89 &    0.91 &   10.55 &     0.1 &    16.6 &     2.0 & $ -1.768\pm  1.017$ & $ -1.008\pm  1.071$ &   -43.0\\
 409 &     0.3570 &  -20.38 &    0.62 &   10.18 &     2.1 &    46.1 &     3.0 & $  1.536\pm  0.924$ & $  2.571\pm  0.871$ &    43.9\\
 410 &     0.5146 &  -20.04 &    0.46 &    9.16 &     3.5 &    41.6 &     3.5 & $  0.184\pm  0.707$ & $ -0.459\pm  0.738$ &    63.9\\
 411 &     0.3668 &  -20.11 &    0.65 &    9.89 &     1.8 &    38.3 &    14.2 & $  0.321\pm  0.161$ & $ -0.375\pm  0.192$ &    97.5\\
 412 &     0.7309 &  -20.99 &    0.53 &   10.11 &     6.1 &    26.5 &     7.1 & $  0.363\pm  0.291$ & $  0.044\pm  0.321$ &   -87.7\\
 413 &     0.4275 &  -20.85 &    1.17 &   10.74 &     0.0 &    41.3 &     9.6 & $  0.065\pm  0.204$ & $  0.727\pm  0.228$ &    -3.1\\
 414 &     0.4382 &  -19.95 &    0.57 &    9.69 &     5.7 &    40.8 &    11.2 & $  0.073\pm  0.188$ & $  0.384\pm  0.168$ &   -93.2\\
 417 &     0.5688 &  -20.56 &    0.87 &   10.21 &     2.9 &    28.0 &     7.2 & $  0.333\pm  0.294$ & $ -0.199\pm  0.290$ &   -12.3\\
 601 &     0.4929 &  -19.28 &    0.51 &    9.15 &     1.3 &    44.3 &    18.4 & $  0.032\pm  0.141$ & $ -0.189\pm  0.136$ &  -114.8\\
 602* &     0.5168 &  -20.16 &    0.66 &    9.87 &     2.5 &    40.5 &    21.6 & $  0.225\pm  0.104$ & $  0.279\pm  0.111$ &    50.5\\
 603 &     0.6970 &  -22.55 &    1.11 &   11.24 &     1.2 &    20.1 &    19.1 & $ -0.320\pm  0.136$ & $  0.339\pm  0.142$ &  -194.7\\
 604 &     0.2503 &  -20.58 &    0.52 &    9.07 &     0.7 &    13.2 &     2.0 & $ -0.296\pm  1.270$ & $ -0.950\pm  1.243$ &  -145.3\\
 606 &     0.3466 &  -19.25 &    0.75 &    9.83 &     1.4 &    39.9 &     8.6 & $  0.536\pm  0.356$ & $  0.625\pm  0.340$ &   -78.1\\
 608 &     0.5036 &  -21.16 &    0.78 &   10.52 &     5.5 &    14.6 &    15.3 & $  1.567\pm  0.179$ & $  1.869\pm  0.186$ &   133.7\\
 609* &     0.4376 &  -19.77 &    0.45 &    9.28 &     7.7 &    25.3 &    76.9 & $  0.471\pm  0.034$ & $  0.419\pm  0.036$ &    70.0\\
 611 &     0.3516 &  -20.86 &    1.26 &   10.85 &     0.2 &    13.2 &     7.3 & $  0.730\pm  0.317$ & $  1.091\pm  0.324$ &    67.9\\
 612* &     0.3608 &  -19.09 &    0.16 &    9.00 &     5.5 &    12.2 &    29.0 & $  1.235\pm  0.080$ & $  0.968\pm  0.087$ &   -31.2\\
 613* &     0.7406 &  -21.06 &    0.61 &   10.11 &    11.0 &    26.2 &    17.2 & $  2.286\pm  0.142$ & $  2.135\pm  0.146$ &    89.7\\
 614 &     0.3636 &  -21.94 &    1.26 &   11.41 &     0.3 &    22.8 &     3.3 & $  1.816\pm  0.985^b$ & $ -0.378\pm  0.740^b$ &  -153.7\\
 617* &     0.4881 &  -21.27 &    0.97 &   10.65 &     2.6 &    37.9 &    15.4 & $  1.300\pm  0.189$ & $  1.043\pm  0.170$ &   166.3\\
 619 &     0.3524 &  -18.88 &    0.78 &    9.62 &     0.6 &    29.9 &     7.1 & $  0.197\pm  0.355$ & $ -0.188\pm  0.316$ &   102.1\\
 621 &     0.3574 &  -20.29 &    0.53 &    9.74 &     5.5 &    47.2 &     6.4 & $  0.718\pm  0.394$ & $  1.335\pm  0.419$ &   104.2\\
 623* &     0.8258 &  -20.51 &    0.57 &    9.76 &     4.4 &    43.3 &    16.3 & $  0.157\pm  0.146$ & $  0.017\pm  0.156$ &   -63.9\\
 624 &     0.3605 &  -20.46 &    0.82 &   10.29 &     2.7 &    21.5 &     7.6 & $  1.504\pm  0.403$ & $  1.493\pm  0.369$ &  -161.8\\
 625* &     0.6035 &  -19.77 &    0.55 &    9.44 &     1.9 &    37.1 &    14.5 & $ -0.046\pm  0.139$ & $ -0.079\pm  0.155$ &    29.0\\
 626 &     0.4249 &  -20.64 &    1.07 &   10.65 &     0.6 &    31.0 &     4.5 & $  1.008\pm  0.508$ & $  1.612\pm  0.550^b$ &    49.6\\
 1101 &     0.4072 &  -20.44 &    0.77 &   10.19 &     3.5 &   136.7 &     1.2 & $  0.851\pm  2.324$ & $  0.866\pm  2.600$ &   137.6\\
 1201* &     0.5653 &  -19.78 &    0.56 &    9.35 &     2.7 &    60.0 &     5.9 & $  0.558\pm  0.402$ & $  0.538\pm  0.415$ &    11.8\\
 1202 &     0.4082 &  -20.30 &    1.33 &   10.89 &     0.1 &   128.8 &     9.0 & $  0.040\pm  0.273$ & $ -0.085\pm  0.314$ &     1.8\\
 1205 &     0.3724 &  -20.92 &    1.26 &   11.03 &     0.1 &    65.2 &    11.9 & $ -0.340\pm  0.239$ & $ -0.027\pm  0.203$ &  -234.8\\
 1207 &     0.7423 &  -21.19 &    0.96 &   10.69 &     3.4 &   131.0 &    25.2 & $  2.038\pm  0.096$ & $  1.953\pm  0.100$ &    29.1\\
 1208* &     0.8357 &  -21.75 &    0.55 &   10.41 &    19.2 &    56.5 &    16.1 & $  1.119\pm  0.145$ & $  1.162\pm  0.136$ &   -23.3\\
 1209 &     0.3710 &  -21.01 &    0.96 &   10.68 &     2.3 &    94.1 &    14.3 & $  0.247\pm  0.179$ & $  0.404\pm  0.175$ &    76.1\\
 1210* &     0.5957 &  -21.16 &    0.99 &   10.68 &     3.5 &    55.5 &     7.6 & $  1.264\pm  0.331$ & $  0.761\pm  0.320$ &   -13.6\\
 1300 &     0.5851 &  -19.92 &    0.77 &    9.71 &     1.8 &   101.9 &    22.7 & $  0.269\pm  0.110$ & $ -0.073\pm  0.116$ &   -84.4\\
 1402 &     0.5418 &  -21.71 &    0.73 &   10.56 &    12.2 &    72.2 &    11.3 & $ -0.383\pm  0.220$ & $  0.145\pm  0.200$ &   -70.7\\
 1600 &     0.4971 &  -20.52 &    0.69 &   10.06 &     5.3 &    16.1 &    19.8 & $  2.005\pm  0.154$ & $  1.742\pm  0.151$ &   154.7\\
 1601* &     0.5359 &  -20.05 &    0.62 &    9.72 &     2.9 &    58.1 &    28.4 & $  2.529\pm  0.095^b$ & $  1.066\pm  0.100^b$ &     8.9\\
 1602 &     0.3643 &  -21.81 &    1.22 &   11.23 &     0.2 &    94.0 &    12.9 & $  0.316\pm  0.208$ & $  0.212\pm  0.198$ &    87.7\\
 1603 &     0.3669 &  -20.72 &    0.88 &   10.51 &     4.4 &    79.0 &    15.6 & $  1.617\pm  0.183^b$ & $  2.432\pm  0.211^b$ &   -51.6\\
 1604 &     0.4971 &  -20.52 &    0.69 &   10.06 &     5.3 &    33.1 &    10.6 & $  1.519\pm  0.257$ & $  1.610\pm  0.267$ &   130.0\\
 1605* &     0.3761 &  -21.51 &    0.94 &   10.89 &    13.9 &    86.3 &    36.7 & $  0.769\pm  0.073^b$ & $  1.128\pm  0.072^b$ &  -604.8\\
 1606* &     0.3717 &  -18.83 &    0.59 &    9.15 &     0.8 &    58.6 &     4.5 & $  0.673\pm  0.577$ & $  0.636\pm  0.527$ &   -75.2\\
 1608 &     0.4383 &  -19.86 &    0.65 &    9.57 &     1.7 &    81.4 &     2.3 & $ -1.315\pm  1.072$ & $  1.714\pm  1.052$ &   257.8\\
 1609 &     0.3552 &  -19.74 &    0.71 &    9.46 &     0.7 &    83.0 &     0.6 & $ -1.537\pm  4.421$ & $  1.108\pm  4.582$ &  -122.6\\
 1610* &     0.7309 &  -20.48 &    0.43 &    9.53 &     5.2 &    96.5 &    41.4 & $ -0.031\pm  0.058$ & $  0.048\pm  0.059$ &    50.3\\
 1611 &     0.3738 &  -18.89 &    0.52 &    9.49 &     1.4 &    92.6 &     3.1 & $  2.069\pm  0.760$ & $ -1.311\pm  0.865$ &    -0.7\\
 1612 &     0.3604 &  -19.10 &    0.76 &    9.40 &     0.7 &    99.6 &    11.7 & $ -0.266\pm  0.229$ & $ -0.749\pm  0.233$ &   -60.6\\
 1613 &     0.4399 &  -21.20 &    1.20 &   10.74 &     0.2 &    89.5 &     4.5 & $  1.982\pm  0.743^b$ & $  1.121\pm  0.648$ &   -71.5\\
\enddata
\tablecomments{Pair IDs marked with an asterisk indicate pairs with a broad-line AGN in the background.}
\tablenotetext{a}{Rest-frame photometric measurements are not available from \citet{Moustakas2013} for objects with blank entries in this column. }
\tablenotetext{b}{Marks transitions affected by blending with \ion{Fe}{2} absorption associated with the background galaxy, or with the Ly$\alpha$ forest. }
\end{deluxetable*}

\tabletypesize{\footnotesize}
\begin{deluxetable*}{llccccccc}
\tablewidth{290pt}
\tablecolumns{9}
\tablecaption{$W_{2796}$ Dependence on Intrinsic Galaxy Properties \label{tab.galpropsplit}}
\tablewidth{0pt}
\tablehead{   & & All  & \multicolumn{2}{c}{$\log M_*/M_{\odot}$}   &
  \multicolumn{2}{c}{SFR ($M_{\odot}~\rm yr^{-1}$)}  &
  \multicolumn{2}{c}{$\log \rm sSFR ~\rm / yr^{-1} $}} 
\startdata
\multicolumn{9}{c}{\underline{PRIMUS Sample}}\\
Subsample &  &  & $< 9.9$ & $> 9.9$ & $<2.5$ & $>2.5$ & $< -9.46$ & $> -9.46$\\
$R_{\perp} < 30$ kpc  & $N$ & 19  &    8 & 11                          &  6         &13                     &  8   &   11\\
                   & Median $W_{2796}$ (\AA)    &   1.30             &  0.90 & 1.54                  &  0.70& 1.73                      &0.50     &   1.30   \\
                                  & Probability         &       & \multicolumn{2}{c}{0.892} & \multicolumn{2}{c}{0.049}  & \multicolumn{2}{c}{0.222}\\
30 kpc $< R_{\perp} < 50$ kpc  & $N$ & 25  &    15  & 10                         &  13         &12                     &  14   &   11\\
                  & Median $W_{2796}$ (\AA)    &   0.27             &                  0.20  &          1.06         &  0.12 &          0.62           & 0.25    &    0.36 \\
                                 & Probability         &    0.003   &                           \multicolumn{2}{c}{0.006} &    \multicolumn{2}{c}{0.306}  &       \multicolumn{2}{c}{0.973}\\
\hline\\
\multicolumn{9}{c}{\underline{QSO-Galaxy Comparison Sample}}\\
Subsample &  &  & $< 10.1$ & $> 10.1$ & $<1.3$ & $>1.3$ & $< -10.0$ & $> -10.0$\\
$R_{\perp} < 30$ kpc  & $N$ & 19  &    13 & 6                        &     13     &         6            &  6   &   13\\
                   & Median $W_{2796}$ (\AA)    &     0.80          &                   0.87 & 0.75                  &  0.76   &      0.80                 &  0.71   &   0.87   \\
                                  & Probability         &       & \multicolumn{2}{c}{0.726} & \multicolumn{2}{c}{0.661}  & \multicolumn{2}{c}{0.792}\\
30 kpc $< R_{\perp} < 50$ kpc  & $N$ & 31  &   6   &          25               &      12     &       19           &   15  &  16 \\
                  & Median $W_{2796}$ (\AA)    &      0.43        &      0.20           &       0.55           &    0.31    &         0.71           &  0.39  &   0.48  \\
                                 & Probability         &   0.024    &             \multicolumn{2}{c}{0.017} &    \multicolumn{2}{c}{0.003}  &       \multicolumn{2}{c}{0.736}\\
\enddata
\tablecomments{Two-sample comparisons use Gehan's generalized Wilcoxon
  test of the probability that the two $W_{2796}$ distributions in question are drawn from the same parent population. }
\end{deluxetable*}

\clearpage

\appendix

\section{Keck/LRIS and VLT/FORS2 Background Galaxy Spectroscopy}\label{sec.appendix}
Figure~\ref{fig.primus_allmgii_p1} shows our spectroscopy of all
PRIMUS background sightlines in the region surrounding \ion{Mg}{2} in the
rest frame of the foreground galaxy.  The ID number of each pair is
indicated in the upper left corner of each panel.

\begin{figure*}
\begin{center}
\includegraphics[angle=0,width=6.4in,trim=0 40 0 0,clip=]{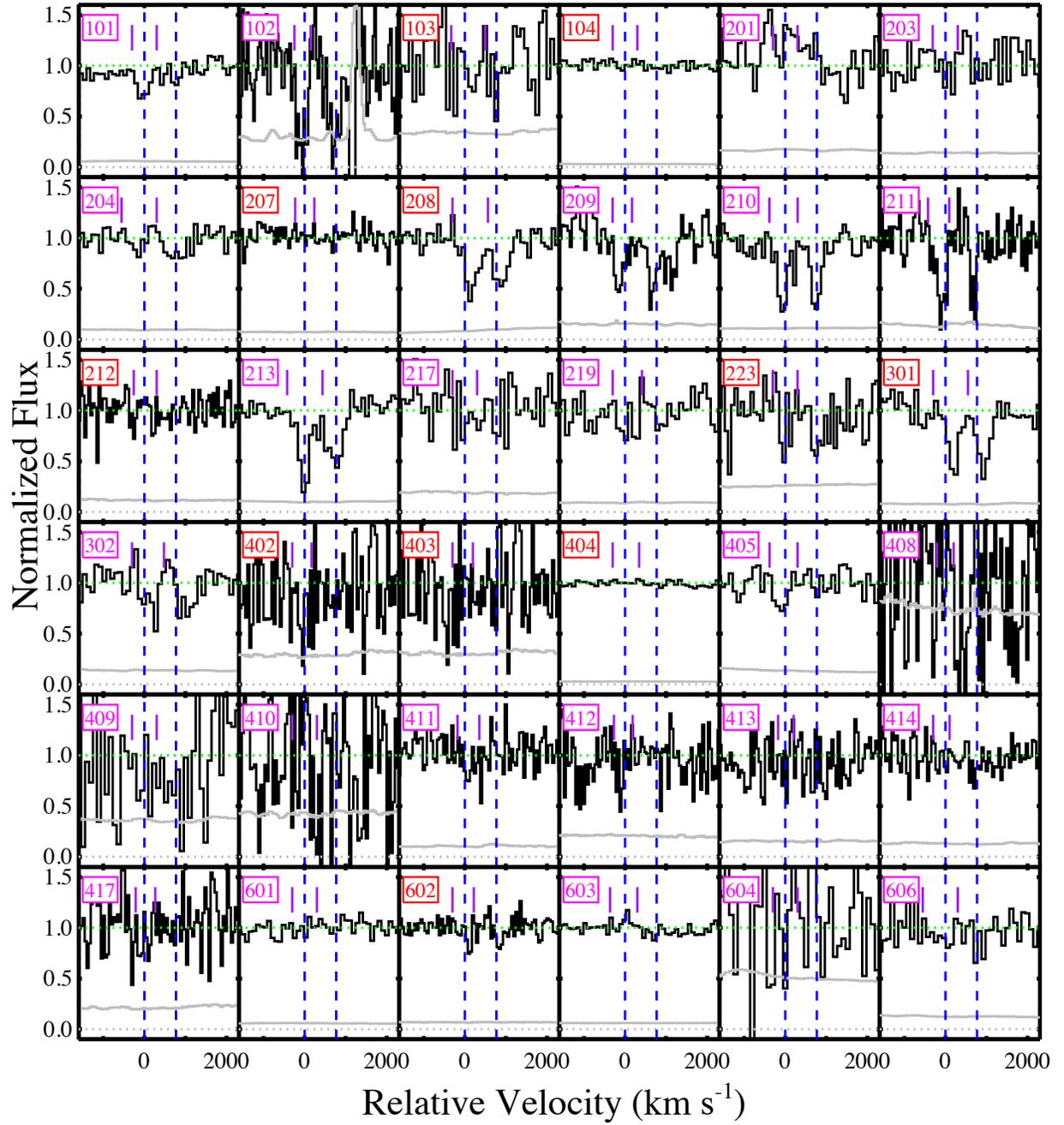}
\caption{Background galaxy spectroscopy covering the \ion{Mg}{2}
  $\lambda \lambda 2796, 2803$ doublet at the systemic velocity of the
  paired foreground galaxy.  The pair ID is printed in red for those pairs with b/g galaxies hosting
  broad-line AGN,  and is printed in magenta for the remaining pairs.  
 The relative velocity is $0\mkms$ at the wavelength of the
 \ion{Mg}{2} 2796 transition at 
 $z_{\rm f/g}$, with the blue vertical dashed
  lines indicating the velocities of both doublet transitions.  The gray
  histogram shows the error in each spectral pixel, and the green
  dotted curve marks the continuum level.  The two vertical purple
  hashes indicate the velocity range adopted in our calculation of the
  boxcar $W_{2796}$.
\label{fig.primus_allmgii_p1}}
\end{center}
\end{figure*}

\setcounter{figure}{17}
\begin{figure*}
\begin{center}
\includegraphics[angle=0,width=6.4in,trim=0 40 0 0,clip=]{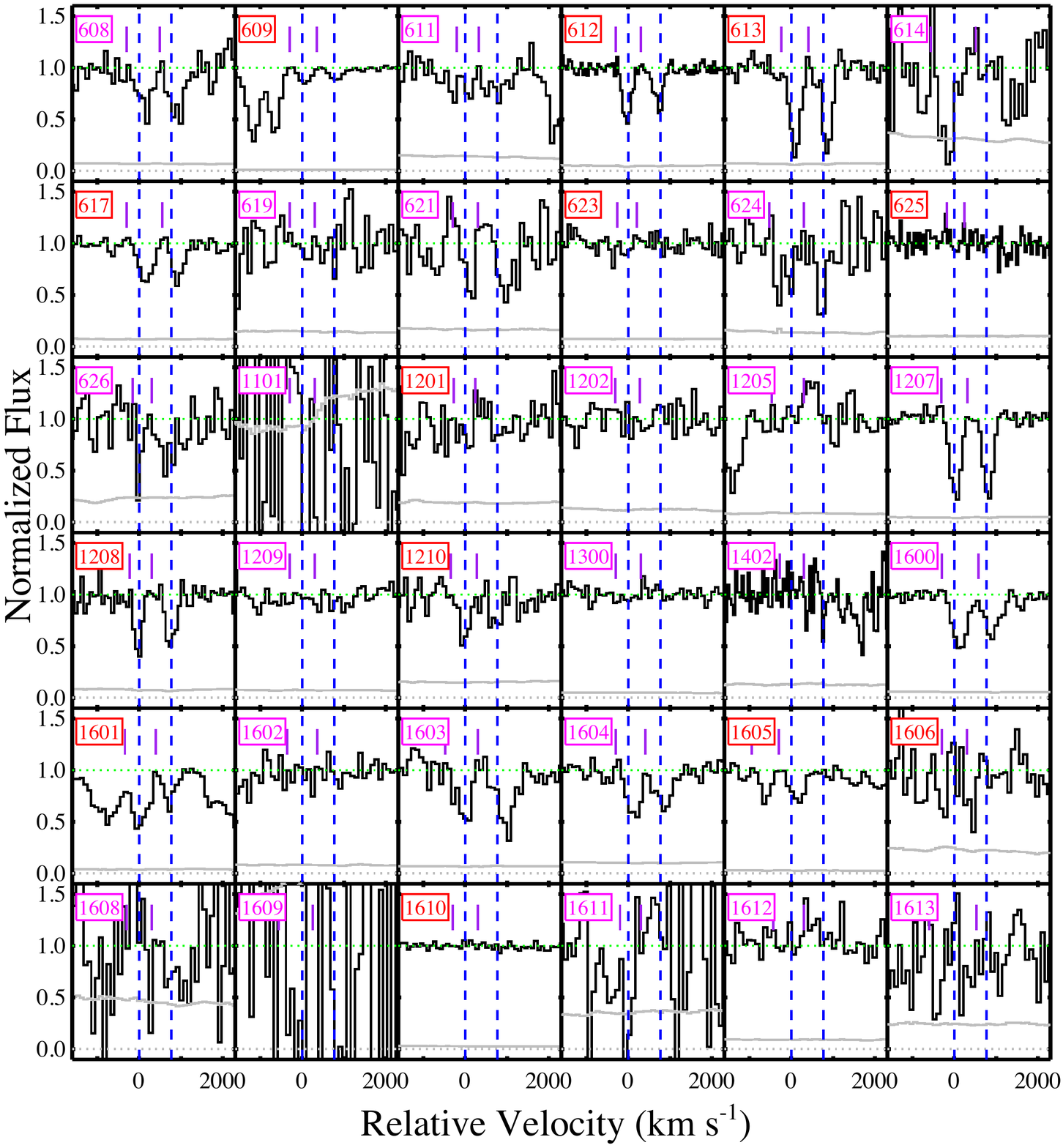}
\caption{ -- continued
\label{fig.primus_allmgii_p2}}
\end{center}
\end{figure*}

\end{document}